\documentclass[aps,reprint,amsmath,amssymb,nofootinbib,superscriptaddress]{revtex4-2}
\usepackage{graphicx}   
\usepackage{bm}
\usepackage{orcidlink}
\usepackage{braket}
\usepackage{booktabs}
\usepackage{float}
\usepackage[mathlines]{lineno}

\providecommand{\aap}{A\&A}
\providecommand{\apj}{ApJ}

\providecommand{\jcap}{JCAP}
\providecommand{\mnras}{MNRAS}
\providecommand{\physrep}{Phys.~Rep.}
\providecommand{\pra}{Phys.~Rev.~A}
\providecommand{\prd}{Phys.~Rev.~D}

\usepackage{xcolor}
\newif\ifrevhl \revhlfalse
\newcommand{\edited}[1]{\ifrevhl\textcolor{blue!60!black}{#1}\else#1\fi}

\newcommand{\be}{\begin{equation}}
\newcommand{\ee}{\end{equation}}
\newcommand{\bea}{\begin{eqnarray}}
\newcommand{\eea}{\end{eqnarray}}
\newcommand{\diff}[1]{d#1}

\newcommand{\ii}{\mathrm{i}}
\newcommand{\sachsvec}{X}
\newcommand{\sachssaddle}{X^{\scriptscriptstyle\mathrm{(sa)}}}
\newcommand{\Dsachs}{\mathcal{X}}
\newcommand{\RespField}{\Tilde{\mathcal{X}}}
\newcommand{\src}{\boldsymbol{\phi}}
\newcommand{\Dsrc}{\varphi}
\newcommand{\direc}{\hat{\Omega}}
\newcommand*\funcdiff{\mathop{}\!\mathrm{D}}
\newcommand{\funcprob}[1]{P\left[#1\right]}
\newcommand{\funccondprob}[2]{P\!\left[#1 \mid #2\right]}
\newcommand{\funcDiracD}[1]{\delta\left[#1\right]}
\newcommand{\Action}{S}
\newcommand{\ActionRef}{S_0}
\newcommand{\ActionInt}{S_{\rm int}}
\newcommand{\RespOp}{\mathcal{R}}
\newcommand{\CorrOp}{\mathcal{C}}

\newcommand{\cumulant}{\mathcal{K}}

\newcommand{\Texp}{\mathcal{T}\!\exp}
\newcommand{\sachi}{a}
\newcommand{\sachj}{b} 
\newcommand{\sachk}{c}
\newcommand{\sachl}{d}

\newcommand{\screenvec}{Z}

\begin{document}

\preprint{}   

\title{Statistical Field Theory for Weak Gravitational Lensing}

\author{Zheng Zhang\,\orcidlink{0000-0002-9154-2803}}
\thanks{Corresponding author.}
\email{zheng.zhang@manchester.ac.uk}
\affiliation{Jodrell Bank Centre for Astrophysics, University of Manchester, Manchester, M13 9PL, United Kingdom}

\author{Philip Bull\,\orcidlink{0000-0001-5668-3101}}
\affiliation{Jodrell Bank Centre for Astrophysics, University of Manchester, Manchester, M13 9PL, United Kingdom}
\affiliation{Department of Physics and Astronomy, University of the Western Cape, Cape Town 7535, South Africa}

\author{Chris Clarkson}
\affiliation{Astronomy Unit, School of Physical and Chemical Sciences, Queen Mary University of London, London E1 4NS, United Kingdom}
\affiliation{Department of Physics and Astronomy, University of the Western Cape, Cape Town 7535, South Africa}

\author{Andrina Nicola}
\affiliation{Jodrell Bank Centre for Astrophysics, University of Manchester, Manchester, M13 9PL, United Kingdom}

\begin{abstract}
    Standard weak-lensing calculations treat lensing as a linear remapping of the matter field along the line of sight. We instead formulate lensing as a stochastic field theory for the Sachs optical scalars, driven by random Ricci-focusing and Weyl-shearing fields.
    With convergence and shear expressed linearly in the Sachs-scalar fluctuations, the resulting path integral generates a diagrammatic expansion for their $n$-point correlation functions, organised into linear response, nonlinear propagation, and driving-field cumulants. The conventional calculation emerges as the lowest-order, linear-propagation limit.
    Beyond it, nonlinear Sachs evolution couples to driving-field non-Gaussianity, mixing the matter cumulant hierarchy into the lensing hierarchy.
    A selection rule governs the couplings: an $n$-point observable receives a direct contribution from the $n$-point driving-field cumulant, and its leading hierarchy-mixing correction from the $(n+1)$-point cumulant via one nonlinear Sachs interaction, with higher cumulants entering only at higher order.
    The two-point function, for instance, receives corrections from three-point cumulants of Ricci focusing and Weyl shearing through nonlinear Sachs evolution.
    Rather than a restrictive approximation scheme, the formalism is a paradigm shift: a unified framework naturally accommodating path corrections, higher-order matter statistics, stochasticity, and small-scale effects.
\end{abstract}

\maketitle

\section{Introduction}
\label{sec: intro}

Weak gravitational lensing converts the tidal history of light rays into
cosmological observables.  Current cosmic-shear analyses already constrain
large-scale structure at sub-percent statistical precision~\citep{asgari2021kids,amon2022dark},
and at Stage-IV precision~\citep{ivezic2019lsst,laureijs2011euclid,spergel2015wide} the accuracy of the
light-propagation model becomes a limiting systematic in its own right.  The
standard calculation treats lensing as a projection: matter correlations are
filtered along the line of sight and read out as convergence and shear
statistics~\citep{bartelmann2001weak,kilbinger2015cosmology,mandelbaum2018weak}.
In the Born limit this gives the familiar hierarchy in which the
convergence two-point function is derived from the matter power spectrum, the
convergence bispectrum inherits the matter bispectrum, and so on.

This projection picture is useful, but it packages together approximations that
must be refined for next-generation wide, deep, high-precision weak-lensing
surveys such as LSST, Euclid, and Roman~\citep{ivezic2019lsst,laureijs2011euclid,spergel2015wide}.
Post-Born calculations include corrections from evaluating lensing fields along
perturbed rays and from coupling successive lensing
distortions (lens-lens coupling)~\citep{cooray2002second,dodelson2006reduced,
bernardeau2010full,krause2010weak,PrattenLewis2016,fabbian2018cmb,marozzi2018cmb}; 
higher-order perturbation theory and
phenomenological matter models account for non-Gaussian structure
formation~\citep{bernardeau2010full,fry1994gravity,scoccimarro2001bispectrum,
jeong2009primordial,halder2021integrated}; beyond-Limber
treatments restore long line-of-sight correlations~\citep{LemosEtAl2017,FangEtAl2020,
KitchingNonLimber2017}. These ingredients are usually developed as separate
corrections to the same baseline projection calculation, 
often quantified one assumption at a time~\citep{2018MNRAS.477..741C}.
High-fidelity ray tracing can combine several of them
numerically~\citep{HilbertEtAl2009RayTracing,PetriEtAl2013,fabbian2018cmb,
TakahashiEtAl2017Nbody}, but its accuracy is ultimately bounded by finite
resolution and volume, sampling variance, and lens-plane artifacts, at
substantial computational cost; and it does not by itself identify which
physical mechanism controls a given residual. Our aim is to determine how
nonlinear optical evolution combines with non-Gaussian gravitational-field
statistics to generate lensing correlations, while keeping their contributions
explicit.

We build such a framework from the Sachs optical equations~\citep{sachs1961gravitational}, the exact
relativistic equations that govern how the cross-section of a thin bundle of
neighbouring light rays is focused and sheared as it propagates through an
arbitrary spacetime. We treat the beam as a test congruence: spacetime curvature
drives its optical evolution, with no feedback on matter or geometry. For a given
metric, the optical expansion and shear evolve deterministically. We prescribe
the Ricci focusing scalar $\Phi_{00}$ and the Weyl shearing scalar $\Psi_0$ as
random driving fields along the rays, making the optical problem stochastic
throughout propagation. We
promote the Sachs trajectory to a path
integral~\citep{MartinSiggiaRose1973,Janssen1976,Krommes2002Review}, which
encodes the nonlinear Sachs evolution together with the statistics
of the driving fields in a single probability functional, so that in principle the statistics
of observables reconstructed from the Sachs trajectory follow from it.
The screen-space Jacobi map carries a small separation of observed directions
to the transverse separation of neighbouring rays at the source. Its distortion
relative to the reference beam defines convergence and shear. We use linearized
relations between these observables and the Sachs scalars
(Section~\ref{subsec: from sachs to wl kernels}), while retaining nonlinear Sachs
evolution in the path integral.
The resulting diagrammatic expansion resolves every $n$-point
correlation function ($n$PCF) of convergence and shear into three ingredients:
the linear response, nonlinear propagation, and the connected
cumulants of the driving fields.  For later reference we label these last two
mechanisms $F$ (nonlinear propagation) and $K^{(n)}$ (an $n$-point cumulant of
the driving fields); in field-theory language they play the role of
interaction vertices.

The statistics of these driving fields depend on the matter model.
Standard perturbation theory predicts matter correlations from gravitational
dynamics~\citep{Bernardeau2002Review}. Effective descriptions account for
unresolved small-scale physics through effective stresses, including stochastic
contributions~\citep{2012JCAP...07..051B}.
Vlasov perturbation theory incorporates velocity dispersion and higher
cumulants of the local velocity distribution generated by orbit
crossing~\citep{2023PhRvD.107f3539G}. These cumulants describe particle
velocities at a given position. These matter models provide correlations of
matter and gravitational fields from which the Ricci and Weyl driving-field
statistics along light paths can be constructed.

Related diagrammatic and path-integral calculations of large-scale structure
predict matter correlations by evolving random initial density and velocity
fields under deterministic gravitational dynamics~\citep{2006PhRvD..73f3519C,
2008A&A...484...79V}. These examples use Gaussian initial statistics, with
extensions to primordial non-Gaussianity~\citep{2007PhRvD..76h3517I}.
Our optical problem has fixed observer boundary conditions for the Sachs fields
and random Ricci and Weyl driving fields acting throughout propagation. For
each driving-field realisation, Sachs evolution is deterministic; averaging
over their joint histories predicts correlations of convergence and shear.
The driving fields may correlate different observed directions and affine
parameters and need not be Gaussian. The companion paper~\citep{SFTwick2026} develops the general
prescription adopted here to calculate how such driving statistics are
transformed by nonlinear evolution into the statistics of the driven fields.
It also automates the resulting diagrammatic calculations.
The perturbative order of a matter model and the order of our optical diagrams
thus refer to different expansions.

At leading order, this optical expansion recovers the conventional
lensing-efficiency projection. The key identity is that
propagating an $n$-point driving-field cumulant through the free Sachs response,
with no nonlinear Sachs ($F$) insertion, is
equivalent to weighting that cumulant by $n$ copies of the lensing-efficiency
window, the familiar geometric weight that sets how strongly structure at a
given distance lenses the source.  Thus the leading branch of the $n$-point observable is the usual
line-of-sight projection of the $n$-point driving-field cumulant: the free
two-point sector reproduces the standard two-point kernel, the leading
three-point sector projects the driving-field three-point cumulant, and so forth.

With convergence and shear expressed as linear integrals of Sachs-scalar
fluctuations, nonlinear Sachs evolution and gravitational
non-Gaussianity combine to mix the statistical hierarchies, so that the order of
a lensing observable and the order of the driving-field cumulant that sources it
need not coincide: their interplay is governed by a selection rule.  For an
$n$-point lensing observable, the direct cumulant
contribution comes from the $n$-point driving-field cumulant; the first
hierarchy-mixing correction is sourced by the $(n+1)$-point cumulant through one
nonlinear Sachs correction, and higher cumulants are absent at that order.
Equivalently, a driving-field three-point cumulant contributes directly to a
three-point lensing observable, but enters a two-point observable through
one nonlinear interaction vertex, $F$.  The observable hierarchy and the
Ricci/Weyl driving-field hierarchy are therefore mixed, rather than tied by a
strict one-to-one projection, with each step between them mediated by a
nonlinear Sachs correction.  This nonlinear-Sachs hierarchy mixing is present
regardless of the Born limit or the perturbative matter setup.

As a demonstration, we work out this structure for \edited{the two-point
functions of lensing observables}.  At Order-$0$ the formalism reproduces the
conventional weak-lensing kernels for the four two-point functions considered.
At the next order \edited{each of them acquires} two physically distinct
\edited{corrections}: FF, from nonlinear optical propagation, and
FK, in which Ricci/Weyl three-point cumulants enter the two-point statistic through
one nonlinear Sachs correction.  \edited{Under the Limber approximation adopted in the demonstration, the
three-point cumulant enters in a collapsed configuration with two coincident
legs, coupling large separations to small-scale driving-field fluctuations.
Its amplitude depends on the matter model, source redshift, angular scale,
and multipole cutoff (Section~\ref{subsec: cutoff}). For the linear spectra
and tree-level bispectrum used here, at $z_s=5$ and
$\ell_{\max}=15360$ the FK contribution raises the convergence correlation
by $1.5$--$1.6\%$ over the sampled separations within $2'$--$12'$, the band
considered in the discussion of current cosmic-shear
analyses~\citep{asgari2021kids,amon2022dark}. This increases from
$0.86$--$0.94\%$ at $z_s=1$ (Fig.~\ref{fig: multiz kappa}).}

The demonstration scenario adopted in this paper is deliberately narrow: scalar
perturbations with $\Phi=\Psi$, unperturbed path, a linear matter power spectrum,
and a three-point cumulant evaluated from the tree-level bispectrum.  
The matter model determines the curvature statistics. Relaxing the
unperturbed-path approximation changes both
where the curvature is evaluated and how it is projected onto the ray and screen.
At the level of the Sachs equations, these effects are encoded in the joint
statistics of the resulting driving fields as functions of observed direction
and affine parameter. The example neglects effects including
shell crossing~\citep{2023PhRvD.107f3539G} and
nonlinear baryonic feedback~\citep{vanDaalen2011}, so its numerical corrections
illustrate the mechanism without providing calibrated small-scale predictions.
We assume that each source
is seen as a single, weakly distorted image. Caustics and multiple imaging
invalidate this single-image description and cannot be accounted for by changing
the driving-field cumulants alone.

In particular, the non-Gaussianity leakage into the 2PCF depends on the
collapsed three-point cumulant, where the tree-level bispectrum
underestimates the nonlinear small-scale clustering\edited{; a nonlinear
evaluation, however, requires more than substituting a nonlinear bispectrum.
It calls for $N$-body simulations with baryonic corrections, or for a
separate-universe treatment of the squeezed response \citep{wagner2015squeezed, li2014supersample, barreira2017responses}; and it calls for a scale
below which the driving field is treated as unresolved, together with the
counterterms that smoothing the field generates in the nonlinear propagation.
We leave both to future work.}
Within the single-image regime, perturbed-path
corrections~\citep{krause2010weak,HilbertEtAl2009RayTracing},
nonlinear matter statistics~\citep{Bernardeau2002Review,scoccimarro2001bispectrum},
primordial non-Gaussianity~\citep{jeong2009primordial}, and small-scale
physics~\citep{vanDaalen2011} modify the driving-field statistics entering the
Sachs evolution. For perturbed paths, these must be the statistics of curvature
projected along the corrected rays and onto their screen bases. Changes to the mapping from Sachs fields
to measured observables must be treated separately.

Because the starting point is the Sachs system rather than a
cosmology-specific projection operator, the same construction can also be
adapted to other light-cone observables, including the lensing of the
stochastic gravitational-wave background in the geometric-optics
limit~\citep{DiegoEtAl2021}.

This paper is organised as follows.  
Section~\ref{sec: sachs dynamics} formulates the Sachs evolution as a centred
stochastic dynamical system.  Section~\ref{sec: path integral} builds the
path integral for the Sachs equations and derives the diagrammatic rules and
their selection rule.
Section~\ref{sec: cosmological setup} specialises the driving fields and the
free Sachs response to a perturbed Friedmann--Lemaitre--Robertson--Walker
(FLRW) spacetime in Poisson gauge and establishes the lensing-kernel
correspondence. Section~\ref{sec: insights} presents the Order-0 and FF/FK
two-point results. Section~\ref{sec: conclusion} summarises the worked
example and its three structural insights.  The appendices collect conventions,
Jacobi-map relations, driving-field third cumulants, and numerical validations.

\section{Nonlinear Evolution of Null Geodesics Congruence}
\label{sec: sachs dynamics}

\begin{figure}[t]
    \centering
    \includegraphics[width=\linewidth]{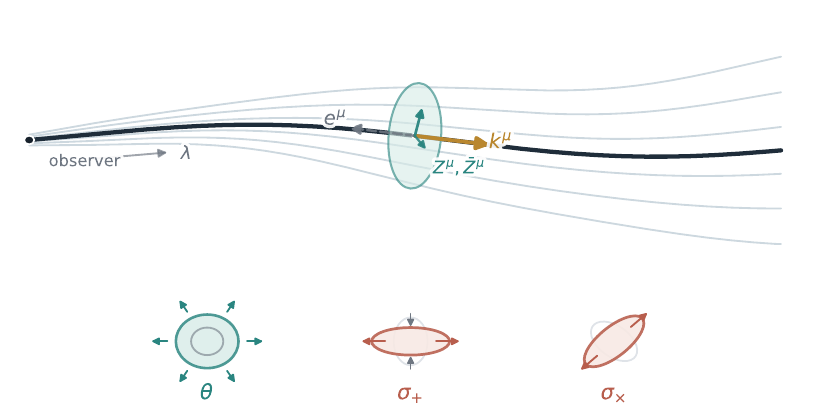}
    \caption{Schematic of a null geodesic congruence evolved along a
    past-directed affine parameter, the local Newman--Penrose tetrad, and the
    Sachs scalar degrees of freedom $(\theta,\sigma_+,\sigma_\times)$ on the
    screen.}
    \label{fig:null-geodesic-congruence}
\end{figure}

We describe the observed light beam as a two-dimensional congruence of null
geodesics~\citep{sachs1961gravitational,1967ApJ...150..737G,ellis2012relativistic}, labelled by an affine parameter $\lambda$ and an observed sky
direction $\direc$. At each point on a ray we choose a Newman--Penrose null
tetrad~\citep{1962JMP.....3..566N}
\begin{equation}
    \{k^\mu, e^\mu, \screenvec^{\mu}, \Bar{\screenvec}^{\mu}\},
\end{equation}
where $k^\mu$ is tangent to the ray, $e^\mu$ is the auxiliary null direction,
and $\screenvec^\mu$, $\Bar{\screenvec}^\mu$ span the screen plane with
$k^\mu e_\mu=-1$ and
$\screenvec^\mu\Bar{\screenvec}_\mu=1$. The complete construction and the
spin-weight conventions are collected in
Appendix~\ref{append: NP formalism}; here we keep only the ingredients used
below. Figure~\ref{fig:null-geodesic-congruence} summarises the local
geometry and the optical degrees of freedom used in this section.

The two screen-projected contractions of $\nabla_\nu k_\mu$ are the
Newman--Penrose spin coefficients
\begin{align}
    \rho & = -\nabla_\nu k_\mu \screenvec^{\mu} \Bar{\screenvec}^\nu,
    &
    \sigma & = \nabla_\nu k_\mu \screenvec^{\mu} \screenvec^{\nu}.
\end{align}
For an affinely parametrised congruence with a parallel-transported screen
basis, their propagation equations are~\citep{sachs1961gravitational}
\bea
\frac{\diff{\rho}}{\diff{\lambda}} & = \rho^2 + \sigma \Bar{\sigma} - \Phi_{00}, \\
\frac{\diff{\sigma}}{\diff{\lambda}} & = (\rho + \Bar{\rho}) \sigma  + \Psi_{0}.
\eea
The two source fields are the Ricci focusing scalar and the Weyl shear scalar,
\begin{align}
    \Phi_{00}
    &= -\frac{1}{2} R_{\mu \nu } k^{\mu} k^{\nu},
    \label{eq: Phi 00 from Ricci} \\
    \Psi_{0}
    &=
    -C_{\alpha \beta \gamma \delta} k^\alpha \screenvec^{\beta}
    k^\gamma \screenvec^{\delta}
    =
    -R_{\alpha \beta \gamma \delta} k^\alpha \screenvec^{\beta}
    k^\gamma \screenvec^{\delta}.
    \label{eq: Psi0}
\end{align}
Here $C_{\alpha\beta\gamma\delta}$ is the Weyl tensor. The second equality in
Eq.~(\ref{eq: Psi0}) follows because the Ricci and trace parts of the Riemann
tensor vanish under the contraction
$k^\alpha \screenvec^\beta k^\gamma \screenvec^\delta$.
In natural units, the Einstein equations also give
\begin{equation}
    R_{\mu \nu } - \frac{1}{2} R g_{\mu \nu} = 8 \pi G \, T_{\mu\nu}
    \quad\Longrightarrow\quad
    \Phi_{00} = -4 \pi G \, T_{\mu\nu}\, k^{\mu} k^{\nu},
    \label{eq: Phi00}
\end{equation}
where the $R\,g_{\mu\nu}$ contraction drops out because $k^\mu$ is null.
These equations provide the geometric-to-statistical bridge used throughout
the paper: $\Phi_{00}$ drives focusing, while the real and imaginary parts
of $\Psi_0$ control the two shear polarisations. Section~\ref{sec: cosmological setup}
evaluates these source fields explicitly in Poisson gauge.

\subsection{Field Equations of Sachs Scalars}

The complex scalars $\rho$ and $\sigma$ are in one-to-one correspondence with the more 
familiar real-valued optical scalars through the mapping
\begin{align}
    \rho & = -(\theta + \ii \, \omega) \, , &
    \sigma &= \sigma_+ + \ii \, \sigma_{\times}
\end{align}
where $\{\theta, \sigma_+, \sigma_{\times}, \omega\}$ are the Sachs (or optical)
scalars. Geometrically, $\theta$ measures the expansion of the congruence,
while the two real shear components describe the elliptic distortion of the
beam cross-section: $\sigma_+$ stretches and compresses along the screen-basis
axes $\mathbf{x}$ and $\mathbf{y}$, and $\sigma_{\times}$ gives the
corresponding distortion along axes rotated by $45^{\circ}$. The remaining
scalar $\omega$ is the optical twist of the null-ray congruence, the
antisymmetric part of its local screen-projected deformation rate. Together, these scalars
form the basis of the standard weak-lensing formalism.

In terms of the optical scalars, the Newman--Penrose field equations reduce to
the familiar Sachs equations. Concretely, we view the congruence as a
two-dimensional field of null rays,
each labelled by its sky direction $\direc$ and evolving along its affine parameter 
$\lambda$. On the smooth part of the observer's past light cone, the rays
are normal to wavefronts and the optical twist vanishes by Frobenius' theorem,
$\omega=0$~\citep{1984ucp..book.....W}. The Sachs (or Raychaudhuri) system
then governs the evolution of the three non-trivial components:
\bea
\Dot{\theta} &=& - \theta^2 - \sigma_+^2 - \sigma_{\times}^2 + \Phi_{00}, \\[2pt]
\Dot{\sigma_+} &=& - 2 \theta \, \sigma_+ + \mathcal{W}_1, \\[2pt]
\Dot{\sigma_{\times}} &=& - 2 \theta \, \sigma_{\times} + \mathcal{W}_2, 
\eea
where the dot denotes differentiation with respect to $\lambda$, $\Phi_{00}$ is 
the Ricci focusing scalar introduced above, and $\mathcal{W}_{1,2}$ denote the 
real and imaginary parts of $\Psi_0$.

The Sachs equations combine quadratic optical couplings with the Ricci and
Weyl driving fields, whose statistics encode the matter and curvature
fluctuations along and across the light paths.
For convenience, we collect the relevant fields into the vectors
\be
\sachsvec \equiv (\theta, \sigma_+, \sigma_{\times}),
\quad
\src \equiv (\Phi_{00}, \mathcal{W}_1, \mathcal{W}_2).
\ee
In this notation, the evolution equations take the compact form
\be
\Dot{\sachsvec}_\sachi (\direc, \lambda)
=
 F_{\sachi \sachj \sachk} \, \sachsvec_\sachj(\direc, \lambda) \, \sachsvec_\sachk(\direc, \lambda) + \src_\sachi(\direc, \lambda) 
\label{eq: dot sachsvec}
\ee
where $F_{\sachi \sachj \sachk}$ is a constant coupling array whose non-zero entries 
are listed in Table~\ref{tab: Fabc}. Throughout the remainder of this paper we adopt 
the convention that repeated Latin indices labelling the three field components are 
implicitly summed over $\{1,2,3\}$.
\begin{table}[H]
    \centering
    \begin{tabular}{cccccc}
        \toprule
        $\sachi \sachj \sachk$ & 111 & 122 & 133 & 212 & 313\\
        \midrule
        $F_{\sachi \sachj \sachk}$ & -1 & -1 & -1 & -2 & -2\\
        \bottomrule
    \end{tabular}
    \caption{Entries of $F_{\sachi \sachj \sachk}$ that are not zero.}
    \label{tab: Fabc}
\end{table}

To connect this local optical evolution to image distortions, we first
distinguish the different notions of rotation.

\subsubsection*{Optical twist and other notions of rotation}
Optical twist, image rotation, matter vorticity and screen-basis rotation
describe different quantities despite their similar terminology.
As noted above, optical twist vanishes on the observer's smooth light cone;
nonzero optical twist requires a congruence that is not normal to a family
of wavefronts.

Image rotation is defined by the antisymmetric part of the accumulated
screen-space Jacobi map relative to parallel-transported screen bases.
Zero optical twist does not require this map to be symmetric. The first-order
Jacobi map is symmetric even for general linear
perturbations~\citep{2022JCAP...08..052F}, but scalar lens-lens coupling can
generate image rotation at second order in gravitational
perturbations~\citep{cooray2002second}. This order counts powers of the
gravitational fields, whereas our Order-0 and Order-2 labels count $F$ and $K$
insertions in the two-point functions.

Matter vorticity describes the rotation of a matter velocity field. For
collisionless dark matter, a coarse-grained mean velocity field can develop
vorticity after shell crossing~\citep{2009PhRvD..80d3504P}. This is distinct
from optical twist: matter evolution changes the Ricci and Weyl driving
fields, while the observer's smooth light-cone congruence remains normal to
wavefronts.

Screen-basis rotation concerns the orientation of the screen frame rather
than the deformation of the beam. The standard Sachs and screen-space Jacobi
equations do not by themselves determine the basis orientation relative to a
chosen reference frame; this requires tracking the basis transport
separately~\citep{2022JCAP...08..052F}. We do not evaluate it in this work.

\subsubsection*{Nonlinear Sachs evolution and conversion to observables}
The distinction between local deformation rates and accumulated image
distortions also helps separate two sources of nonlinearity: the quadratic
couplings in the Sachs equations and the relations that convert optical
deformation rates into convergence and shear. Here nonlinear propagation
refers to the former. Our demonstration retains these Sachs couplings but
linearizes the conversion to observables.
For convergence,
\begin{equation*}
    \kappa(\direc,\lambda_s)
    = -\int_0^{\lambda_s}\delta\theta(\direc,\lambda')\,\diff{\lambda'}
      + \mathcal{O}(\varepsilon^2),
\end{equation*}
where $\delta\theta=\theta-\theta^{\scriptscriptstyle\mathrm{(sa)}}$, and
$\varepsilon$ counts powers of deformation-rate fluctuations in the observable
map. Quadratic and higher-order contributions to these observable relations
are omitted in this demonstration. They include both symmetric
convergence/shear corrections and antisymmetric image rotation. The
path-ordered reconstruction in
Section~\ref{subsec: from sachs to wl kernels} displays the quadratic terms explicitly. Our example therefore
retains nonlinear Sachs evolution with a linearized observable map and does
not provide a complete second-order weak-lensing calculation.
The reference expansion $\theta^{\scriptscriptstyle\mathrm{(sa)}}$ is defined
below; Section~\ref{subsec: from sachs to wl kernels} gives the linear convergence and
shear relations in detail.

\subsection{Mean Field Expansion}

To organise a perturbative treatment of Eq.~(\ref{eq: dot sachsvec}), we first
isolate the evolution driven by the mean source. We solve the nonlinear Sachs
equations with the ensemble-mean driving field $\Bar{\src}(\lambda)$ and the
same observer boundary conditions as the stochastic Sachs fields:
\be
\Dot{\sachsvec}_{\sachi}^{\scriptscriptstyle\mathrm{(sa)}} (\lambda)
=
F_{\sachi \sachj \sachk} \, \sachssaddle_{\sachj}(\lambda) \, \sachssaddle_{\sachk}(\lambda) + \Bar{\src}_{\sachi}(\lambda) .
\label{eq: dot sachssaddle}
\ee
The superscript $\mathrm{(sa)}$ indicates that the source is averaged before
solving the nonlinear equations, so $\sachssaddle$ need not equal the ensemble
mean of $\sachsvec$. Statistical isotropy and the observer boundary conditions
make this solution independent of sky direction. In the flat FLRW application
below, it is the homogeneous optical solution with zero shear and background
distance $\bar D=a\chi$ (Appendix~\ref{append: D equals a chi}).

We define fluctuations by subtracting this solution and the mean driving field:
\bea
\Dsachs(\direc, \lambda) &=& \sachsvec(\direc, \lambda) - \sachssaddle(\lambda), \\
\Dsrc(\direc, \lambda) &=& \src(\direc, \lambda) - \Bar{\src}(\lambda) .
\eea
In the flat FLRW application, the first subtraction removes the nonzero
homogeneous optical expansion, while the second removes the mean
Ricci-focusing source.
Subtracting Eq.~(\ref{eq: dot sachssaddle}) from Eq.~(\ref{eq: dot sachsvec}) yields the evolution equation for $\Dsachs$:
\be
\Dot{\Dsachs}_{\sachi} = A_{\sachi \sachj}\, \Dsachs_{\sachj}
+ F_{\sachi \sachj \sachk}\, \Dsachs_{\sachj} \Dsachs_{\sachk}
+ \Dsrc_{\sachi}
\label{eq: centred SPDE}
\ee
where
\be
A_{\sachi \sachj } (\lambda)   = (F_{\sachi \sachj \sachk} + F_{\sachi \sachk \sachj}) \,  \sachssaddle_{\sachk}(\lambda) . 
\ee
The Sachs fluctuation vanishes at the observer, and the driving-field
fluctuation has zero mean. The linear evolution set by $A$, together with the
source two-point cumulant, defines the Gaussian reference used in
Section~\ref{sec: path integral}. The quadratic Sachs term and higher source
cumulants are treated as interactions.

For an isotropic cosmological mean source, the Weyl components vanish and
$\Bar{\src}(\lambda) = \bigl(\bar{\Phi}_{00}(\lambda),\, 0,\, 0\bigr)$.
The observer conditions then give zero shear for the source-averaged solution,
and the linear coefficient takes the diagonal form
\be
    \label{eq: explicit A matrix}
    A_{\sachi \sachj } (\lambda)
    =
    -2 \, \delta_{\sachi \sachj } \, \theta^{\scriptscriptstyle\mathrm{(sa)}}(\lambda),
\ee
where $\delta_{\sachi \sachj }$ is the Kronecker delta.
Appendix~\ref{append: D equals a chi} gives the Jacobi equation and vertex
conditions underlying $\theta^{\scriptscriptstyle\mathrm{(sa)}}$.

\section{Path Integral of Sachs Trajectory}
\label{sec: path integral}

We now specialise the companion formulation~\citep{SFTwick2026} to the
centred Sachs equation [see Eq.~(\ref{eq: centred SPDE})].
The response-field construction has also been used in path-integral
calculations of matter correlations~\citep{2008A&A...484...79V,2007PhRvD..76h3517I}.
Here, averaging over the prescribed driving fields gives the probability
functional of the Sachs fields. The linear evolution and two-point driving
cumulant define the Gaussian reference, whose correlation and response
propagators supply the diagram lines. Nonlinear Sachs evolution and third
and higher driving cumulants supply the interaction vertices.

\subsection{Probability Functional for Stochastic Light Propagation}
\label{subsec: probability functional}

Equation~(\ref{eq: centred SPDE}) describes the evolution of the Sachs fields
along the affine parameter, with sky direction labelling the rays.
For a given realisation of the Ricci and Weyl driving fields and fixed observer
boundary conditions, this evolution is deterministic. Its stochastic character
comes from the ensemble of externally prescribed driving fields, whose values
can be correlated across rays and affine parameters. We enforce this
deterministic evolution in the path integral through a functional $\delta$-constraint:
\be
    \funccondprob{\Dsachs}{\Dsrc}
    = \prod_{\sachi}
    \funcDiracD{\Dot{\Dsachs}_{\sachi} -  A_{\sachi \sachj} \, \Dsachs_{\sachj}
    - F_{\sachi \sachj \sachk} \, \Dsachs_{\sachj} \, \Dsachs_{\sachk} - \Dsrc_{\sachi}} .
\ee
For brevity, we suppress overall normalisation factors, which cancel against the 
partition function in any expectation value.

The unconditional probability functional for the $\Dsachs$-trajectory then 
follows by marginalising over the driving field $\Dsrc$,
\be
    \funcprob{\Dsachs}
    = \int \funccondprob{\Dsachs}{\Dsrc} \funcprob{\Dsrc} \funcdiff{\Dsrc}.
    \label{eq: prob Dsachs 1}
\ee
The $\delta$-functional in $\funccondprob{\Dsachs}{\Dsrc}$ is made tractable by trading it for an 
auxiliary functional integral through its Fourier representation. Equation~(\ref{eq: prob Dsachs 1}) 
thereby takes the form
\be
\begin{split}
    \funcprob{\Dsachs}
    &= \int \funcdiff{\Dsrc} \funcdiff{\RespField} \,
    \funcprob{\Dsrc} \,
    \exp\biggl[
    - \int \ii \, \RespField_{\sachi} \\
    &\qquad \times
    \left( \Dot{\Dsachs}_{\sachi} - A_{\sachi \sachj}\Dsachs_{\sachj} - F_{\sachi \sachj \sachk}\Dsachs_{\sachj}\Dsachs_{\sachk} - \Dsrc_{\sachi}\right)
    \diff{\direc} \diff{\lambda}
    \biggr] ,
\end{split}
\ee
where $\RespField_{\sachi}$ is the so-called response field, conjugate to the full argument 
of the $\delta$-functional, namely $\bigl(\Dot{\Dsachs}_{\sachi} - A_{\sachi \sachj}\Dsachs_{\sachj} 
- F_{\sachi \sachj \sachk}\Dsachs_{\sachj}\Dsachs_{\sachk} - \Dsrc_{\sachi}\bigr)$.
The remaining functional integral over the stochastic driving field $\Dsrc$ is precisely the cumulant 
generating function (CGF) of $\Dsrc$ evaluated at $\ii\RespField$, so that the probability functional 
reduces into
\be
    \label{eq: trajectory probability}
    \funcprob{\Dsachs}
    = \int \funcdiff{\RespField} \,
    e^{- \Action[\Dsachs, \RespField]}
\ee
where $\Action$ is the `action' given by
\be
\begin{split}
    \Action[\Dsachs, \RespField]
    &=
    \int \ii \, \RespField_{\sachi}
    \left( \Dot{\Dsachs}_{\sachi}
    - A_{\sachi\sachj}\Dsachs_{\sachj}
    - F_{\sachi\sachj\sachk}\Dsachs_{\sachj}\Dsachs_{\sachk}\right)
    \diff{\direc} \diff{\lambda} \\
    &\quad
    -
    W_{\Dsrc}[\ii \RespField],
\end{split}
    \label{eq: action 1}
\ee
and $W_{\Dsrc}$ is the CGF of the stochastic driving field,
\be
    W_{\Dsrc}[\ii \RespField] = \sum_{n=2}^{\infty} W_{\Dsrc}^{\scriptstyle{(n)}}[\ii \RespField]
\ee
where the sum starts at $n=2$ because the first cumulant (the mean) vanishes by construction, 
and each term is given by
\begin{multline}
    W_{\Dsrc}^{\scriptstyle{(n)}}[\ii \RespField] = \frac{\ii^n}{n!} \sum_{\{a_k\}}
    \int \prod_{m=1}^n \diff{\bm{z}_m} \, \\
    \RespField_{a_1}(\bm{z}_1) \dots \RespField_{a_n}(\bm{z}_n) \,
    \cumulant^{\scriptstyle{(n)}}_{a_1 \dots a_n}(\bm{z}_1; \cdots; \bm{z}_n)
    \label{eq: src cumulants}
\end{multline}
where $a_k \in \{1,2,3\}$ for $k=1,\dots,n$, the shorthand $\bm{z}_k \equiv \{\direc_k, \lambda_k\}$ 
collects the $i$-th affine and directional coordinates, and
\begin{multline}
    \cumulant^{\scriptstyle{(n)}}_{a_1 \dots a_n}(\direc_1, \lambda_1; \dots; \direc_n, \lambda_n)
    \equiv \\
    \left\langle
    \Dsrc_{a_1}(\direc_1, \lambda_1)\dots \,\Dsrc_{a_n}(\direc_n, \lambda_n)
    \right\rangle_c
\end{multline}
are the connected $n$-point functions of $\Dsrc$.
The full statistics of the stochastic sources are thus encoded in the cumulant hierarchy. 
This representation is particularly convenient: the first cumulant vanishes by centring, 
and for purely Gaussian sources all cumulants of order $n \geq 3$ vanish as well. 
The hierarchy can therefore be truncated at an order dictated by the assumed statistical model.

This hierarchy can be recast as the formal split
used in the perturbative expansion below.  After centring fixes the one-point
source cumulant to zero, the effective linear Sachs operator and the connected
two-point source cumulant define the Gaussian-reference theory: the
linear operator generates the response propagator, while the two-point source
cumulant supplies the noise kernel entering the correlation propagator.  The
remaining non-quadratic pieces, namely the nonlinear Sachs evolution and the
connected source cumulants with $n\geq3$, are treated as interaction vertices.
At this stage no choice of driving-field channels or cumulant truncation is
assumed.

\subsection{Free Propagators and Interaction Vertices}
\label{subsec: propagators and vertex}

A direct evaluation of the path integral is, in general, intractable. 
Following standard practice in field theory, we therefore adopt a perturbative 
approach: the action is split into a free (quadratic) part and an interacting 
(higher-order) part, and observables are computed as convergent or asymptotic
series in the interaction.

Concretely, we decompose the action $\Action$ in Eq.~(\ref{eq: action 1}) 
into a quadratic reference action $\ActionRef$ and an interaction action 
$\ActionInt$ that collects all non-quadratic contributions,
\bea
    \Action &=& \ActionRef + \ActionInt, \\[4pt]
    \ActionRef &=& \ii \int \, \RespField_{\sachi} \left( \edited{\Dot{\Dsachs}_{\sachi}} - A_{\sachi \sachj}\Dsachs_{\sachj} \right) \, \diff{\direc} \diff{\lambda}
    - W_{\Dsrc}^{\scriptstyle{(2)}}[\ii \, \RespField], \label{eq: S0 definition}\\ 
    \ActionInt &=& - \ii \int \, F_{\sachi \sachj \sachk}  \, \RespField_{\sachi} \Dsachs_{\sachj} \Dsachs_{\sachk} \, \diff{\direc} \diff{\lambda}
    - \sum_{k=3}^{\infty} W_{\Dsrc}^{\scriptstyle{(k)}}[\ii \, \RespField] ,
    \label{eq: action int definition}
\eea
where $W_{\Dsrc}^{\scriptstyle{(k)}}$ is defined in Eq.~(\ref{eq: src cumulants}).
The reference action $\ActionRef$ encodes the linear dynamics together with 
the Gaussian statistics of the driving field, whereas $\ActionInt$ collects 
both the nonlinear dynamics and all non-Gaussian contributions of the driving field.

For any observable $\mathcal{O}$ of the Sachs fields, the expectation value with 
respect to the full action is
\be
    \langle \mathcal{O} \rangle_{\Action}
    \equiv
    \int \funcdiff{\Dsachs}\funcdiff{\RespField} \, \mathcal{O} \,
    e^{- \Action[\Dsachs, \RespField]}
    .
\ee
Expanding the factor $e^{-\ActionInt}$ order by order, the expectation value 
reduces to a series of Gaussian (reference-theory) averages:
\be
    \label{eq: perturbation}
    \langle \mathcal{O} \rangle_{\Action}
    =
    \sum_{n=0}^{N} \frac{(-1)^n}{n!}\,
    \langle {\mathcal{O}\,\ActionInt^n} \rangle_{\ActionRef}
\ee
Physical observables are typically expressed as moments (correlation functions) 
of the Sachs fields, and $\ActionInt$ itself is a polynomial in the physical 
field $\Dsachs$ and the response field $\RespField$. Every term in Eq.~(\ref{eq: perturbation}) 
is therefore a moment evaluated in the Gaussian reference theory defined by $\ActionRef$, 
for which Wick's theorem applies: odd-order moments vanish, while even-order moments factorise 
into products of two-point functions. Two independent two-point functions suffice:
\begin{align}
    \left\langle
    \Dsachs_{\sachi}(\direc, \lambda) \, \Dsachs_{\sachj}(\direc', \lambda')
    \right\rangle_{\ActionRef}
    & \equiv
    \CorrOp_{\sachi \sachj}(\direc, \lambda; \direc', \lambda') \\
    \left\langle
    \RespField_{\sachi}(\direc, \lambda) \, \Dsachs_{\sachj}(\direc', \lambda')
    \right\rangle_{\ActionRef}
    & \equiv 
    - i \,
    \RespOp_{\sachi \sachj}(\direc, \lambda; \direc', \lambda') 
    \label{eq: moment example 2}    
\end{align}
where $\CorrOp$ is the correlation propagator of the physical fields, 
and $\RespOp$ is the response propagator, which measures how $\Dsachs$ reacts to an 
infinitesimal source conjugate to $\RespField$. The auto-correlation of the response 
field itself vanishes identically,
\be
\langle \RespField_{\sachi} \RespField_{\sachj} \rangle_{\ActionRef} =0.
\ee

More explicitly, the response propagator is defined as
\begin{equation}
\begin{split}
    \RespOp_{\sachi \sachj}&(\direc, \lambda; \direc', \lambda')
    \equiv
    \delta(\direc' - \direc) \,
    \Theta(\lambda' - \lambda) \\
    & \qquad\times
    \left\{ \Texp \left( \int_{\lambda}^{\lambda'} \mathbf{A}(\tau) \diff{\tau} \right) \right\}_{\sachi \sachj} \\
    & = \delta_{\sachi \sachj } \, \delta(\direc' - \direc) \,
    \Theta(\lambda' - \lambda) \,
    e^{-2 \,  \int_{\lambda}^{\lambda'} \, \theta^{\scriptscriptstyle\mathrm{(sa)}}(\tau) \diff{\tau} }
    \label{eq: explicit resp op}
\end{split}
\end{equation}
where $\Theta$ is the Heaviside step function and $\Texp$ denotes the $\tau$-ordered 
exponential along the congruence (propagators evaluated at larger $\tau$ are placed 
to the left); and the second equality uses the explicit form of $\mathbf{A}$ from 
Eq.~(\ref{eq: explicit A matrix}).
The correlation propagator is given by
\begin{multline}
    \CorrOp_{\sachi \sachj}(\direc_1, \lambda_1; \direc_2, \lambda_2) 
    = \int \diff{\direc'}\diff{\direc''}\diff{\lambda'}\diff{\lambda''} \\
     \RespOp_{\sachi \sachl}(\direc', \lambda'; \direc_1, \lambda_1) \,
    \cumulant^{\scriptstyle{(2)}}_{\sachl \sachk}(\direc', \lambda'; \direc'', \lambda'') \,
    \RespOp_{\sachj \sachk}(\direc'', \lambda''; \direc_2, \lambda_2) .
    \label{eq: Corr Op}
\end{multline}
Substituting Eq.~(\ref{eq: explicit resp op}) into Eq.~(\ref{eq: Corr Op}) yields
\begin{multline}
    \CorrOp_{\sachi \sachj}(\direc_1, \lambda_1; \direc_2, \lambda_2)
    = \int_0^{\lambda_1} \diff{\lambda'}\int_0^{\lambda_2}\diff{\lambda''} \\
    \cumulant^{\scriptstyle{(2)}}_{\sachi \sachj}(\direc_1, \lambda'; \direc_2, \lambda'') \,
    e^{-2 \,  \left( \int_{\lambda'}^{\lambda_1} + \int_{\lambda''}^{\lambda_2} \right) \,
    \theta^{\scriptscriptstyle\mathrm{(sa)}}(\tau) \diff{\tau} } .
    \label{eq: Corr Op 2}
\end{multline}
For statistics that are spatially homogeneous and isotropic on the celestial sphere, 
the two-point cumulant of the driving field depends on the two directions only 
through their relative angle and admits the Legendre expansion
\be
\cumulant^{(2)}_{\sachi \sachj}(\hat r_1,\lambda';\hat r_2,\lambda'')
=
\sum_{\ell}
\frac{2\ell+1}{4\pi}
C_\ell^{\sachi \sachj}(\lambda',\lambda'') \,
P_\ell(\hat r_1\!\cdot\!\hat r_2)\,,
\label{eq: src cumulant Legendre}
\ee
where $P_\ell$ is the Legendre polynomial of order $\ell$ and 
$C_\ell^{\sachi\sachj}(\lambda',\lambda'')$ \edited{collects the Legendre coefficients of the 
corresponding correlator} between the driving-field components $\sachi$ and $\sachj$ at affine 
parameters $\lambda'$ and $\lambda''$\edited{, taken in the pair-aligned screen basis fixed in 
Sec.~\ref{sec: cosmological setup}. Each such correlator depends only on the angle between the 
two directions, so the expansion is exact for the spin-2 components as well as the scalar one; 
the spin weight is carried by the basis, not by the expansion}. 
Substituting this multipole decomposition into Eq.~(\ref{eq: Corr Op 2}) 
yields the angular expansion of the correlation propagator,
\begin{multline}
    \CorrOp_{\sachi \sachj}(\direc_1, \lambda_1; \direc_2, \lambda_2) 
    = 
    \sum_{\ell} \frac{2\ell+1}{4\pi} P_\ell(\direc_1\!\cdot\! \direc_2) \, \\
    \int_0^{\lambda_1} \diff{\lambda'}\int_0^{\lambda_2}\diff{\lambda''} \,
    C_\ell^{\sachi \sachj}(\lambda',\lambda'') \,
    e^{-2 \,  \left( \int_{\lambda'}^{\lambda_1} + \int_{\lambda''}^{\lambda_2} \right) \,
    \theta^{\scriptscriptstyle\mathrm{(sa)}}(\tau) \diff{\tau} } .
    \label{eq: Corr Op multipole}
\end{multline}
Equation~(\ref{eq: Corr Op multipole}) thus expresses the correlation propagator 
entirely in terms of the angular power spectra of the stochastic driving field, 
$C_\ell^{\sachi\sachj}(\lambda',\lambda'')$, which serve as the only statistical 
input to the Gaussian reference theory.

In summary, we have (i) split the action into a Gaussian reference part $\ActionRef$ 
and a non-Gaussian interaction part $\ActionInt$, (ii) identified the two independent 
free propagators of the reference theory, $\RespOp_{\sachi\sachj}$ and $\CorrOp_{\sachi\sachj}$, 
and (iii) expressed them in closed form in terms of the source-averaged expansion scalar
$\theta^{\scriptscriptstyle\mathrm{(sa)}}$ and the angular power spectra 
$C_\ell^{\sachi\sachj}$ of the driving field. 
With these ingredients in hand, the companion package \textsc{sft-wick}~\cite{SFTwick2026}
automates the perturbative expansion of a given observable into Feynman diagrams built
from $\RespOp$, $\CorrOp$, and evaluates the resulting integrals numerically.

\subsection{Diagrammatic Expansion of Correlation Functions}
\label{subsec: diagrammatic expansion}

The perturbative expansion in Eq.~(\ref{eq: perturbation}) can be
organised graphically as Feynman diagrams. Each diagram represents an integral
contribution, with its propagator factors, vertex coefficients and
combinatorial weight fixed by the expansion. For an observable
$\mathcal{O}=\Dsachs_{\sachi_1}(\bm{z}_1)\cdots
\Dsachs_{\sachi_p}(\bm{z}_p)$, the external vertices of the diagram denote the
Sachs-scalar insertions.  
Each edge is evaluated in the Gaussian reference theory using the free propagators derived in
Eqs.~(\ref{eq: explicit resp op}) and~(\ref{eq: Corr Op}).  
Diagrammatically, we draw the correlation propagator $\CorrOp$ as an undirected solid line, 
while a response contraction is drawn as a directed dashed line from a response-field endpoint
to a Sachs-field endpoint and carries the factor $-\ii\RespOp$.
There is no line joining two response fields, since
$\langle \RespField_{\sachi}\RespField_{\sachj}\rangle_{\ActionRef}=0$.
Closed directed response cycles therefore vanish by causality.

The interaction action supplies the internal vertices.  Each internal vertex
stands for an integration over the spacetime point at which its attached
legs meet, weighted by the corresponding interaction coefficient.
The deterministic Sachs non-linearity gives the local cubic vertex
\be
    \mathcal{V}_{F}
    =
    -\ii
    \int
    F_{\sachi\sachj\sachk} \,
    \RespField_{\sachi}(\bm{z})
    \Dsachs_{\sachj}(\bm{z})
    \Dsachs_{\sachk}(\bm{z})
    \,\diff{\bm{z}},
    \label{eq: sachs vertex}
\ee
so a square vertex carries one response leg and two physical legs,\footnote{All 
legs share the same integration variable, hence the term local vertex.}
together with
the component tensor $F_{\sachi\sachj\sachk}$ and an integration over the
internal spacetime point.  Non-Gaussianity in the driving fields produces
additional cumulant vertices through the terms
$-W_{\Dsrc}^{(n)}[\ii\RespField]$ in Eq.~(\ref{eq: action int definition}).
Equivalently, for $n\geq3$,
\begin{multline}
    \mathcal{V}_{\cumulant_n}
    =
    -\frac{\ii^n}{n!}
    \sum_{\{a_r\}}
    \int
    \prod_{r=1}^{n}\diff{\bm{z}_r}\, \\
    \RespField_{a_1}(\bm{z}_1)\cdots
    \RespField_{a_n}(\bm{z}_n)\,
    \cumulant^{\scriptstyle{(n)}}_{a_1\cdots a_n}
    (\bm{z}_1;\cdots;\bm{z}_n),
    \label{eq: driving cumulant vertex}
\end{multline}
where each index $a_r$ labels a component of the driving field. In this
notation, the interaction action of Eq.~(\ref{eq: action int definition})
collapses to the compact form
$\ActionInt = \mathcal{V}_{F} + \sum_{n\geq 3}\mathcal{V}_{\cumulant_n}$.

The two vertex families differ structurally. The cumulant vertex
$\mathcal{V}_{\cumulant_n}$ is non-local: its $n$ response legs are weighted
by $\cumulant^{\scriptstyle{(n)}}_{a_1\cdots a_n}(\bm{z}_1;\dots;\bm{z}_n)$
and integrated over $n$ independent affine-direction arguments
$\bm{z}_r=(\direc_r,\lambda_r)$. By contrast, the Sachs vertex
$\mathcal{V}_{F}$, drawn as an $F$ node, is local: its three legs share a
single $\bm{z}=(\direc,\lambda)$.

The diagrammatic rule is then just Wick's theorem applied to the expanded
path integral.  To compute a connected $p$-point function to a fixed order, one
inserts the required number of $\mathcal{V}_{F}$ and
$\mathcal{V}_{\cumulant_n}$ vertices, contracts all physical (Sachs) and response fields
with the free propagators above, drops any subgraph that is disconnected
from the external points, and integrates over every internal argument.  Wick
contractions that have the same connectivity are collected into a single
topology with the corresponding symmetry and multiplicity factors.  This is
the bookkeeping automated by \textsc{sft-wick}: the algebraic expression for a
given topology is obtained by multiplying all propagators, tensors, cumulants,
and vertex signs associated with that graph, and then summing over all internal
Sachs indices.

\subsubsection*{Example: 2-point correlation}

\begin{figure*}[t]
    \centering
    \includegraphics[width=\textwidth]{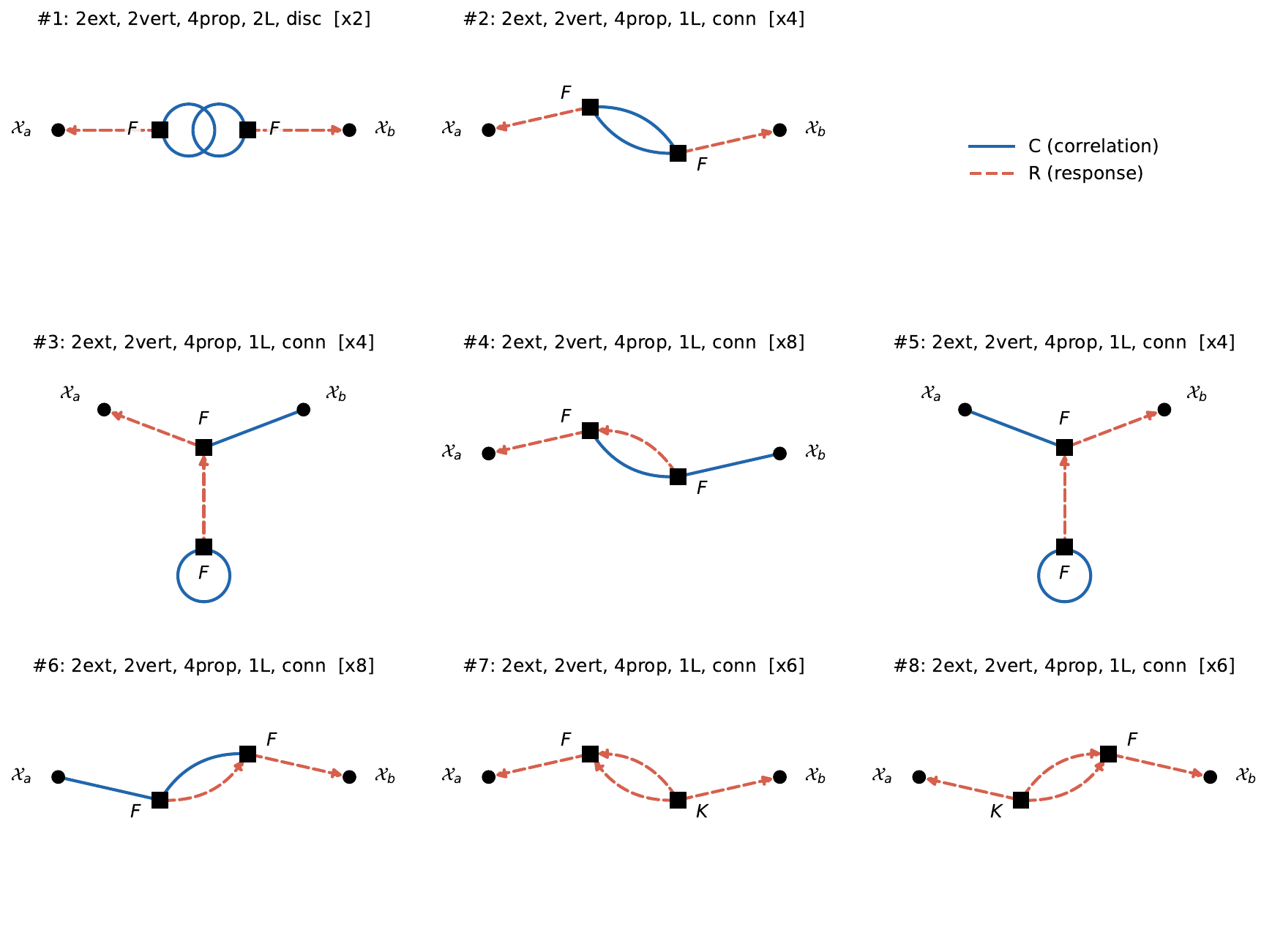}
    \caption{Second-order diagrams for the Sachs two-point correlation
    function, each containing two interaction vertices. Circular external
    nodes denote Sachs fields. Square nodes labelled $F$ represent the local
    interaction generated by nonlinear Sachs evolution
    [Eq.~(\ref{eq: sachs vertex})]. Nodes labelled $K$ represent the non-local
    vertex associated with the third connected cumulant of the Ricci and Weyl
    driving fields [Eq.~(\ref{eq: driving cumulant vertex}) with $n=3$].
    Blue solid lines carry $\CorrOp$, the Sachs-field two-point correlation
    in the Gaussian reference theory [Eq.~(\ref{eq: Corr Op})].
    Red directed dashed lines carry $-\ii\RespOp$, where $\RespOp$ is the
    linear response propagator [Eq.~(\ref{eq: explicit resp op})].
    The label ``disc'' marks diagrams in which the two external points
    are not connected.}
    \label{fig: feyn diag 2pt order2}
\end{figure*}

As a first example, consider the two-point correlation function. 
The zeroth-order contribution is the single free correlation line
$\CorrOp_{\sachi\sachj}(\bm{z}_1,\bm{z}_2)$. At first order, no single
vertex insertion can contribute to a connected two-point function: the
response-pairing rule of the reference action,
$\langle\RespField\RespField\rangle_{\ActionRef}=0$, prevents any
contraction in which the response legs ($\RespField$) of the inserted vertex outnumber
the available Sachs partners ($\Dsachs$), and in every other Wick channel an odd
number of fields is left over. We refer the reader to \cite{SFTwick2026}
for a systematic catalogue of these vanishing conditions. The leading
interaction correction therefore appears at second order; concretely, the
second-order expansion prior to Wick contraction reads
\begin{multline}
    \frac12
    \left\langle
    \Dsachs_{\sachi}(\direc, \lambda) \, \Dsachs_{\sachj}(\direc', \lambda')
    \ActionInt^2
    \right\rangle_{\ActionRef}
    = \\
    \frac12 \left\langle
    \Dsachs_{\sachi}(\direc, \lambda) \, \Dsachs_{\sachj}(\direc', \lambda')
    \left( \mathcal{V}_{F}^2 + 2\,\mathcal{V}_{\cumulant_3}\mathcal{V}_{F}  \right)
    \right\rangle_{\ActionRef} \,.
\end{multline}
On the right-hand side we have already dropped every quadratic combination
of $\ActionInt$ whose Wick contractions would force a response-response
pair: any product $\mathcal{V}_{F}\mathcal{V}_{\cumulant_n}$ with $n\geq 4$
or $\mathcal{V}_{\cumulant_n}\mathcal{V}_{\cumulant_{n'}}$ with $n,n'\geq 3$
carries more response legs than the available Sachs partners can absorb,
and so collapses through
$\langle\RespField\RespField\rangle_{\ActionRef}=0$. 
The corresponding
Wick diagrams are collected in Fig.~\ref{fig: feyn diag 2pt order2}: each
diagram stands for the corresponding internal integrations and the sum
over Sachs component labels, and entries marked ``disc'' are topologies
whose Wick graph is disconnected, with the two external points sitting in
different connected components.
For brevity, hereafter we use the shorthands
$F\equiv\mathcal{V}_{F}$ and $K\equiv\mathcal{V}_{\cumulant_3}$ for the
local cubic Sachs vertex and the leading non-local source-cumulant vertex,
matching the node labels in the figures.
The two surviving
topologies are therefore FF (two local cubic vertices) and
FK (one local $F$ paired with one non-local $K$). 

\subsubsection*{Example: 3-point correlation}

\begin{figure*}[t]
    \centering
    \includegraphics[width=\textwidth]{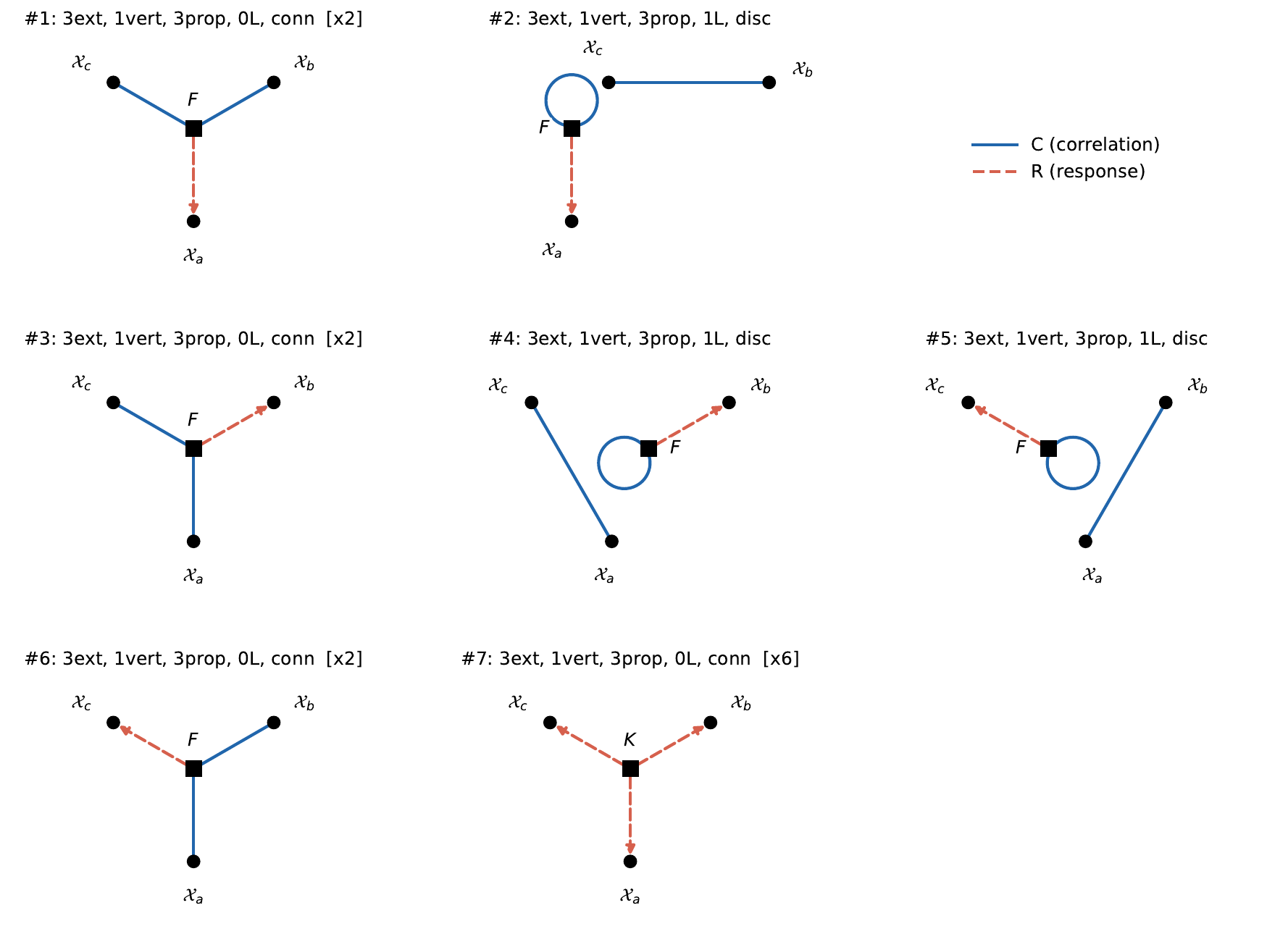}
    \caption{First-order diagrams for the Sachs three-point correlation
    function generated by either one local cubic vertex $\mathcal{V}_{F}$ or
    one non-local source-cumulant vertex
    $\mathcal{V}_{K}\equiv\mathcal{V}_{\cumulant_3}$.  The diagrammatic
    conventions are the same as in Fig.~\ref{fig: feyn diag 2pt order2}.}
    \label{fig: feyn diag 3pt order1}
\end{figure*}

For the three-point correlation function, the zeroth-order contribution vanishes
identically. At first order, both the local cubic vertex $\mathcal{V}_{F}$ and the
non-local cumulant vertex $\mathcal{V}_{\cumulant_3}$ contribute:
\begin{multline}
    -
    \left\langle
    \Dsachs_{\sachi}(\direc, \lambda) \, \Dsachs_{\sachj}(\direc', \lambda') \, \Dsachs_{\sachk}(\direc'', \lambda'')
    \ActionInt
    \right\rangle_{\ActionRef} = \\
    - \left\langle
    \Dsachs_{\sachi}(\direc, \lambda) \, \Dsachs_{\sachj}(\direc', \lambda') \, \Dsachs_{\sachk}(\direc'', \lambda'')
    \left( \mathcal{V}_{F} + \mathcal{V}_{\cumulant_3} \right)
    \right\rangle_{\ActionRef} \,,
\end{multline}
whereas the
higher driving field cumulant vertices $\mathcal{V}_{\cumulant_n}$ with $n\geq 4$ are
excluded by the same vanishing conditions invoked above.
The local $F$ vertex gives propagation-induced
non-Gaussianity: the three external Sachs fields are contracted with the two
Sachs legs and the response leg of the cubic vertex through the free
propagators.  The non-local $K$ vertex instead represents the intrinsic
three-point cumulant of the driving fields, propagated to the Sachs scalars by
three response lines.  The resulting topologies are displayed in
Fig.~\ref{fig: feyn diag 3pt order1}.  

The diagrams in this subsection are diagrams for correlation functions of the
Sachs scalars themselves.  The lensing observables studied in
Section~\ref{sec: case studies} are obtained by applying the linearized
line-of-sight relations to the external Sachs insertions; the
internal propagators and interaction vertices are left unchanged.

\subsection{The Selection Rule}
\label{subsec: selection rule}

\begin{figure*}[t]
    \centering
    \includegraphics[width=0.7\textwidth]{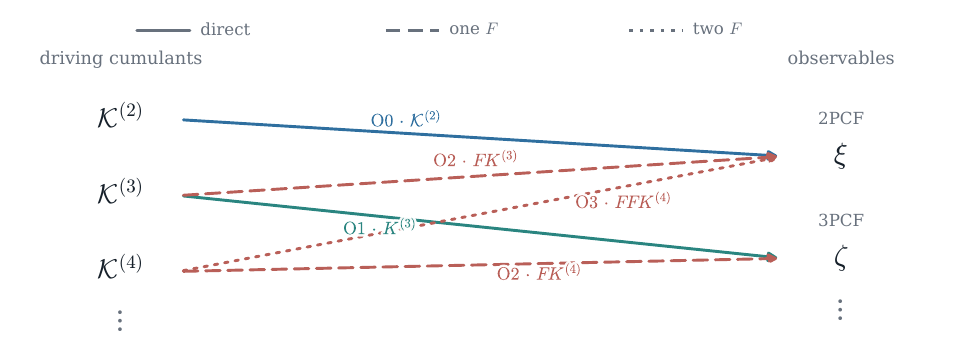}
    \caption{The selection rule of Eq.~(\ref{eq: selection rule}) at a glance:
    which driving-field cumulant $\cumulant^{(n)}$ feeds which observable, and at
    what leading order.  Line style counts the nonlinear-propagation vertices
    $F$.}
    \label{fig: selection rule}
\end{figure*}

The vanishing arguments invoked in the two examples above can be
collected into a single selection rule, summarised in
Fig.~\ref{fig: selection rule}. A connected diagram contributing
to a $p$-point Sachs correlator, built from $k_F$ insertions of
$\mathcal{V}_F$ and $k_n$ insertions of each $\mathcal{V}_{\cumulant_n}$,
vanishes unless
\be
    \;\sum_{n\geq 3} n\, k_n \,\leq\, p + k_F\;
    \qquad \text{and} \qquad
    p + k_F + \sum_{n\geq 3} n\, k_n \in 2\mathbb{Z}.
    \label{eq: selection rule}
\ee
The first inequality says that the response legs supplied by the cumulant
vertices cannot exceed the Sachs legs available to absorb them through
$\RespOp$; the parity condition reflects the fact that Wick's theorem can
only pair fields. Both restrictions descend from the single identity
$\langle\RespField_{\sachi}\RespField_{\sachj}\rangle_{\ActionRef}=0$:
every response leg must find a Sachs leg. The same
identity forbids any closed response cycle, consistent with the causality
remark below Eq.~(\ref{eq: explicit resp op}). One can verify that
Eq.~(\ref{eq: selection rule}) reproduces every vanishing case discussed
above, in particular the collapse of $\ActionInt^2$ to
$\mathcal{V}_F^2 + 2\,\mathcal{V}_{\cumulant_3}\mathcal{V}_F$ at second
order in the two-point function, and the exclusion of any first-order
$\mathcal{V}_{\cumulant_{n\geq 4}}$ contribution to the three-point
function. We refer the reader to \citet{SFTwick2026} for the
generalisation to correlators with external response insertions and for
the enumeration of allowed topologies at arbitrary order.

\section{Cosmological Weak Lensing}
\label{sec: cosmological setup}

The formalism of Sections~\ref{sec: sachs dynamics}--\ref{sec: path integral}
is geometry-agnostic: the Sachs-scalar dynamics couple to an arbitrary
stochastic spacetime through the Ricci focusing scalar $\Phi_{00}$ and the
Weyl shear scalar $\Psi_0$, and the path-integral expansion holds with no
commitment to any particular background. We now specialise to a perturbed
Friedmann--Lema\^{\i}tre--Robertson--Walker (FLRW) universe in Poisson gauge,
a coordinate choice that extends conformal Newtonian gauge to retain vector
and tensor perturbations~\citep{1996clss.conf..273B}. The metric is
\be
\begin{split}
    \mathrm{d}s^{2} = a^{2}(\tau)\Bigl[
      -(1+2\Psi)\,\mathrm{d}\tau^{2}
      - 2\,\edited{B_{i}}\,\mathrm{d}\tau\,\mathrm{d}x^{i} \\
      {} + \bigl((1-2\Phi)\,\delta_{ij} + h_{ij}\bigr)\,\mathrm{d}x^{i}\,\mathrm{d}x^{j}
    \Bigr],
\end{split}
    \label{eq: poisson gauge}
\ee
where $a(\tau)$ is the scale factor, $\tau$ the conformal time, $\Psi$ the
Newtonian gravitational potential perturbation, $\Phi$ the spatial curvature
perturbation, \edited{$B_i$ the transverse shift vector}, and $h_{ij}$ the
transverse-traceless tensor perturbation. Poisson gauge removes the scalar
part of the shift and the scalar and vector parts of the traceless spatial
metric perturbation. The remaining shift $B_i$ is transverse,
$\partial^i B_i = 0$, and $h_{ij}$ is transverse and traceless,
$\partial^i h_{ij} = 0$ and $\delta^{ij}h_{ij} = 0$,
with spatial indices raised using $\delta^{ij}$. At first order in the
metric perturbations, this form lets us compute separately the scalar,
vector and tensor contributions to the Ricci and Weyl fields that drive
the Sachs equations.

Throughout this paper we adopt
the flat $\Lambda$CDM fiducial cosmology used in the numerical pipeline,
matching the Planck-2015 baseline~\citep{planck2015parameters}:
$\Omega_c h^2=0.12029$, $\Omega_b h^2=0.02207$, $h=0.6711$,
$n_s=0.97$, and $\sigma_8=0.81$, giving $\Omega_m=0.31609$.
These numerical choices enter the distance--redshift map, power spectra,
and bispectra used in the figures and comparisons; the formal Sachs
construction itself remains independent of this fiducial model. We use
geometric units with $c=1$, so that length and time are measured in a common
unit and the speed of light never appears explicitly; the potentials $\Phi$
and $\Psi$ are then dimensionless (in physical units, the usual Newtonian
potentials divided by $c^{2}$), and the Hubble rate $H_{0}$ is correspondingly
measured as an inverse comoving length. All four channels ($\Psi$, $\Phi$, \edited{$B_i$}, $h_{ij}$) are
retained throughout this section; the simplifying assumptions used by the
numerical analyses of Section~\ref{sec: insights} (scalar-only, $\Phi=\Psi$,
Born-limit path) are not yet imposed. The treatment proceeds in three pieces.

\emph{Light-cone coordinate map (Section~\ref{subsec: affine to redshift}).}
We relate the affine, line-of-sight (LOS) parametrisation $(\lambda,\direc)$
used by the Sachs dynamics to the cosmological coordinates $(\chi,\hat r)$ and
observed redshift $z$. The Born-limit map (i.e., the unperturbed path) collapses 
to an isotropic one-to-one correspondence and is used throughout the rest of 
the paper; the generic perturbed-path coordinate Jacobian is recorded in
Appendix~\ref{append: jacobian blocks} for completeness.

\emph{Driving fields (Section~\ref{subsec: driving fields}).}
Given the Poisson-gauge potentials, the Sachs sources $\Phi_{00}$
[Eq.~(\ref{eq: Phi 00 from Ricci})] and $\Psi_0$
[Eq.~(\ref{eq: Psi0})] are evaluated to first order on the
metric~(\ref{eq: poisson gauge}), across the scalar, \edited{vector (shift)}, and tensor sectors.
Their connected two- and three-point cumulants enter the present work as
inputs to the later analysis; the technical details of their construction and
numerical implementation are presented in the companion
paper~\citep{canoes2026}.

\emph{Free Sachs propagators (Section~\ref{subsec: from sachs to wl kernels}).}
With the coordinate map and the driving-field cumulants in hand, the free
Sachs response and correlation operators on the FLRW background reduce to
closed forms in $\bar D(\lambda) = a(\lambda)\chi(\lambda)$, the background
angular-diameter distance. 
These forms feed directly into the path-integral
diagrams of Section~\ref{sec: insights}; as a consistency check, integrating
the correlation operator along $\lambda$ recovers the standard textbook
weak-lensing window.

\subsection{Light-cone coordinate map}
\label{subsec: affine to redshift}

The Sachs dynamics of Section~\ref{sec: sachs dynamics} describe each ray by 
$(\lambda,\direc)$, where $\lambda$ is the affine parameter and $\direc\in S^2$ 
denotes the LOS direction in the observer rest frame. Cosmological theory and 
N-body simulations, however, are typically formulated in comoving coordinates 
$(\chi,\hat r)$, where $\chi$ is the comoving radial distance and $\hat r$ the 
spatial direction from the observer. In practice, $\chi$ is often replaced by 
the corresponding background redshift.
Observations, on the other hand, label points along the past light cone and 
therefore inherit $\direc$, but use the observed redshift $z$ rather than the 
comoving distance $\chi$. A complete cosmological interpretation therefore also 
requires the redshift evolution along each ray. 
In the unperturbed or Born limit, both the coordinate map and the redshift evolution
reduce to isotropic one-to-one correspondences.

\subsubsection*{Coordinate-map Jacobian}
Given the ray's spatial trajectory $\vec x(\lambda, \direc)$ labelled by the affine 
parameter $\lambda$ and the observer-today sky direction $\direc$, they are defined by
\be
    \chi(\lambda, \direc) \;\equiv\; |\vec x(\lambda, \direc)|,
    \qquad
    \hat r^{i}(\lambda, \direc) \;\equiv\; \frac{x^{i}(\lambda, \direc)}{\chi(\lambda, \direc)} .
    \label{eq: chi nhat def}
\ee
To construct the ray $\vec x(\lambda, \direc)$ one needs to integrate the null
geodesic equation 
\be
k^\mu = \frac{\diff{x^\mu}}{d\lambda},
\qquad
k^\nu\nabla_\nu k^\mu = 0,
\ee
from the observer with initial data fixed by $\direc$. 
Each past-directed photon that reaches the observer admits, both
at the observer and along the ray, the decomposition
\begin{align}
    k^\mu &= E(-u^\mu + n^\mu), &
    n^\mu u_\mu &= 0, &
    n^\mu n_\mu &= 1, &
    E &= u_\mu k^\mu ,
    \label{eq: photon 4-mom}
\end{align}
where $u^\mu$ is the four-velocity of the observer, with respect to which
$n^\mu$ is the perceived four-vector of the outward line-of-sight
direction, i.e., the unit spacelike
vector in that observer's rest frame pointing from the observer along
the photon's spatial trajectory (with $n^\mu|_{\lambda = 0}$ fixed by
$\direc$), and $E$ is the photon energy in observer's frame.
\footnote{The sign in $E = u_\mu k^\mu$
differs from the common future-directed convention. Here $\lambda$
grows into the past, so $k^\tau = \diff\tau/\diff\lambda < 0$.}

To formulate the cosmological coordinate mapping, we identify the observer 
with the comoving observer at the ray event, namely the fundamental Hubble-flow 
observer, whose four-velocity is
\be
    u^\mu = (u^0, 0, 0, 0), \quad u^\mu u_\mu = -1
    \quad \Rightarrow \quad
    u^0 = \frac{1}{a \sqrt{1 + 2\Psi}} .
    \label{eq: obs 4-vel}
\ee
For the physically observed redshift evolution, however, the relevant observer 
is instead the local peculiar-flow frame intersected by the ray, allowing the 
tracer dynamics to manifest directly in observational cosmology.

The initial condition $k^\mu({\lambda = 0, \direc})$ in Eqs.~(\ref{eq: photon 4-mom}) 
and~(\ref{eq: obs 4-vel}) is fixed by the observer, and integrating the null geodesic 
equation then determines the ray trajectory and redshift evolution. 
Rather than tracking the ray itself, we instead work with the $3\times 3$
Jacobian of the coordinate map $(\lambda, \direc) \mapsto (\chi, \hat r)$, 
which encodes the local deformation of the coordinate mapping:
\be
    J(\lambda, \direc) \;=\;
    \begin{pmatrix}
        \partial_{\lambda}\chi & \partial_{\direc}\chi \\[2pt]
        \partial_{\lambda}\hat r & \partial_{\direc}\hat r
    \end{pmatrix} .
    \label{eq: coord jacobian}
\ee
At first order, $J = J_{(0)} + \delta J$ acquires a radial-rate correction
$\delta J^{\chi}_{\phantom{\chi}\lambda} = \partial_{\lambda}\delta\chi$, an angular-deflection 
rate $\delta J^{\hat r}_{\phantom{\hat r} \lambda } = \partial_{\lambda}\delta\hat r$, and a transverse
$2\times 2$ block $\delta J^{\hat r}_{\phantom{\hat{r}} \direc} = \partial_{\direc}\delta\hat r$; the
explicit line-of-sight integral form of every block, and the scalar-sector
reduction to the standard weak-lensing deflection and amplification matrix,
are derived in Appendix~\ref{append: jacobian blocks}. 
The transverse block
coincides, up to an observer-sphere $\leftrightarrow$ screen-frame rotation,
with the standard Sachs Jacobi matrix $\mathcal{J}^{A}_{\phantom{A}B}(\lambda)$
that maps initial observation angles to screen-space deviation vectors, and
controls the familiar cosmic-shear convergence and shear.
With
$\hat r = \direc$ at background, the first-order corrections
$\delta\chi(\lambda,\direc)$ and $\delta\hat r^{i}(\lambda,\direc)$ are
driven by line-of-sight integrals of transverse potential gradients
(schematically $\int\partial_{\perp}(\Phi+\Psi)\,\diff\chi'$ for $\delta\hat r$,
i.e.\ the weak-lensing deflection) and take the form of the standard Sachs
geodesic-deviation integrals.

We note that, here and after, any quantity $X$ decomposes as
$X = X^{(0)} + \varepsilon X^{(1)} + \mathcal{O}(\varepsilon^{2})$, with the
parenthesised index $(0)$ [$(1)$] denoting the background [first-order]
piece in the metric perturbations $\{\Psi, \Phi, \edited{B_i}, h_{ij}\}$; the index may
appear as a sub- or superscript depending on typographic convenience. 

\subsubsection*{Hubble-flow redshift evolution}
Redshift is an observer-dependent quantity; throughout this subsection
it is defined relative to the fundamental Hubble-flow observer
[four-velocity $u^\mu$ of Eq.~(\ref{eq: obs 4-vel})] at each ray event.
The redshift of the local comoving frame is defined through
\begin{equation}
    1 + z(\lambda) \;\equiv\; \frac{E(\lambda)}{E_{0}}
    \label{eq: z definition}
\end{equation}
with $E_{0} = E(\lambda = 0)$ the observer-today energy.
For convenience we set $E_{0} = 1$ throughout; this choice merely fixes 
the rescaling freedom of the affine parameter on the null geodesic and 
leaves every physical observable unchanged.
Differentiating $E = u_\mu k^\mu$ along the ray
\begin{equation}
    \frac{\diff{E}}{\diff{\lambda}}
    \;=\; k^\nu \nabla_\nu (u_\mu k^\mu)
    \;=\; k^\mu k^\nu \nabla_\nu u_\mu ,
    \label{eq: dE dlambda}
\end{equation}
where the second equality uses $k^\nu \nabla_\nu k^\mu = 0$, 
one obtains the redshift evolution
\begin{equation}
    \frac{\diff{z}}{\diff{\lambda}}
    \;=\; \frac{1}{E_{0}}\,k^\mu k^\nu \nabla_\nu u_\mu .
    \label{eq: dz dlambda master}
\end{equation}
Inserting the Poisson-gauge metric~(\ref{eq: poisson gauge}), the
Hubble-flow $u^\mu$~(\ref{eq: obs 4-vel}), and $k^\mu$~(\ref{eq: photon 4-mom})
into Eq.~(\ref{eq: dz dlambda master}) and expanding to first order, the
symbolic derivation yields
\begin{equation}
\begin{split}
    \left.\frac{\diff{z}}{\diff{\lambda}}\right|^{(1)}_{\direc}
    &=
    \frac{E_{0}\,(1+z)^{2}}{a}
    \biggl\{
        \mathcal{H}
        + \mathcal{H}\,\bigl[\,- \Psi + \edited{\hat n^{i} B_{i}}\,\bigr]
        - \partial_{\tau}\Phi \\
    &\qquad
        - \hat n^{i}\partial_{i}\Psi
        + \edited{\partial_{\tau}\bigl(\hat n^{i} B_{i}\bigr)}
        + \tfrac{1}{2}\,\hat n^{i} \hat n^{j}\partial_{\tau} h_{ij}
    \biggr\},
\end{split}
\label{eq: dz dlambda full}
\end{equation}
where $\mathcal{H}\equiv a'/a$ is the conformal Hubble rate
(${}'\equiv\partial_\tau$) and all fields are evaluated along the ray
of $\direc$ at affine parameter $\lambda$. 
%
%
The first term on the right-hand side is the background rate; the
remaining terms collect the standard Sachs--Wolfe, Doppler-gradient,
and integrated-Sachs--Wolfe contributions familiar from
cosmological-perturbation theory; see, e.g.,
\citet{clarkson2012mis}.

\subsubsection*{Born-limit coordinates and background redshift}

\begin{figure*}[t]
    \centering
    \includegraphics[width=\textwidth]{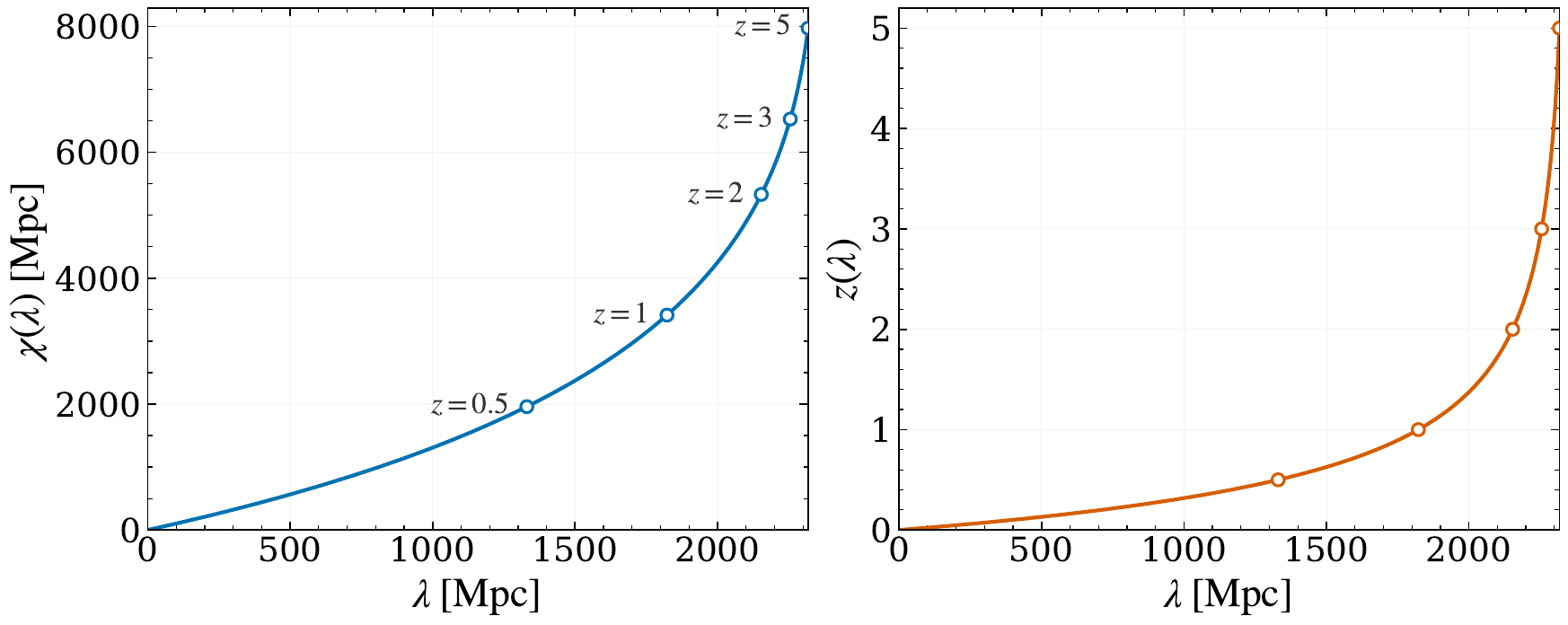}
    \caption{The two scalar branches of the one-to-one light-cone
    coordinate map. \textbf{Left}: comoving distance $\chi(\lambda)$
    obtained from Eq.~(\ref{eq: chi of z}) composed with
    Eq.~(\ref{eq: lambda of z}). \textbf{Right}: redshift $z(\lambda)$
    along the past light cone, the inverse of
    Eq.~(\ref{eq: lambda of z}). Both panels share the affine-parameter
    $\lambda$ axis; open circles mark $z\in\{0.5,1,2,3,5\}$ as
    orientation references. Curves use the fiducial cosmology
    ($\Omega_{m}=0.31609$, $h=0.6711$) adopted throughout the paper and
    the observer-frame normalisation $E_{0}=1$ (so that $\lambda$ has
    units of length); they reflect the exact radial grid the analyses
    of Sec.~\ref{sec: insights} consume.}
    \label{fig: coord maps}
\end{figure*}

In the following we adopt the Born approximation, in which first-order
metric perturbations are evaluated along the unperturbed (background)
null geodesic. This delivers the light-cone coordinate parameterisation
used throughout the rest of this section.

\paragraph*{Coordinate-map Jacobian.}
A symbolic evaluation on the Poisson-gauge background yields a
block-diagonal Jacobian,
\be
    J_{(0)} \;=\;
    \begin{pmatrix}
        E/a & 0 \\[2pt]
        0 & \delta^{B}_{A}
    \end{pmatrix} ,
    \qquad
    \det J_{(0)} \;=\; \frac{E}{a(\tau)} \;>\; 0 ,
    \label{eq: jacobian bg}
\ee
whose non-trivial zeroth-order entries read
\be
    \left.\frac{\diff\chi}{\diff\lambda}\right|_{(0)}
    \;=\; \frac{E}{a(\tau)} ,
    \qquad
    \hat r\big|_{(0)} \;=\; \direc .
    \label{eq: dchi dlambda bg}
\ee
Since $\det J = \det J_{(0)}\bigl(1 + \mathcal{O}(\varepsilon)\bigr)$
is nowhere zero as long as the perturbations stay small, the
coordinate map is a smooth diffeomorphism throughout the weak-lensing
regime; caustics ($\det J \to 0$) signal the onset of strong lensing
and lie outside the scope of this work. The first-order Born
corrections $\delta\chi$ and $\delta\hat r$ are the line-of-sight
integrals of transverse potential gradients derived in
Appendix~\ref{append: jacobian blocks}.

\paragraph*{Background redshift.}
For the observed redshift we adopt the background (zeroth-order)
evolution, matching the convention of the general-relativistic $N$-body
code \textsc{gevolution} used for cross-validation. Dropping the
first-order corrections in Eq.~(\ref{eq: dz dlambda full}) gives
\be
    \left.\frac{\diff{z}}{\diff{\lambda}}\right|_{(0)}
    \;=\; E_{0}\,H_{\rm c}(z)\,(1+z)^{2} ,
    \label{eq: dz dlambda bg}
\ee
with $H_{\rm c} \equiv (da/dt)/a = \mathcal{H}/a$ the cosmic-time Hubble
rate ($t$ the cosmic time, related to conformal time by $dt = a\,d\tau$). Dividing Eq.~(\ref{eq: dchi dlambda bg}) by
Eq.~(\ref{eq: dz dlambda bg}) gives the standard
conformal-distance--redshift relation
\be
    \frac{\diff\chi}{\diff z} \;=\; \frac{1}{H_{\rm c}(z)},
    \qquad
    \chi(z) \;=\; \int_{0}^{z}\!\frac{\diff z'}{H_{\rm c}(z')} ,
    \label{eq: chi of z}
\ee
and, equivalently, the affine parameter along the past light cone as a
function of redshift,
\be
    \lambda(z) \;=\; \frac{1}{E_{0}}\int_{0}^{z}\!\frac{\diff z'}{H_{\rm c}(z')\,(1+z')^{2}} .
    \label{eq: lambda of z}
\ee
The full Born-level first-order correction $\delta z$, obtained as a
line-of-sight integral of Eq.~(\ref{eq: dz dlambda full})'s first-order
right-hand side along the background ray, is deferred to future work.

\subsection{Ricci and Weyl driving fields in Poisson gauge}
\label{subsec: driving fields}

\begin{figure*}[t]
    \centering
    \includegraphics[width=\textwidth]{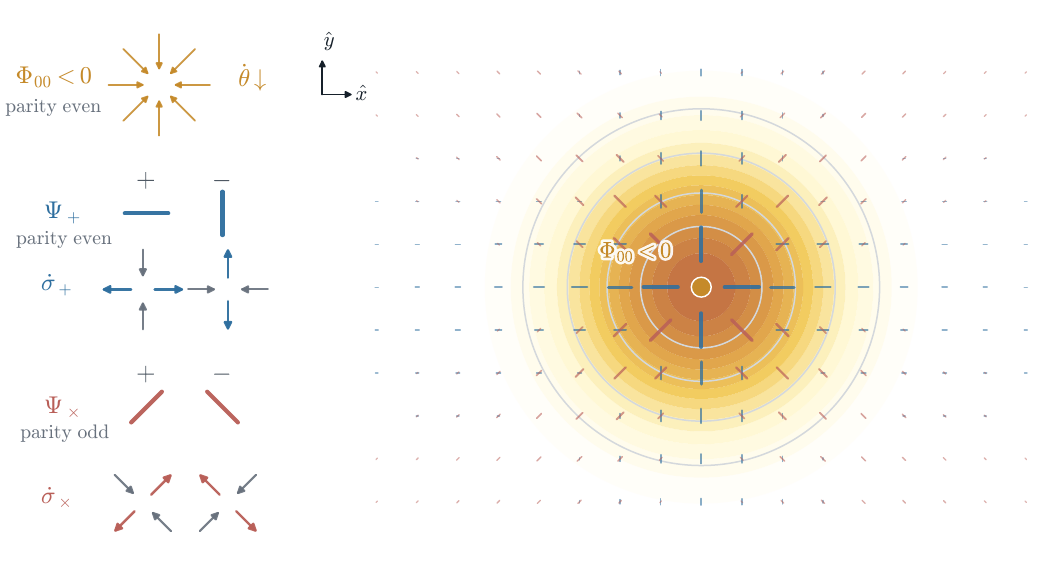}
    \caption{Pair-aligned component dictionary and field pattern.  The inset
    fixes the shared screen basis.  The left dictionary relates the driving
    components to the corresponding Sachs-equation rates
    $(\dot\theta,\dot\sigma_+,\dot\sigma_\times)$; the $+/-$ marks define
    the line-sample sign convention; the component tags show the parity class
    under reflection about $\hat x$.  Blue horizontal (vertical) segments
    encode positive (negative) source for $\dot\sigma_+$, while red $+45^\circ$
    ($-45^\circ$) segments encode positive (negative) source for
    $\dot\sigma_\times$.  The right field pattern applies this convention to an
    isolated focusing patch with $\Phi_{00}<0$; line length indicates component
    amplitude.  The dictionary arrows show rate trends, not accumulated shear
    shapes.}
    \label{fig:spin2-pattern}
\end{figure*}

Before entering the Poisson-gauge transfer functions, it is useful to fix the
real component basis used for the two-point statistics.  Figure~\ref{fig:spin2-pattern}
complements Fig.~\ref{fig:null-geodesic-congruence}: the latter introduced the
local screen and the optical scalars, while here we only fix the pair-aligned
component convention.  With $\hat x$ chosen along the separation of the two
directions, the screen-projected optical tidal matrix
$\mathcal{R}_{ab}$, $a,b\in\{\hat x,\hat y\}$, splits into three real pieces:
$\Phi_{00}$ is the trace part, $\Psi_+\equiv{\rm Re}\,\Psi_0$ is the plus-type
trace-free part, and $\Psi_\times\equiv{\rm Im}\,\Psi_0$ is the cross-type
trace-free part.  Figure~\ref{fig:spin2-pattern}
matches these components to the source-driven rates
$(\dot\theta,\dot\sigma_+,\dot\sigma_\times)$ that appear in the Sachs
equations and to the line-segment convention used for the real and imaginary
parts of $\Psi_0$.  A reflection across the separation axis leaves the trace
and plus pieces unchanged but flips the cross piece.
With the convention of Eq.~(\ref{eq: Phi 00 from Ricci}), a positive matter
overdensity gives a negative Ricci-focusing source, $\Phi_{00}<0$, which lowers
$\dot\theta$ relative to the no-overdensity case.

\begin{figure*}[t]
    \centering
    \includegraphics[width=\textwidth]{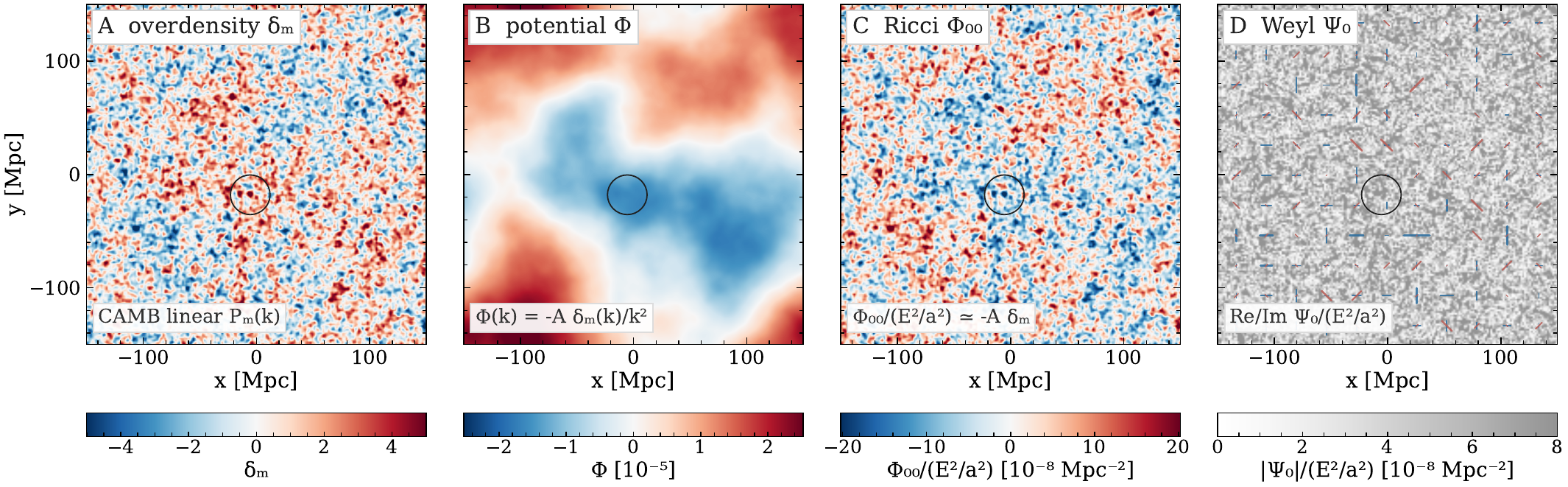}
    \caption{Map-level relation between matter fluctuations and the Sachs
    driving fields in the scalar sector.  Panel A shows a slice through a
    synthetic overdensity field $\delta_m$ drawn from a CAMB linear matter power
    spectrum at $z=0.7$.  Panel B shows the corresponding Poisson-gauge scalar
    potential $\Phi=\Psi$ for this scalar-only snapshot, obtained from
    $\Phi(k)=-\mathcal{A}\delta_m(k)/k^2$ with
    $\mathcal{A}=\tfrac{3}{2}\Omega_m H_0^2/a$; in the geometric units adopted
    here ($c=1$) the Hubble rate is an inverse comoving length,
    $H_0\simeq2.24\times10^{-4}\,{\rm Mpc}^{-1}$, so that $\mathcal{A}$ carries
    dimensions of ${\rm Mpc}^{-2}$ and the dimensionless $\Phi$ follows from the
    dimensionless $\delta_m$.  Panel C shows the rescaled
    Ricci-focusing trace, with the common local $E^2/a^2$ prefactor removed,
    approximately $-\mathcal{A}\delta_m$; panel D shows the correspondingly
    rescaled $\Psi_+$ and $\Psi_\times$ from the trace-free screen
    Hessian of the scalar lensing potential $\Phi+\Psi$; the short blue and red
    line segments use the component convention illustrated in
    Fig.~\ref{fig:spin2-pattern}, with length tracking the corresponding local
    component amplitude.  The colour bars are scaled to expose the relevant
    orders of magnitude in this snapshot: $\Phi$ is at the $10^{-5}$ level, while
    the displayed rescaled Ricci and Weyl fields are at the $10^{-8}\,{\rm Mpc}^{-2}$ level. The
    black circle, identical in every panel, tracks the peak overdensity of
    panel~A through the chain.}
    \label{fig:driving-field-portrait}
\end{figure*}

This subsection records the explicit first-order form of the Ricci focusing
$\Phi_{00}$ and the Weyl shear $\Psi_0$ on the Poisson-gauge
metric~(\ref{eq: poisson gauge}), decomposed into scalar, \edited{vector (shift)}, and
tensor sectors. The connected two- and three-point cumulants of
$\{\Phi_{00},\Psi_0\}$ that source the path-integral expansion of
Section~\ref{sec: insights} are then obtained, by linearity, by transferring
the corresponding statistics of the underlying Poisson-gauge perturbation
potentials $\{\Psi,\Phi,\edited{B_i},h_{ij}\}$ through per-multipole linear operators
acting on their multipole amplitudes.
Given these transfer formulas, the numerical construction of the cumulant
functions of $\{\Phi_{00},\Psi_0\}$ from the statistics of the Poisson-gauge
perturbation potentials is carried out in the companion
paper~\citep{canoes2026}. The two-point cumulants follow directly from the
power spectra of the perturbation potentials, whereas the three-point
cumulants, sourced by non-Gaussianity, require additional treatment. As a
concrete example, the analyses of Section~\ref{sec: insights} use a
three-point cumulant built from the tree-level matter bispectrum; this model
is summarised in Appendix~\ref{append: driving-field spectra}.

The null tetrad $\{k^\mu, e^\nu, \screenvec^{\mu}, \bar{\screenvec}^{\mu}\}$ and its real
orthonormal underpinnings $\{\mathbf{x}^\mu,\mathbf{y}^\mu\}$ were introduced in
Section~\ref{sec: sachs dynamics} on a generic curved spacetime, where they are
conventionally parallel-transported along the null geodesic, e.g.,
$k^\nu\nabla_\nu\screenvec^\mu = 0$. 
However, for the first-order cosmological evaluation below, we do
not need to solve the parallel-transport equation explicitly. It is enough that
the tetrad is \emph{first-order orthonormal} in the Poisson-gauge metric
[Eq.~(\ref{eq: poisson gauge})] around the Hubble-flow observer $u^\mu$
[Eq.~(\ref{eq: obs 4-vel})]: any two such bases differ by a first-order rotation
$\mathbf x^\mu \!\to\! \mathbf x^\mu + \varepsilon\,\beta\,\mathbf y^\mu$,
$\mathbf y^\mu \!\to\! \mathbf y^\mu - \varepsilon\,\beta\,\mathbf x^\mu$,
as a result $\screenvec^\mu\to\screenvec^\mu e^{-\ii\varepsilon\beta}$
and $\Psi_0\to e^{-2\ii\varepsilon\beta}\Psi_0$; the induced first-order shift
$-2\ii\varepsilon\beta\,\Psi_0^{(0)}$ vanishes because $\Psi_0^{(0)} = 0$ on
the (conformally flat) FLRW background.
The parallel-transported tetrad and the locally orthonormalised tetrad 
used below therefore give the same first-order $\Psi_0$.
By contrast, $\Phi_{00}$ has no spin and is immune to the screen-basis
rotation. First-order changes in the screen angle do not affect the first-order
Ricci and Weyl driving fields used here. We do not evaluate screen-basis
rotation in this demonstration.

Let $\{\hat x^i, \hat y^i\}$ be a pair of Cartesian coordinate
unit 3-vectors orthonormal with respect to the Euclidean metric
$\delta_{ij}$ and transverse to the line of sight, i.e.,
$\delta_{ij}\hat x^i\hat x^j = \delta_{ij}\hat y^i\hat y^j = 1$
and $\delta_{ij}\hat x^i\hat y^j = 0$. Promoting them to four-vectors that
are orthonormal in the full perturbed metric to first order in the
Poisson potentials, namely enforcing
$\mathbf{x}^\mu u_\mu = \mathbf{y}^\mu u_\mu = 0$,
$\mathbf{x}^\mu \mathbf{x}_\mu = \mathbf{y}^\mu \mathbf{y}_\mu = 1$,
and $\mathbf{x}^\mu \mathbf{y}_\mu = 0$,
the screen four-vectors read
\begin{align}
    \mathbf{x}^\mu_{(1)} &= \Bigl(-\tfrac{1}{a}\,\edited{\hat x^{i} B_i}\, ,\,
                     \tfrac{1}{a}\hat x^{i} + \tfrac{1}{a}(\Phi\,\hat x^{i} - \tfrac{1}{2} h^{i}{}_{j}\hat x^{j})
                     \Bigr), \\
    \mathbf{y}^\mu_{(1)} &= \Bigl(-\tfrac{1}{a}\,\edited{\hat y^{i} B_i}\, ,\,
                     \tfrac{1}{a}\hat y^{i} + \tfrac{1}{a}(\Phi\,\hat y^{i} - \tfrac{1}{2} h^{i}{}_{j}\hat y^{j})
                     \Bigr),
    \label{eq: poisson screen tetrad}
\end{align}
and $\screenvec_{(1)}^\mu = \left(\mathbf{x}_{(1)}^\mu + \ii\,\mathbf{y}_{(1)}^\mu\right)/\sqrt{2}$ inherits the
corrections through its real and imaginary parts. These corrections are not cosmetic:
dropping them (i.e., using $\mathbf{x}^\mu = (0,\hat x^i/a)$ in the
perturbed metric) would break the normalisation 
$\mathbf{x}^\mu \mathbf{x}_\mu = 1$ and cause a spurious shift in
$\Psi_0^{(1)}$, 
because the screen-plane rotation invariance of the previous paragraph 
holds only within the class of first-order-orthonormal bases.

With $k^\mu$, $u^\mu$, and $\screenvec^\mu$ in hand, the driving fields defined in
Eqs.~(\ref{eq: Phi 00 from Ricci}) and~(\ref{eq: Psi0}) reduce to polynomials in the metric
perturbations once $R_{\mu\nu}$ and $R_{\alpha\beta\gamma\delta}$ are expanded to first order.
For later convenience, we now introduce the line-of-sight directional 
derivative $\hat n^i\partial_i$, 
the transverse Laplacian $\nabla^2_\perp = (\delta^{ij}-\hat n^i \hat n^j)\partial_i\partial_j$, 
and the null-like differential operator
\be
    D \equiv \partial_\tau - \hat n^i \partial_i,
\ee
related to the affine-parameter derivative along the background null geodesic by
$k^\mu\partial_\mu = -(E/a)\,D$. Tensor perturbation components projected onto the
flat triad $\{\hat n^i,\hat x^i,\hat y^i\}$ are denoted
\be
\begin{gathered}
    {}_0 h \equiv \hat n^i \hat n^j h_{ij}, \qquad
    {}_1 h \equiv \hat{\screenvec}^i \hat n^j h_{ij}, \\
    {}_2 h \equiv \hat{\screenvec}^i \hat{\screenvec}^j h_{ij},
    \qquad \hat{\screenvec}^i \equiv (\hat x^i + \ii\,\hat y^i)/\sqrt{2},
\end{gathered}
\ee
where the left-subscript $s = 0, 1, 2$ labels the spin weight of each 
projection under rotations of the screen basis $\{\hat x^i, \hat y^i\}$. 
Projecting onto the screen basis makes ${}_s h_{\ell m}$ coincide directly
with the standard cosmological tensor-mode $E$/$B$ amplitudes, facilitating
polarisation analyses in weak lensing.
In the expressions below we use the exact relation $E = E_{0}(1+z)$;
in the first-order prefactors, $(1+z)$ is evaluated at its background value.

\paragraph{Background.}
On the unperturbed FLRW spacetime the Weyl tensor vanishes and the Ricci tensor reduces to
$R^{(0)}_{\tau\tau}=-3\mathcal{H}'$, $R^{(0)}_{ij}=(\mathcal{H}'+2\mathcal{H}^2)\delta_{ij}$. Contracting with
$k^{\mu}_{(0)} = (E/a)(-1,\,\hat n^i)$ gives
\be
    \Phi_{00}^{(0)} = -\frac{E_{0}^{2}(1+z)^{2}}{a^{2}}\,(\mathcal{H}^2 - \mathcal{H}'), \qquad
    \Psi_{0}^{(0)} = 0,
    \label{eq: driving bg}
\ee
so the Weyl shear is intrinsically a linear-order quantity.

\paragraph{Scalar sector.}
Keeping only $\Phi$ and $\Psi$, the first-order contributions to the driving fields are
\begin{align}
    \Phi_{00}^{(1,s)}
    &= -\frac{E_{0}^{2}(1+z)^{2}}{a^{2}}
    \Bigl[
        2(\mathcal{H}' - \mathcal{H}^2)\Psi
        + \mathcal{H}(\partial_\tau\Psi - \partial_\tau\Phi) \notag \\
    &\qquad\quad
        - 2 \mathcal{H}\,\hat n^i\partial_i \Psi
        + \tfrac{1}{2}\nabla^2_\perp (\Phi + \Psi)
        + D^2 \Phi
    \Bigr],
    \label{eq: Phi00 scalar} \\
    \Psi_{0}^{(1,s)}
    &= -\frac{E_{0}^{2}(1+z)^{2}}{a^{2}}\, \hat{\screenvec}^i \hat{\screenvec}^j \partial_i\partial_j (\Phi + \Psi) .
    \label{eq: Psi0 scalar}
\end{align}
Expanding $D^2\Phi = \partial_\tau^2\Phi - 2\,\hat n^i\partial_i\partial_\tau\Phi + \hat n^i \hat n^j\partial_i\partial_j\Phi$ and
$\nabla^2_\perp f = \nabla^2 f - \hat n^i \hat n^j\partial_i\partial_j f$, Eq.~(\ref{eq: Phi00 scalar}) can be
rearranged into a form that isolates the well-known lensing potential $\Phi+\Psi$ and the
line-of-sight projection:
\begin{equation}
\begin{split}
    \frac{a^{2}}{E_{0}^{2}(1+z)^{2}}\,\Phi_{00}^{(1,s)}
    &=
    -\tfrac{1}{2}\nabla^2(\Phi+\Psi)
    - \tfrac{1}{2}\,\hat n^i \hat n^j\partial_i\partial_j(\Phi - \Psi) \\
    &\quad
    - \partial_\tau^2\Phi + 2\,\hat n^i\partial_i\partial_\tau\Phi \\
    &\quad
    - \mathcal{H}(\partial_\tau\Psi - \partial_\tau\Phi)
    + 2 \mathcal{H}\,\hat n^i\partial_i\Psi \\
    &\quad
    - 2(\mathcal{H}'-\mathcal{H}^2)\,\Psi.
\end{split}
\label{eq: Phi00 scalar expanded}
\end{equation}
The leading term $\tfrac{1}{2}\nabla^2(\Phi+\Psi)$ is the familiar Born-approximation sub-horizon
lensing source; the remaining contributions are sub-dominant on sub-Hubble scales but become
relevant for integrated Sachs--Wolfe-like effects and for full-sky calculations.
Equation~(\ref{eq: Psi0 scalar}) is the complex spin-$2$ Laplacian of the lensing potential
on the screen, recovering the standard flat-sky expression
$\Psi_0 = \tfrac{1}{2}(\mathcal{R}_{\hat x\hat x}-\mathcal{R}_{\hat y\hat y})
 + \ii\, \mathcal{R}_{\hat x\hat y}$ once $\hat{\screenvec}^i = (\hat x^i + \ii\,\hat y^i)/\sqrt{2}$ is
expanded.

Figure~\ref{fig:driving-field-portrait} illustrates this scalar-sector chain
at the map level.  A matter overdensity first sources the scalar perturbation
potential through Poisson's equation; the Ricci focusing $\Phi_{00}$ is the
trace part of the resulting optical tidal field, while the Weyl scalar
$\Psi_0$ is its spin-$2$ trace-free part.  The same figure also shows the
scale separation that underlies the weak-lensing expansion.  In this snapshot,
the potential has $\Phi \sim 10^{-5}$, and the displayed rescaled Ricci and
Weyl fields, with the common local $E^2/a^2$ prefactor removed, are of order
$10^{-8}\,{\rm Mpc}^{-2}$.  Thus even regions that are comparatively dense in
the cosmological sense still induce weak curvature driving.  Away from
strong-field environments such as the immediate vicinity of black holes, the
dimensionless combinations built from $\Phi_{00}$ and $\Psi_0$ are therefore
far smaller than unity, making the perturbative treatment well controlled.

\paragraph{\edited{Vector (shift, $B_i$)} sector.}
\edited{The transverse shift vector $B_i$, with $\partial^i B_i = 0$,} contributes to both
driving fields via
\begin{align}
    \Phi_{00}^{(1,B)} &= -\frac{E_{0}^{2}(1+z)^{2}}{a^{2}}
    \Bigl[
        \edited{2(\mathcal{H}^2 - \mathcal{H}')\, \hat n^i B_i
        + \mathcal{H}\, \hat n^i \hat n^j\partial_i B_j}
        \notag\\
    &\qquad
        \edited{- \tfrac{1}{2}\nabla^2 \bigl(\hat n^k B_k\bigr)
        + \tfrac{1}{2}\,\partial_\tau\bigl(\hat n^i \hat n^j\partial_i B_j\bigr)}
    \Bigr],
    \label{eq: Phi00 B} \\
    \Psi_{0}^{(1,B)} &= \frac{E_{0}^{2}(1+z)^{2}}{a^{2}}
    \Bigl[
        \edited{\hat{\screenvec}^i \hat{\screenvec}^j \partial_i\partial_j \bigl(\hat n^k B_k\bigr)
        + D\,\hat{\screenvec}^i\partial_i \bigl(\hat{\screenvec}^j B_j\bigr)}
    \Bigr].
    \label{eq: Psi0 B}
\end{align}
\edited{Here $D \equiv \partial_\tau - \hat n^i\partial_i$. The gauge condition
$\partial^i B_i = 0$ removes the longitudinal part of the shift, not $B_i$
itself, so these terms are not switched off by the gauge choice: they are the
frame-dragging vector mode. Transversality does simplify the focusing, trading
the transverse divergence for $-\hat n^i \hat n^j\partial_i B_j$ and leaving
$\Phi_{00}^{(1,B)}$ dependent on $B_i$ only through $\hat n^k B_k$, as is also
true of the redshift evolution~(\ref{eq: dz dlambda full}). The shear is not
reduced this way: beside the spin-$2$ screen Hessian of $\hat n^k B_k$ it
carries $D\,\hat{\screenvec}^i\partial_i(\hat{\screenvec}^j B_j)$, built from the
spin-$1$ projection $\hat{\screenvec}^j B_j$, which a longitudinal shift cannot
supply on its own. Structurally this is Eq.~(\ref{eq: Psi0 tensor}) with the
spin-$2$ slot absent, as a vector source requires. Vector modes have no growing
linear mode in $\Lambda$CDM, so Section~\ref{sec: insights} keeps the scalar
sector alone; the channel is retained because relativistic $N$-body output
supplies it slot for slot~\citep{adamek2016gevolution}.}

\paragraph{Tensor sector.}
Keeping only the transverse-traceless tensor mode $h_{ij}$ and imposing
$h^i{}_i = 0$, $\partial^i h_{ij} = 0$, the contribution of gravitational waves reduces to
the compact forms
\begin{align}
    \Phi_{00}^{(1,t)}
    &= -\frac{E_{0}^{2}(1+z)^{2}}{4\,a^{2}}\,\bigl[\,\partial_\tau^2 {}_0 h + 2 \mathcal{H}\partial_\tau {}_0 h - \nabla^2 {}_0 h\bigr],
    \label{eq: Phi00 tensor} \\
    \Psi_{0}^{(1,t)}
    &= \frac{E_{0}^{2}(1+z)^{2}}{2\,a^{2}}\,\Bigl[
        D^2 {}_2 h + 2\,D\,(\hat{\screenvec}^i\partial_i)\, {}_1 h + (\hat{\screenvec}^i\partial_i)^2 {}_0 h
    \Bigr].
    \label{eq: Psi0 tensor}
\end{align}
Equation~(\ref{eq: Phi00 tensor}) is proportional to the FLRW wave operator
$\square \equiv \partial_\tau^2 + 2 \mathcal{H}\partial_\tau - \nabla^2$ acting on the
line-of-sight projection of $h_{ij}$: for free gravitational waves obeying
$\square h_{ij} = 0$, the Ricci focusing contribution of tensor modes vanishes \emph{``on-shell''}.
This is the expected behaviour, since gravitational waves source Weyl rather than Ricci
curvature. 
The Weyl contribution in Eq.~(\ref{eq: Psi0 tensor}) couples the tensor
polarisations ${}_2 h, {}_1 h, {}_0 h$ through powers of the
transverse complex screen gradient, $\hat{\screenvec}^i\partial_i$. The mapping onto the standard
$E$-mode/$B$-mode decomposition (equivalently, the real-space $+$/$\times$
polarisation basis) and the on-shell reduction for a wave whose wavevector is
aligned with the line of sight are recorded in \cite{canoes2026}.

\paragraph{Full first-order expression.}
Summing the three sectors, the complete first-order driving fields in the Poisson gauge are
\begin{align}
    \Phi_{00}^{(1)} &= \Phi_{00}^{(1,s)} + \Phi_{00}^{(1,B)} + \Phi_{00}^{(1,t)},
    \label{eq: Phi00 full} \\
    \Psi_{0}^{(1)}  &= \Psi_{0}^{(1,s)}  + \Psi_{0}^{(1,B)}  + \Psi_{0}^{(1,t)} .
    \label{eq: Psi0 full}
\end{align}
Equations~(\ref{eq: driving bg})--(\ref{eq: Psi0 full}) are the explicit realisations of
Eqs.~(\ref{eq: Phi 00 from Ricci}) and~(\ref{eq: Psi0}) on the metric~(\ref{eq: poisson gauge}),
and serve as the input to the per-multipole transfer operators that connect the angular power
spectra of the Poisson-gauge perturbation fields $\{\Psi, \Phi, \edited{B_i}, h_{ij}\}$ to those of
the Ricci focusing $\Phi_{00}$ and the Weyl shear $\Psi_0$. Technical details of the implementations of those
transfer operators and of the resulting two- and three-point driving-field cumulants are
presented in the companion paper~\citep{canoes2026}.

\subsection{Sachs propagators in cosmology and the standard weak-lensing kernel}
\label{subsec: from sachs to wl kernels}

In this section we first record the explicit form taken by the free Sachs
response and correlation propagators on the flat-FLRW background. We then
use that explicit form to show, from first principles, that the direct
contribution of a connected $n$-point driving-field cumulant to linearized
weak-lensing observables is its standard lensing-efficiency projection:
the 2PCFs and angular power spectra inherit this projection
already at Order-0 (free theory), while the single-$K$ contribution to the
3PCFs and bispectra inherits it at Order-1. These results connect
the path-integral formalism of Sections~\ref{sec: sachs
dynamics}--\ref{sec: path integral} to the comoving-distance kernel
formulation for the same driving fields, without invoking Limber.
Recovering the usual density-based lensing statistic also requires
specializing the Ricci source to the standard transverse lensing source.

\subsubsection*{Response and correlation propagators}
\begin{figure}[t]
    \centering
    \includegraphics[width=\columnwidth]{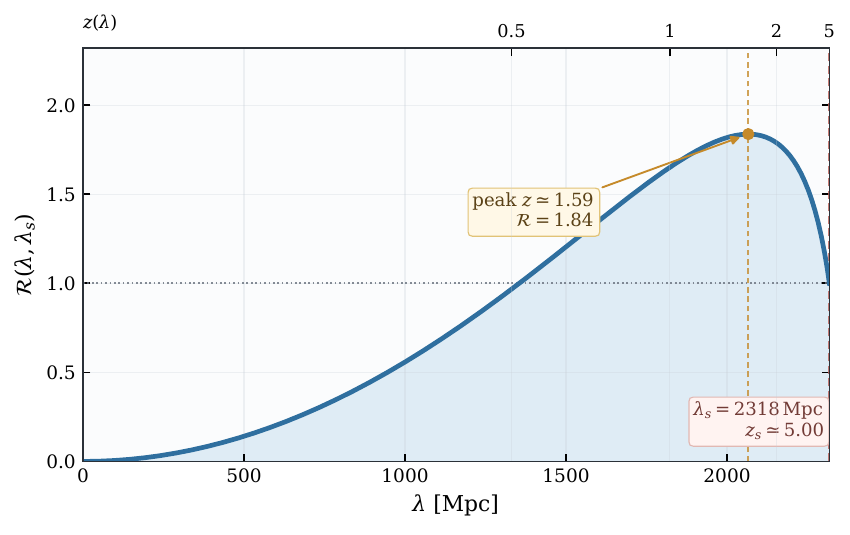}
    \caption{FLRW response propagator [Eq.~(\ref{eq: Resp Op cosmo})] for $\lambda_s=2318\,{\rm Mpc}$
    ($z_s\simeq5$).}
    \label{fig:response-operator-cosmo}
\end{figure}

\begin{figure*}[t]
    \centering
    \includegraphics[width=\textwidth]{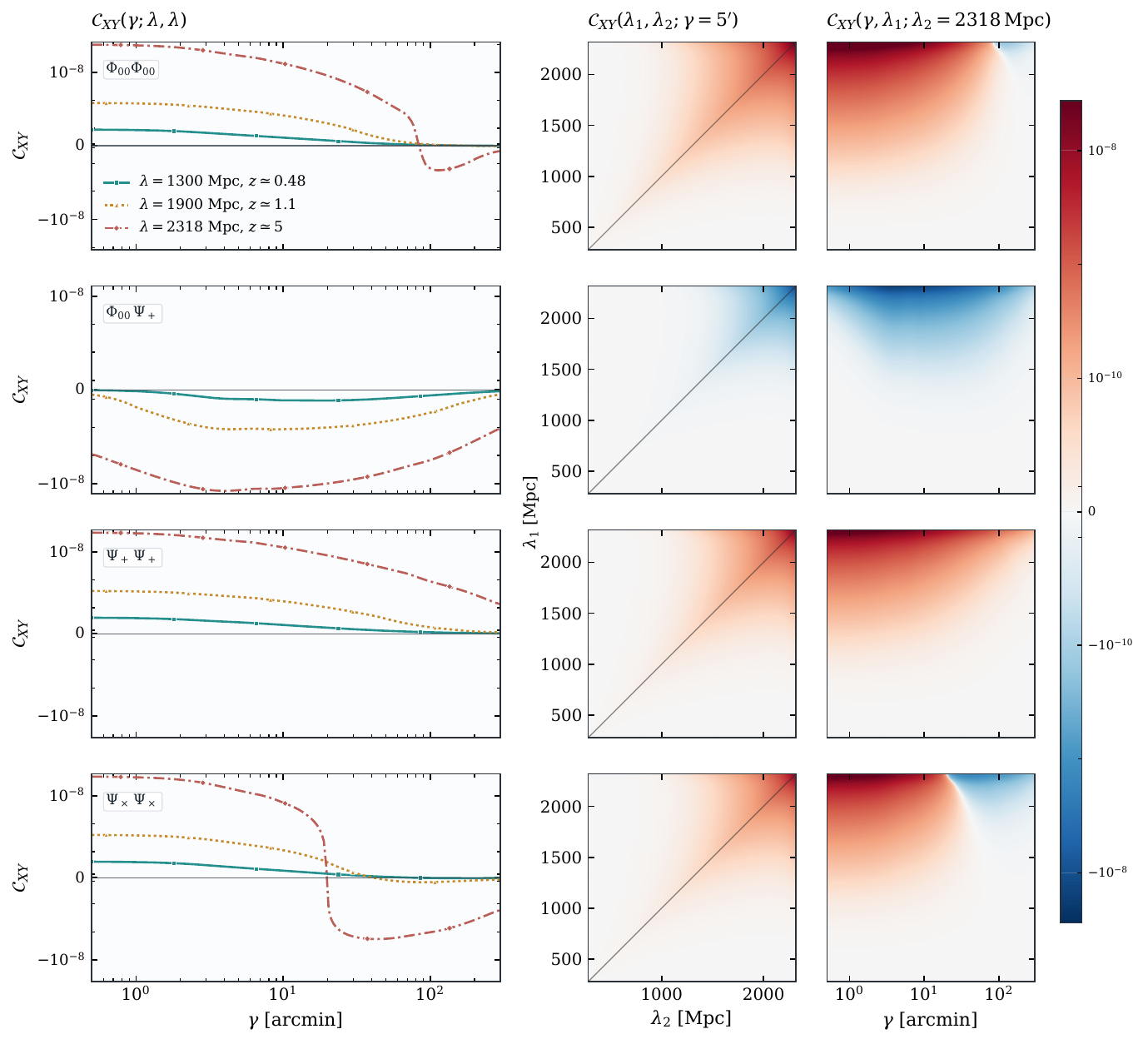}
    \caption{Slices of different components of the correlation propagator:
    $\mathcal{C}_{XY}(\gamma;\lambda,\lambda)$,
    $\mathcal{C}_{XY}(\lambda_1,\lambda_2;\gamma=5')$, and
    $\mathcal{C}_{XY}(\gamma,\lambda_1;\lambda_2=\lambda_s)$.
    The omitted $\Phi_{00}\Psi_\times$ and
    $\Psi_+\Psi_\times$ components each contain one parity-odd
    $\Psi_\times$ insertion; any nonzero numerical residual is negligible
    compared with the displayed components.}
    \label{fig:corr-operator-slices}
\end{figure*}

On the unperturbed FLRW background, $\Phi_{00}^{(0)}$ 
[Eq.~(\ref{eq: driving bg})] sources
the Jacobi equation for the background angular-diameter distance, whose
closed-form solution is $\bar D(\lambda) = a(\lambda)\,\chi(\lambda)$
(Appendix~\ref{append: D equals a chi}). The free response operator
$\RespOp$ and correlation operator $\CorrOp$ of
Section~\ref{sec: sachs dynamics} then collapse to compact forms. The response
propagator is
\be
    \RespOp_{\sachi \sachj}(\direc, \lambda; \direc', \lambda')
    =
    \begin{cases}
        \delta_{\sachi \sachj } \, \delta(\direc - \direc') \, \left( \dfrac{\bar D(\lambda)}{\bar D(\lambda')} \right)^{\!2},
        &   \lambda < \lambda', \\
        0 , & \text{otherwise,}
    \end{cases}
    \label{eq: Resp Op cosmo}
\ee
A fixed-source cut is shown in Fig.~\ref{fig:response-operator-cosmo}.
The correlation propagator is given by
\begin{multline}
    \CorrOp_{\sachi \sachj}(\direc_1, \lambda_1; \direc_2, \lambda_2)
    = \int_0^{\lambda_1}\!\!\diff{\lambda'} \int_0^{\lambda_2}\!\!\diff{\lambda''}\, \\
    \times
    \left( \frac{\bar D(\lambda')}{\bar D(\lambda_1)} \right)^{\!2}
    \cumulant^{\scriptstyle{(2)}}_{\sachi \sachj}(\direc_1, \lambda';\, \direc_2, \lambda'')
    \left( \frac{\bar D(\lambda'')}{\bar D(\lambda_2)} \right)^{\!2}.
    \label{eq: Corr Op cosmo}
\end{multline}
Representative slices of different components of this correlation propagator are shown in
Fig.~\ref{fig:corr-operator-slices}.  The slices use the pair-aligned screen
basis: $\hat x$ lies along the angular separation of the two lines of sight,
while $\hat y$ is perpendicular to the ray-pair plane.  A reflection through
that plane leaves $\Phi_{00}$ and $\Psi_+$ even, but sends
$\Psi_\times$ to minus itself, as summarised in Fig.~\ref{fig:spin2-pattern}.
For the parity-even scalar statistics used here,
components with a single $\Psi_\times$ insertion vanish in the pair-aligned
full-sky continuum; any nonzero numerical residual is negligible compared with
the displayed parity-even components.
Equation~(\ref{eq: Resp Op cosmo}) is the response propagator by which the Sachs
fluctuations propagate from a source vertex at $(\direc,\lambda)$ to a
target vertex at $(\direc',\lambda')$;
Eq.~(\ref{eq: Corr Op cosmo}) is the
double-$\lambda$ `convolution' of the driving-field two-point cumulant through two such
response kernels. 

\subsubsection*{Convergence and shear as Sachs-scalar line-of-sight integrals}
For a null geodesic labelled by $(\direc, \lambda)$, the convergence is
defined at linear order by the perturbation of the angular-diameter distance
about the background trajectory,
\be
    D(\direc, \lambda) = \bar{D}(\lambda)\bigl[\,1 - \kappa(\direc, \lambda)\,\bigr] + \mathcal{O}(\kappa^2),
    \label{eq: kappa definition}
\ee
with the sign convention $\kappa > 0 \Leftrightarrow D < \bar D$ of the
standard weak-lensing literature. Using $\theta = \dot D/D$ and linearising
Eq.~(\ref{eq: kappa definition}), the Sachs expansion fluctuation
$\Dsachs_1 \equiv \theta - \sachssaddle_1$ reduces to
\be
    \Dsachs_1(\direc, \lambda)
    = \frac{d}{d\lambda}\!\left(\frac{\delta D}{\bar D}\right)
    = -\dot{\kappa}(\direc, \lambda) + \mathcal{O}(\kappa^2),
    \label{eq: Dsachs1 as kappa dot}
\ee
so that integrating from the observer vertex $\lambda=0$, where $\kappa$
vanishes geometrically, gives
\be
    \kappa(\direc, \lambda)
    = -\int_{0}^{\lambda} \Dsachs_1(\direc, \lambda')\,\diff{\lambda'}.
    \label{eq: kappa from Dsachs1}
\ee
The source-averaged reference component $\sachssaddle_1$ is absorbed into $\bar D(\lambda)$
and drops out by construction, so $\kappa$ vanishes on the reference trajectory.
Its statistics are inherited entirely from those of $\Dsachs_1$;
its ensemble mean need not vanish.

The same LOS integration applies to the spin-$2$ Sachs
fluctuations $\Dsachs_{2,3} = (\sigma_+, \sigma_\times)$, the shear-rate
components of the Sachs system that vanish on the background and whose
integrals along the past null cone are the two real shear observables on
the sky,
\be
    \gamma_+(\direc, \lambda) \pm \ii\,\gamma_\times(\direc, \lambda)
    = -\int_{0}^{\lambda}\bigl[\Dsachs_2 \pm \ii\,\Dsachs_3\bigr](\direc, \lambda')\,\diff{\lambda'} .
    \label{eq: gamma from Dsachs23}
\ee
Equations~(\ref{eq: kappa from Dsachs1}) and~(\ref{eq: gamma from Dsachs23})
are the linearized relations used in Section~\ref{sec: insights} to attach external
$\kappa$ and $\gamma_\pm$ legs to the path-integral diagrams: every linear
weak-lensing observable on the sky inherits its statistics from the
spin-aware Sachs 3-vector $\Dsachs_\sachi$ propagated through the Sachs
equations.

Let $\mathcal{D}$ be the screen-space Jacobi matrix, with local Sachs matrix
$\mathsf{S}=\dot{\mathcal{D}}\mathcal{D}^{-1}$ in parallel-transported
screen bases~\citep[Eqs.~(2.13)--(2.14)]{fleury2015theory}.
Define $\delta\mathsf{S}=\mathsf{S}-\theta^{\mathrm{(sa)}}\mathsf{I}$
and $\delta\theta=\theta-\theta^{\mathrm{(sa)}}$, where
$\theta^{\mathrm{(sa)}}=\dot{\bar D}/\bar D$. In our shear convention,
\begin{equation}
\begin{aligned}
\delta\mathsf{S}&=
\begin{pmatrix}
\delta\theta+\sigma_+ & \sigma_\times\\
\sigma_\times & \delta\theta-\sigma_+
\end{pmatrix},\\[3pt]
\mathcal{M}&=\frac{\mathcal{D}}{\bar D}=
\begin{pmatrix}
1-\kappa-\gamma_+ & -\gamma_\times-\varpi\\
-\gamma_\times+\varpi & 1-\kappa+\gamma_+
\end{pmatrix}.
\end{aligned}
\label{eq:r2-sachs-jacobi}
\end{equation}
Here $\varpi$ denotes image rotation, distinct from optical twist $\omega$.
Writing $\delta\mathsf{S}_i=\delta\mathsf{S}(\lambda_i)$,
the solution with $\lim_{\lambda\to0}\mathcal{M}=\mathsf{I}$ is
\begin{align}
\mathcal{M}(\lambda_s)
&=\mathcal{P}\exp\!\left[
\varepsilon\int_0^{\lambda_s}\delta\mathsf{S}_1\,d\lambda_1
\right]\notag\\
&=\mathsf{I}+\varepsilon\int_0^{\lambda_s}
\delta\mathsf{S}_1\,d\lambda_1\notag\\
&\quad+\varepsilon^2\int_0^{\lambda_s}d\lambda_1
\int_0^{\lambda_1}d\lambda_2\,
\delta\mathsf{S}_1\delta\mathsf{S}_2
+\mathcal{O}(\varepsilon^3).
\label{eq:r2-ordered-map}
\end{align}
Here $\varepsilon$ counts factors of Sachs fluctuations and is set to one;
$\mathcal{P}$ places the later matrix on the left. Although each local
matrix is symmetric, noncommuting successive shears can give the accumulated
Jacobi map an antisymmetric part, hence image rotation. This is the
multiple-lens effect~\citep{cooray2002second}; a single scalar lens plane,
or commuting shears, gives no such rotation. Our numerical example omits
these quadratic and higher-order corrections to the observable map,
including image rotation and symmetric convergence/shear terms, and is
therefore not a complete second-order weak-lensing calculation.

The four two-point correlation functions used
throughout this paper, all evaluated as functions of the angular separation
$\gamma = \arccos(\direc_1\cdot\direc_2)$ between two lines of sight and
of a pair of source affine parameters $(\lambda_s, \lambda_s')$, are
\begin{align}
    \xi_\kappa(\gamma; \lambda_s, \lambda_s')
        &\equiv \left\langle \kappa(\direc_1,\lambda_s)\,\kappa(\direc_2,\lambda_s') \right\rangle , \label{eq: def xi kappa}\\
    \xi_\pm(\gamma; \lambda_s, \lambda_s')
        &\equiv \bigl\langle \gamma_t(\direc_1,\lambda_s)\,\gamma_t(\direc_2,\lambda_s')
                \notag\\
        &\qquad\quad \pm\, \gamma_\times(\direc_1,\lambda_s)\,\gamma_\times(\direc_2,\lambda_s') \bigr\rangle , \label{eq: def xi shear}\\
    \xi_{\kappa\gamma_t}(\gamma; \lambda_s, \lambda_s')
        &\equiv \left\langle \kappa(\direc_1,\lambda_s)\,\gamma_t(\direc_2,\lambda_s') \right\rangle ,
    \label{eq: def xi cross}
\end{align}
where $(\gamma_t, \gamma_\times)$ are the tangential and cross
components of the shear obtained by rotating $(\gamma_+, \gamma_\times)$
of Eq.~(\ref{eq: gamma from Dsachs23}) into the basis aligned with the
great circle connecting $\direc_1$ and $\direc_2$ \citep{bartelmann2001weak,
kilbinger2015cosmology}.
In this pair-aligned basis we use $\gamma_t=-\gamma_+$.

\subsubsection*{From the correlation operator to the standard weak-lensing kernel}

By construction, the correlation propagator $\CorrOp_{\sachi\sachj}$ 
of Eq.~(\ref{eq: Corr Op cosmo}) is the free-theory two-point function 
of the Sachs scalars $\Dsachs_\sachi$ on the past light cone. 
Integrating the Sachs scalars along the line of sight via Eqs.~(\ref{eq: kappa from Dsachs1})
and~(\ref{eq: gamma from Dsachs23}) then promotes that propagator to
the free-reference two-point function of the weak-lensing observables
themselves. 
As we show below, the resulting expression coincides exactly
with the textbook weak-lensing efficiency kernel projection of the
driving-field two-point cumulant functions.

As a concrete example, we work through the convergence 2PCF $\xi_\kappa$
in detail. 
The first step substitutes Eq.~(\ref{eq: kappa from Dsachs1}) into
$\xi_\kappa$, which lays the two convergence legs onto the past light cone
as a pair of LOS integrals of the Sachs scalar $\Dsachs_1$. Averaging
under the Gaussian reference action contracts the two $\Dsachs_1$ legs
through the correlation operator of Eq.~(\ref{eq: Corr Op cosmo}):
\be
    \xi_\kappa(\gamma; \lambda_s, \lambda_s')
    = \int_0^{\lambda_s}\!\!\diff{\lambda_1} \int_0^{\lambda_s'}\!\!\diff{\lambda_2} \;
      \CorrOp_{11}(\direc_1, \lambda_1;\, \direc_2, \lambda_2) .
    \label{eq: xi kappa from CorrOp}
\ee
%
The second step substitutes the explicit form of $\CorrOp_{11}$ from
Eq.~(\ref{eq: Corr Op cosmo}), which exposes the real-space driving-field
two-point cumulant $\cumulant^{(2)}_{11}$ that ultimately gets convolved with the
response propagator,
\begin{multline}
    \xi_\kappa(\gamma; \lambda_s, \lambda_s')
    = \int_0^{\lambda_s}\!\!\diff{\lambda_1} \int_0^{\lambda_s'}\!\!\diff{\lambda_2}
      \int_0^{\lambda_1}\!\!\diff{\lambda'} \int_0^{\lambda_2}\!\!\diff{\lambda''}\, \\
      \times
      \left[\frac{\bar D(\lambda')\bar D(\lambda'')}{\bar D(\lambda_1)\bar D(\lambda_2)}\right]^{\!2}
      \cumulant^{(2)}_{11}(\gamma; \lambda', \lambda'') ,
    \label{eq: xi kappa unfolded}
\end{multline}
where $\cumulant^{(2)}_{11}(\gamma; \lambda', \lambda'')$ 
is the two-point cumulant (or connected 2-point correlation) of the Ricci focusing field.

The two outer integrals, over $\lambda_1$ and $\lambda_2$, act only on the
response factors $\bar D^{-2}(\lambda_1)$ and $\bar D^{-2}(\lambda_2)$.
Swapping the order of integration via
\be
    \int_0^{\lambda_s}\!\!\diff{\lambda_1}\!\int_0^{\lambda_1}\!\!\diff{\lambda'}
    = \int_0^{\lambda_s}\!\!\diff{\lambda'}\!\int_{\lambda'}^{\lambda_s}\!\!\diff{\lambda_1},
\ee
and folding the inner $\lambda_1$- and $\lambda_2$-integrals into effective
weights, Eq.~(\ref{eq: xi kappa unfolded}) reduces to the two-fold form
\begin{multline}
    \xi_\kappa(\gamma; \lambda_s, \lambda_s')
    = \int_0^{\lambda_s}\!\!\diff{\lambda'}\!\int_0^{\lambda_s'}\!\!\diff{\lambda''}\, \\
      \times
      K(\lambda',\lambda_s)\,K(\lambda'',\lambda_s')\,
      \cumulant^{(2)}_{11}(\gamma; \lambda', \lambda'') ,
    \label{eq: xi kappa folded}
\end{multline}
with effective affine-parameter kernel
\be
    K(\lambda,\lambda_s)
    \equiv
    \bar D^2(\lambda) \int_\lambda^{\lambda_s} \frac{\diff{\lambda_1}}{\bar D^2(\lambda_1)} .
    \label{eq: K affine kernel}
\ee
Equation~(\ref{eq: K affine kernel}) is the affine-parameter form of the
weak-lensing window.
Substituting $\bar D(\lambda) = a(\lambda)\chi(\lambda)$
and the affine measure $\diff{\lambda} = a^2(\chi)\,\diff{\chi}$ into
Eq.~(\ref{eq: K affine kernel}), the $a^{2}$ factors cancel inside the
integrand and the affine kernel collapses to the closed form
\be
    K(\lambda,\lambda_s)
    = a^2(\chi)\,\chi^{2} \int_\chi^{\chi_s}\!\frac{\diff{\chi_1}}{\chi_1^{2}}
    = a^2(\chi)\,\frac{\chi(\chi_s - \chi)}{\chi_s} .
    \label{eq: K comoving kernel}
\ee
The geometric factor $\chi(\chi_s-\chi)/\chi_s$ is the textbook lensing
efficiency~\citep{Bernardeau2002Review}; the prefactor $a^2(\chi)$ is the
Jacobian that converts the affine-parameter density of the source field to
a comoving-distance density. Because $\cumulant^{(2)}_{11}(\lambda, \lambda')$ is 
essentially a density in $\lambda$, the same change of radial measure
$\diff{\lambda} = a^2(\chi)\,\diff{\chi}$ acts on it as Jacobian factors,
\be
    \cumulant^{(2)}_{11}(\gamma; \chi, \chi')
    \,=\,
    a^2(\chi)\,a^2(\chi')\,
    \cumulant^{(2)}_{11}\!\bigl(\gamma; \lambda(\chi), \lambda(\chi')\bigr) .
    \label{eq: cumulant 2 measure change}
\ee
The same symbol on the two sides denotes the same physical correlation
expressed through two different radial measures; the field $\Phi_{00}$
itself is not rescaled. Substituting into
Eq.~(\ref{eq: xi kappa folded}) and changing the integration variable
from $\lambda$ to $\chi$ then gives
\begin{multline}
    \xi_\kappa(\gamma; \chi_s, \chi_s')
    = \int_0^{\chi_s}\!\diff{\chi}\!\int_0^{\chi_s'}\!\diff{\chi'}\, \\
      \times
      \frac{\chi(\chi_s-\chi)}{\chi_s}\,
      \frac{\chi'(\chi_s'-\chi')}{\chi_s'}\,
      \cumulant^{(2)}_{11}(\gamma; \chi, \chi') .
    \label{eq: xi kappa comoving}
\end{multline}
This is the standard kernel projection: two lensing-efficiency
windows acting on the same Ricci-focusing two-point cumulant, read in comoving-distance
coordinates. 
For continuity with the more familiar harmonic-space formula, the
Legendre transform [see Eq.~(\ref{eq: src cumulant Legendre})] 
of Eq.~(\ref{eq: xi kappa comoving}) gives
\begin{multline}
    C_\ell^{\kappa\kappa}(\chi_s, \chi_s')
    = \int_0^{\chi_s}\!\diff{\chi}\!\int_0^{\chi_s'}\!\diff{\chi'}\, \\
      \times
      \frac{\chi(\chi_s-\chi)}{\chi_s}\,
      \frac{\chi'(\chi_s'-\chi')}{\chi_s'}\,
      C_\ell^{\Phi_{00}\Phi_{00}}(\chi, \chi') ,
    \label{eq: Cell kappa harmonic bridge}
\end{multline}
where $C_\ell^{\Phi_{00}\Phi_{00}}(\chi, \chi')$ is the harmonic
projection of the same driving-field two-point cumulant on each radial slice.
Equation~(\ref{eq: Cell kappa harmonic bridge}) is the harmonic form of
the identity proved above.
For a fixed driving-field cumulant, the equivalence is exact: no Limber approximation, no
equal-shell collapse, and no flat-sky simplification have been invoked.
The same chain runs for every other linear weak-lensing 2PCF: $\xi_+$,
$\xi_-$, $\xi_{\kappa\gamma_t}$ take the form of
Eq.~(\ref{eq: xi kappa folded}) with the appropriate spin-block of
$\cumulant^{(2)}_{\sachi\sachj}$ on the right-hand side (now including
the $\Psi_0$ contribution that drives the shear-rate channels).

\paragraph{A Unified View.}
The derivation above is not specific to the two-point case. 
It is the consequence of the generic ingredients: the FLRW 
response operator [Eq.~(\ref{eq: Resp Op cosmo})], which acts on driving field cumulants. 
For any $n$-point correlation function of the weak-lensing
observables, the direct cumulant contribution in the path-integral expansion folds the
connected $n$-point cumulant of the driving fields through $n$ external response
propagators, and the change of radial measure to $\chi$ applies on each
external propagator. The general statement is therefore the following:
\begin{quote}\itshape
The conventional lensing-efficiency-kernel formulation corresponds to the
leading connected diagram associated with the corresponding driving-field cumulant.
\end{quote}
\noindent In comoving radial coordinates the response operator $\RespOp$
collapses to the textbook lensing-efficiency window
$\chi(\chi_s-\chi)/\chi_s$, with one such collapse per external propagator
giving the $n$-window projection of the same driving-field cumulant.

Two concrete cases bracket the general statement. For the weak-lensing
2PCFs, this result is already the free-theory (order-0)
prediction, since the response operator covers the two-point cumulant case exactly.
For the connected 3PCFs ($\langle\kappa\kappa\kappa\rangle$ and friends), the same result appears
as the single-$K$ term at Order-1 in the path-integral expansion, while
the order-0 contribution vanishes identically: a centred Gaussian free
theory has no three-point function. A single-$F$ contribution also occurs
at Order-1 and generates propagation-induced non-Gaussianity, as shown
in Fig.~\ref{fig: feyn diag 3pt order1}.

The remaining diagrams of the path-integral expansion describe nonlinear
Sachs propagation; Section~\ref{sec: insights} presents illustrative examples.

\section{Observable consequences for weak lensing}
\label{sec: insights}
\label{sec: case studies}  

Expressing convergence and shear as linear integrals of Sachs-scalar
fluctuations, we use the framework of
Sections~\ref{sec: sachs dynamics}--\ref{sec: cosmological
setup} to write weak-lensing correlation functions as sums of diagrams,
organised along two physical axes: how nonlinearly the ray bundle propagates,
carried by the cubic Sachs vertex (the $F$ insertions), and how non-Gaussian
the driving fields are, carried by their connected cumulants (the $K$
insertions). Each diagram is labelled by how many insertions of each kind it
carries, $(n_{F},n_{K})$, with the total order $n=n_{F}+n_{K}$. We use the two-point correlation functions (2PCFs)
as a worked example. Their lowest order, Order-0, is the free theory, with no
insertion of either kind; the leading corrections to it appear at Order-2
through two channels, FF $(2,0)$ and FK $(1,1)$
(Fig.~\ref{fig: feyn diag 2pt order2}). Throughout we follow the four 2PCFs
$\xi_{\kappa}$, $\xi_{+}$, $\xi_{-}$ and $\xi_{\kappa\gamma_{t}}$ on a single
source plane at $z_{s}=5$, and take the driving-field statistics, built from
the linear matter power spectrum and the tree-level matter bispectrum projected
onto the lightcone, from the companion paper~\citep{canoes2026} as tabulated
inputs; the three-point cumulant that seeds the FK channel is constructed in
Appendix~\ref{append: driving-field spectra}. The matter sector is kept linear and tree-level; nonlinear-matter and
perturbed-path refinements would enter by the same construction, and we leave them
to future work. We take Order-0 and the two Order-2 channels in turn.

\subsection{A unified view}
\label{subsec: order zero}

Order-0 of any 2PCF is the free theory: the two external Sachs legs are paired
through the Gaussian reference action $\ActionRef$, with no vertex inserted. 
Section~\ref{subsec: from sachs to wl
kernels} showed analytically that, after the change of radial variable to
comoving distance, this contraction reduces to the comoving double integral
[Eq.~(\ref{eq: xi kappa comoving}) for the
convergence]\footnote{The other three 2PCFs follow with the matching spin-0 or spin-2
two-point cumulant in place of $\cumulant^{(2)}_{11}$ in Eq.~(\ref{eq: xi kappa comoving})}.
This is the lensing-efficiency projection of the same driving-field
cumulant, with no Limber, equal-shell, or flat-sky step.
Specializing the focusing field to the standard transverse lensing source
and applying the Limber approximation then gives the familiar density-based
Born--Limber statistic,
\begin{align}
    \xi_{\kappa}(\gamma)
    \;&=\;
    \frac{1}{2\pi}\!\int\!\diff{\ell}\,\ell\,P_{\ell}(\cos\gamma)\,
        C_{\ell}^{\kappa\kappa,\,{\rm BL}} , \\
    C_{\ell}^{\kappa\kappa,\,{\rm BL}}
    \;&\propto\!\int_{0}^{\chi_{s}}\!\!\diff{\chi}\,\edited{\frac{W^{2}(\chi)}{\chi^{2}}}\,
        P_{m}\!\bigl((\ell+1/2)/\chi,\, \chi\bigr) ,
    \label{eq: BL kkkernel}
\end{align}
with $W(\chi) = \tfrac{3}{2}\Omega_{m} H_{0}^{2}(1+z)\chi(\chi_{s}-\chi)/\chi_{s}$
the density-lensing window. The source specialization and the equal-shell
collapse used in this expression do not change its optical Order-0.
Adding $F$ or $K$ insertions gives the departures studied below.

\begin{figure*}[t]
    \centering
    \includegraphics[width=\textwidth]{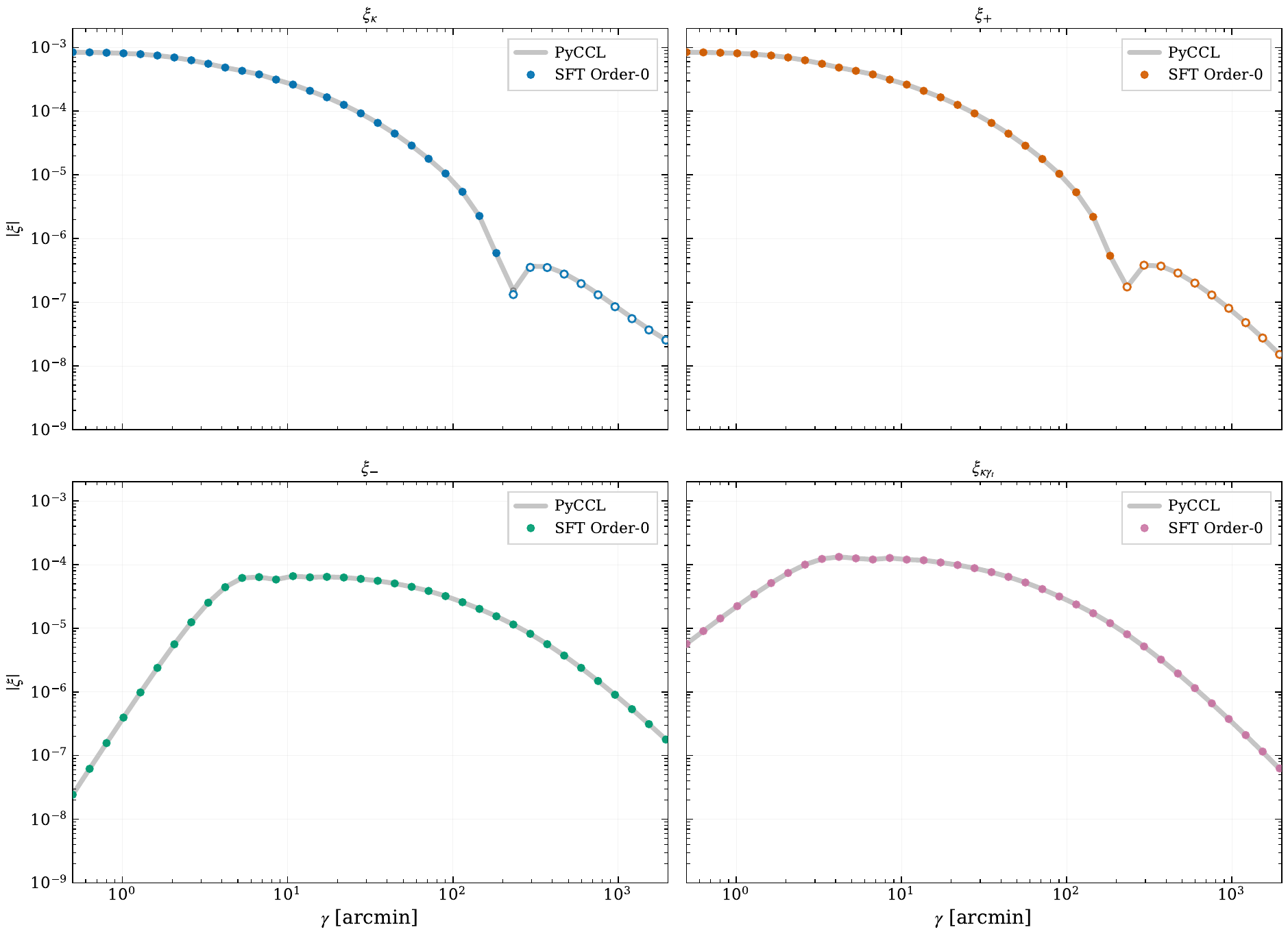}
    \caption{Order-0 prediction compared with an independent
    \textsc{PyCCL} calculation at $z_s=5$, using matched matter statistics,
    background evolution, radial support, and multipole range. The
    non-Limber calculation extends through $\ell=800$, with Limber
    integration at higher multipoles. The panels show $\xi_\kappa$
    (top left), $\xi_+$ (top right), $\xi_-$ (bottom left), and
    $\xi_{\kappa\gamma_t}$ (bottom right). Markers are filled for positive
    values and hollow for negative values. Grey curves show
    \textsc{PyCCL}. The Ricci-source distinction and numerical checks
    are described in the text.}
    \label{fig: BL O0 vs pyccl}
\end{figure*}

Figure~\ref{fig: BL O0 vs pyccl} provides an independent numerical
cross-check against \textsc{PyCCL}~\citep{chisari2019core} at $z_s=5$.
We match the linear matter power spectrum, background evolution, radial
support, and multipole range. \textsc{PyCCL} evaluates the angular power
spectra with the method of Fang, Krause, Eifler and MacCrann
(FKEM)~\citep{FangEtAl2020} through $\ell=800$ and uses Limber integration
at higher multipoles. FKEM retains correlations between different radial
distances. This choice of integration method introduces no $F$ or $K$
insertions and is independent of the optical expansion order.

At low multipoles, our Ricci source retains the radial and time-derivative
terms in Eq.~(\ref{eq: Phi00 scalar expanded}), whereas the standard
\textsc{PyCCL} lensing kernel uses the transverse lensing projection.
\textsc{PyCCL} includes the background expansion but does not retain all
these additional Ricci terms. Above $\ell=800$, our Limber implementation
omits radial-derivative Ricci terms. Over the displayed angular range,
the largest absolute difference is below $0.013\%$ of the corresponding
\textsc{PyCCL} curve's peak amplitude. Tested numerical refinements change
the predicted correlations by up to $0.017\%$ of a channel's peak.
These diagnostics do not isolate the effect of the additional Ricci terms.
The analytic identity in Eq.~(\ref{eq: xi kappa comoving}) concerns the
projection of a fixed driving-field cumulant and holds independently of
Limber.

\subsection{Nonlinear propagation (FF)}
\label{subsec: nonlinear propagation insight}

The first correction beyond the free theory comes from nonlinear propagation
alone. Switching off driving-field non-Gaussianity, so that the driving fields
are Gaussian, and keeping the cubic Sachs vertex leaves a single Order-2
diagram in the 2PCF, FF $(n_{F}=2)$: two propagation insertions tied together by the
Gaussian two-point statistics of the driving fields.
This is the result of the non-linear propagation sourced by the non-linear terms in Sachs equations.
It is a pure
propagation effect, and survives even when the driving fields are perfectly
Gaussian.

\begin{figure*}[t]
    \centering
    \includegraphics[width=\textwidth]{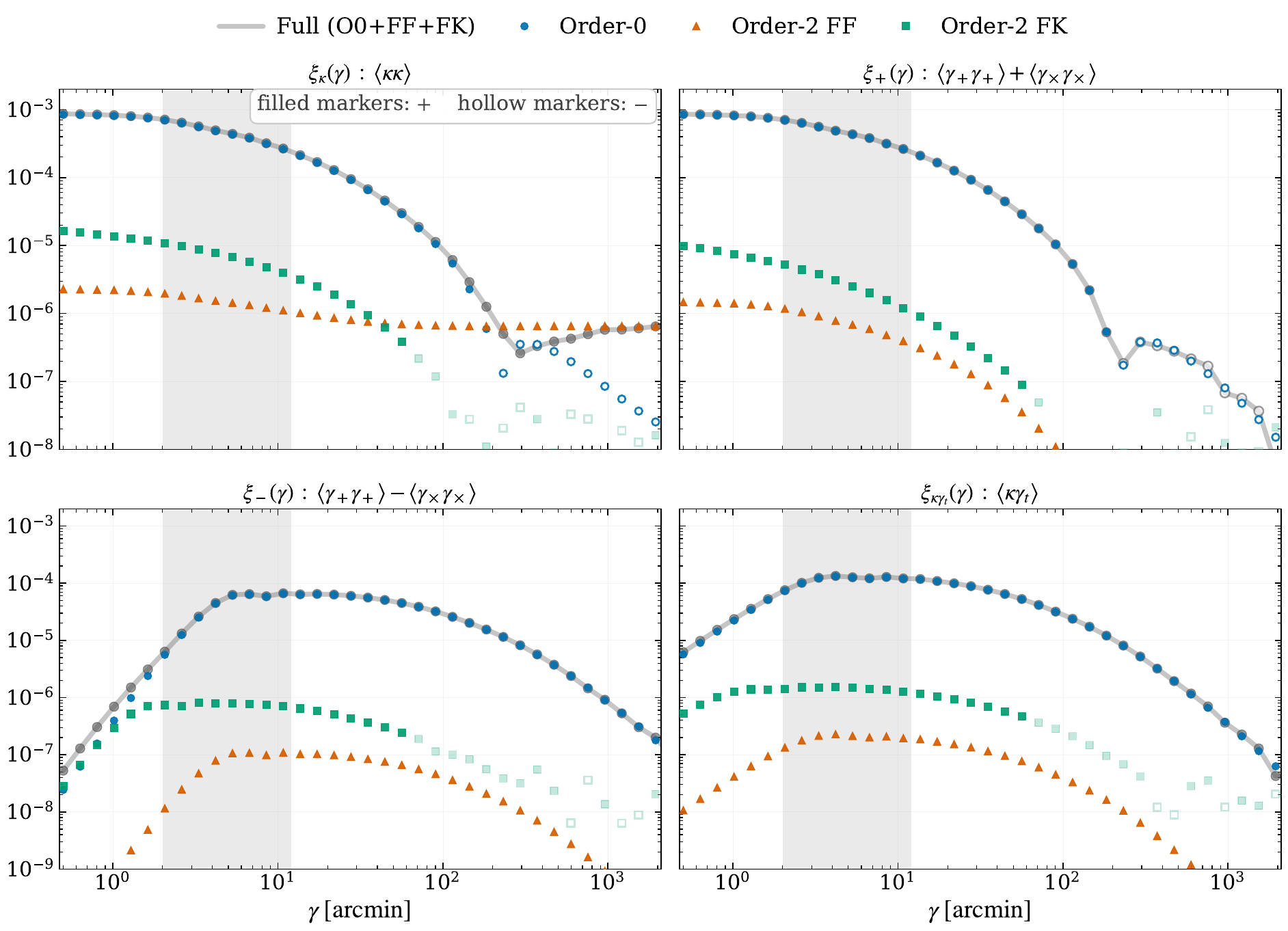}
    \caption{Order-0 and Order-2 decomposition of the four 2PCFs at $z_{s}=5$.
    Order-0 (blue) reproduces Figure~\ref{fig: BL O0 vs pyccl}; the Order-2
    correction splits into FF (orange, nonlinear propagation) and FK (green,
    driving-field non-Gaussianity, Section~\ref{subsec: mixed hierarchy}), and
    the grey curve is their sum. \edited{The band marks $2'$--$12'$, the
    range over which FK is quoted in the text. FK increases
    all four correlations at the sampled points in this band;
    its fractional size depends on the observable.
    As Order-0 decays, FF supplies the large-separation plateau of the
    convergence moment.} \edited{The FK series is drawn faint beyond $\gamma\simeq1^\circ$, where values are excluded from the quantitative discussion (Section~\ref{subsec: cutoff}).} Markers are
    filled where positive, hollow where negative.}
    \label{fig: NLO 2pt FFFK}
\end{figure*}

On the convergence (top-left of Fig.~\ref{fig: NLO 2pt FFFK}) FF reaches
$2.28\times10^{-6}$ at $\gamma=0.5'$, about $0.27\%$ of Order-0 there.
It decreases toward a large-separation plateau of approximately
$6.55\times10^{-7}$, so as the Order-0 signal falls steeply with angle the
FF correction grows in relative importance toward large separations.
In the sampled curve, it overtakes FK at $44'$ and the absolute
Order-0 signal at $183'$.\footnote{\edited{The Sachs vertex gives the convergence a
small mean value at second order. Because $\xi$ is defined as a moment rather
than a covariance, the product of that mean along the two lines of sight is not
subtracted off, and unlike a connected correlation it does not fall away with
separation.}}

\begin{figure*}[t]
    \centering
    \includegraphics[width=\textwidth]{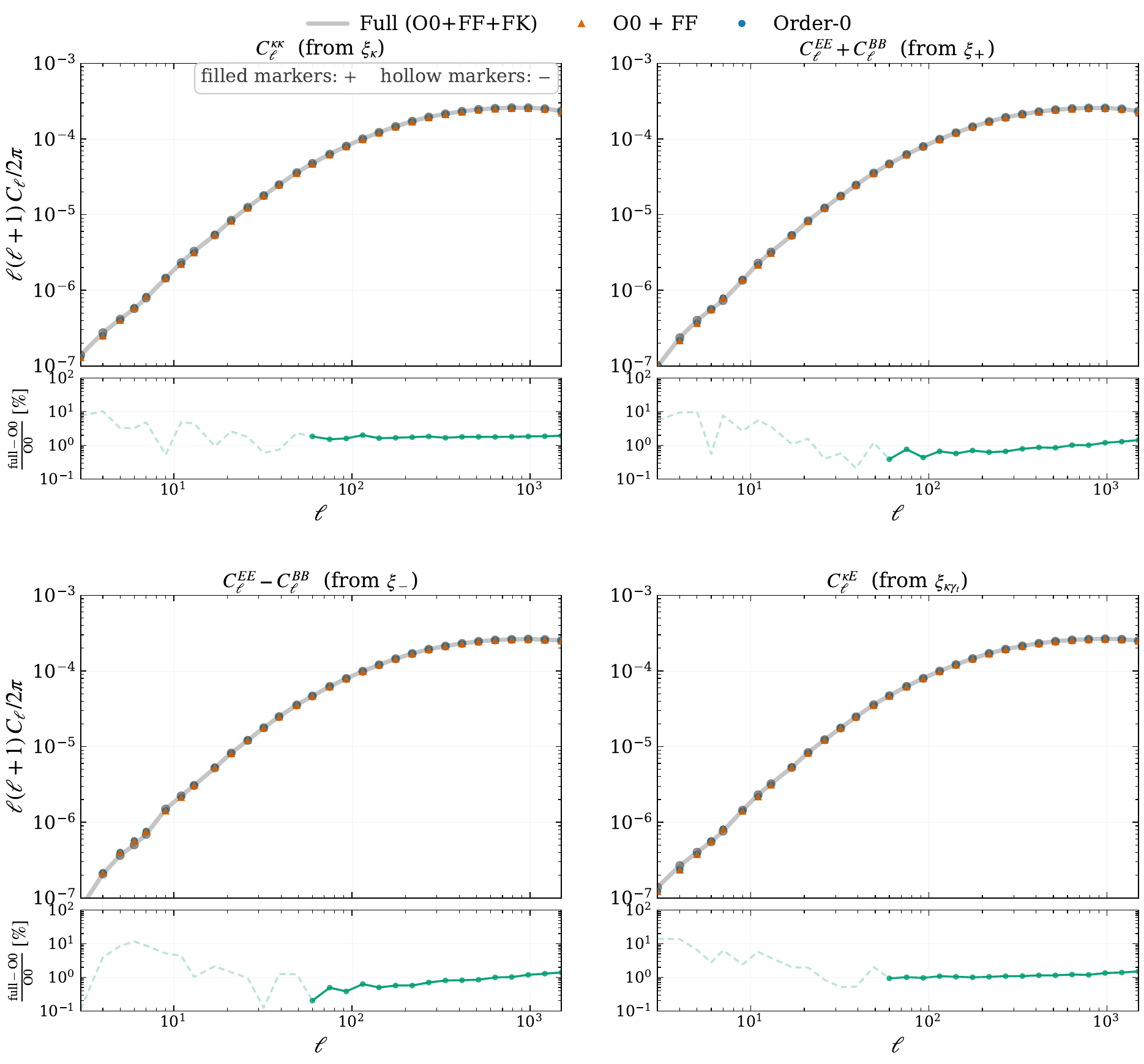}
    \caption{Angular power spectra of the four observables of
    Fig.~\ref{fig: NLO 2pt FFFK}: the full prediction (Order-0$+$FF$+$FK, grey),
    Order-0$+$FF (orange), and Order-0 (blue), as the band power
    $\ell(\ell+1)C_{\ell}/2\pi$ at $z_{s}=5$, from a curved-sky (Wigner-$d$)
    transform of the real-space 2PCFs\edited{; the strip below each panel
    shows the total Order-2 correction as a percentage of Order-0}.
    Before transforming $\xi_\kappa$, its large-separation plateau is
    estimated from the largest sampled separation and subtracted.
    No constant is subtracted from the spin correlations. Finite angular
    coverage and sampling introduce transform residuals, including apparent
    E/B mixing, assessed in Fig.~\ref{fig: cl EB}.
    Markers filled/hollow for positive/negative. Points below
    $\ell\sim50$ are drawn faint and ratios are quoted for
    $\ell\gtrsim50$.}
    \label{fig: NLO 2pt FFFK cl}
\end{figure*}

\subsection{Leakage of driving-field non-Gaussianity (FK)}
\label{subsec: mixed hierarchy}

We now look at the contributions from non-Gaussian statistics of the driving fields. 
At the second order of Feynman diagrammatic expansion,
the component with two cumulant vertices (KK) vanishes (see Fig.~\ref{fig: feyn diag 2pt order2}), but one
nonlinear-propagation vertex F allows another vertex from the driving-field three-point cumulant $\cumulant^{(3)}$, viewing as the leakage of driving field non-Gaussianity into lensing 2PCF as the result of nonlinear Sachs evolution. 
More generally, for this linearized observable map, the hierarchy mixing effect follows the selection rule as stated in Section~\ref{subsec: selection rule}: an $n$-point observable couples directly to the $n$-point
driving-field cumulant, while its first hierarchy-mixing correction is sourced by the
$(n+1)$-point cumulant through one nonlinear-propagation insertion (a single $F$
vertex), and no higher cumulant can reach that order.
For the 2PCF this selection rule singles out the three-point cumulant $\cumulant^{(3)}$: the
four-point cumulant and beyond are forbidden from the leading correction by the
vertex counting itself, not removed by hand.
In other words, which cumulants can appear is fixed by the structure of the expansion, not by any
truncation. 

In the numerical example, the three-point cumulant that seeds FK is represented by
the equal-shell reduction $\zeta_{XYZ}(\gamma,\lambda)$ of the full
cross-shell driving-field cumulant in Limber approximation, 
evaluated on the collapsed (or `squeezed' if beyond limber approximation)
configuration $(1,\cos\gamma,\cos\gamma)$ selected by the 2PCF FK diagram.\footnote{The propagators in the FK diagrams are all given by $\RespOp$, which carries a directional $\delta$-function. Consequently, the three legs attached to a K-vertex must share the directions of the two external points. In the Limber approximation, two legs sharing the same direction correspond to the same point.}
It is built from the tree-level matter bispectrum\citep{Bernardeau2002Review}, 
converted to a bispectrum of the gravitational potential by the
Poisson relation and then transferred to Ricci focusing and Weyl shearing potential statistics, 
which is then projected onto the convergence (spin-0) and shear
(spin-2) patterns seen on the sky; Appendix~\ref{append: driving-field spectra}
gives the construction and the equal-shell reduction it relies on.
Because the configuration is collapsed (or squeezed once the Limber
approximation is relaxed), a long-wavelength separation couples to two local,
small-scale modes, so the FK amplitude depends on the small-scale matter
bispectrum entering this cumulant (Fig.~\ref{fig: zeta slices}).

How large FK is depends on each observable's spin structure. The shear is a
spin-$2$ field, so a pair of shear legs can combine in two ways. The conjugate
pairing in $\xi_+$ has net spin $0$, while the pairing in $\xi_-$ has
spin $4$. At $z_s=5$ and
$\ell_{\max}=15360$, FK raises the convergence $\xi_{\kappa}$ by
$1.5$--$1.6\%$ and the parity-even shear $\xi_{+}$ by $0.5$--$0.8\%$
over the sampled separations within $2'$--$12'$. These correct the propagation;
the mapping from $(\kappa,\gamma)$ to a measured galaxy shape carries its own
reduced-shear correction~\citep{dodelson2006reduced,krause2010weak}.
In the convergence-shear cross $\xi_{\kappa\gamma_{t}}$ it raises the
signal by $1.2$--$2.3\%$ across that range. In the shear difference
$\xi_{-}$ the ratio is not flat: $1.1$--$1.3\%$ at the sampled points
above $5'$, it climbs to $13.2\%$ near the $2'$ edge of the band.
These ratios also depend on the size of the corresponding Order-0 statistic.
A difference between $\xi_+$ and $\xi_-$ does not by itself imply B-mode
power, since the two correlations use different angular kernels even for a
pure E field.

\subsection{\edited{Convergence in the multipole cutoff}}
\label{subsec: cutoff}

\begin{figure}[t]
    \centering
    \includegraphics[width=\columnwidth]{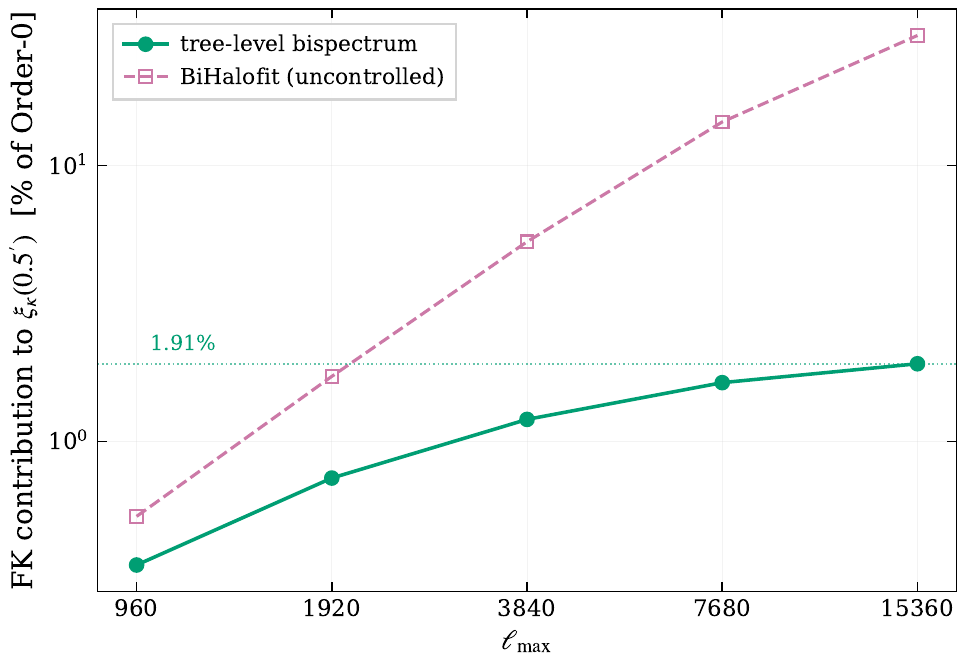}
    \caption{\edited{FK contribution to $\xi_{\kappa}$ at $\gamma=0.5'$,
    as a fraction of Order-0, against the multipole cutoff of the vertex
    table. With the tree-level bispectrum, the final doubling from $7680$
    to $15360$ adds $17\%$; with the BiHalofit bispectrum it adds $106\%$.
    These finite-cutoff changes do not establish an infinite-cutoff limit.}}
    \label{fig: fk cutoff}
\end{figure}

\edited{The FK term depends on the driving-field three-point cumulant in its
collapsed configuration, with two legs at one point. A coincident-point moment
depends on the smallest scales retained, so the multipole cutoff
$\ell_{\max}$ of the vertex table regulates the sum and its
convergence has to be demonstrated rather than assumed.
With the tree-level bispectrum (Fig.~\ref{fig: fk cutoff}), the FK
contribution to $\xi_{\kappa}$ at $\gamma=0.5'$, the smallest separation
computed, rises from $0.36\%$ to $1.91\%$ of Order-0 as $\ell_{\max}$ runs
from $960$ to $15360$; the final doubling adds $17\%$ here, and
$8.5$--$10.8\%$ over the sampled separations within $2'$--$12'$.
These changes measure finite-cutoff sensitivity; they do not establish an
infinite-cutoff value or error bound. Apart from this cutoff comparison, every FK result
in this paper is evaluated at $\ell_{\max}=15360$.}

\edited{The angular dependence also matters when testing the cutoff.
The Legendre factors in Eq.~(\ref{eq: appendix zeta squeezed}) oscillate
at finite separation, so cancellations can make the residual sensitive
to the multipole range. The cutoff test here samples separations below
$14'$. It does not establish convergence of the wide-angle tail.}

\edited{With a nonlinear matter bispectrum the same sum is more sensitive to
small-scale power: the BiHalofit~\citep{takahashi2020bihalofit} curve in
Fig.~\ref{fig: fk cutoff} increases by $106\%$ at $\gamma=0.5'$ when
$\ell_{\max}$ increases from $7680$ to $15360$, compared with $17\%$ for
the tree-level result. A physical nonlinear prediction depends on the
assumed small-scale matter statistics and any smoothing applied to the
driving fields, which is why we quote no calibrated nonlinear amplitude.}

\paragraph*{Validation.} \edited{A Monte-Carlo integration of the stochastic
Sachs equation tests both Order-2 channels with matched inputs (Fig.~\ref{fig: val mc}). The
two channels need different estimators; Appendix~\ref{append: validation}
describes both tests and gives the numbers. The Monte-Carlo is independent of
the expansion, solving the Sachs system directly rather than folding it.
FK agrees within $1\%$ at four tested separations. FF agrees within $0.7\%$
with its continuum Order-$F^2$ reference for the local covariance used by the
simulation; this does not test the nonlocal cosmological FF prediction. A second check
contracts the tabulated cumulant with the FK kernel by plain quadrature,
agreeing within $1\%$. The expansion and the package that evaluates it
are validated more broadly in~\citet{SFTwick2026}.}

\bigskip
\noindent
Across the four 2PCFs the worked example shows a clear pattern. Order-0
reproduces the conventional kernels for matched driving-field cumulants;
the leading corrections then arrive at
Order-2 in two physically distinct channels, nonlinear propagation (FF) and the
leakage of driving-field non-Gaussianity (FK). The four correlations respond
differently to these terms, so keeping them separate preserves the origin of
each correction when relating the two-point statistics to the matter model.
The same structure points beyond the 2PCF, to higher-point
lensing statistics, which we take up in the conclusion.

\section{Conclusions and Discussions}
\label{sec: conclusion}

We have reformulated weak gravitational lensing as a stochastic field theory
for the Sachs optical scalars of a ray bundle, driven by the random
Ricci-focusing and Weyl-shearing fields of the intervening matter. With the
linearized observable map used here, promoting the Sachs trajectory to a path
integral turns $n$-point functions of convergence and shear into diagrammatic
expansions, ordered by how nonlinearly the bundle propagates and how
non-Gaussian the driving fields are.
Working the two-point correlation functions as a worked example brings out
three structural insights.

First, the direct contribution of a specified driving-field cumulant
projects it through the usual lensing-efficiency windows. For the two-point
function this is the free theory, with no $F$ or $K$ insertion, and the
projection identity holds independently of Limber. The standard
density-based statistic additionally uses the transverse lensing source
and, in its Born--Limber form, an equal-shell collapse. Optical Order-0
therefore remains the baseline whether the radial projection is evaluated
with or without Limber.

Second, for this linearized observable map, a selection rule decides which
driving-field cumulants can reach a
given observable: only the next cumulant up the hierarchy enters the leading
correction, and it does so through a single nonlinear-propagation insertion. For
the two-point function this singles out the three-point cumulant, with the
four-point cumulant and beyond shut out. The cutoff is structural, fixed by the
diagram counting rather than imposed by hand.

\begin{figure*}[t]
    \centering
    \includegraphics[width=\textwidth]{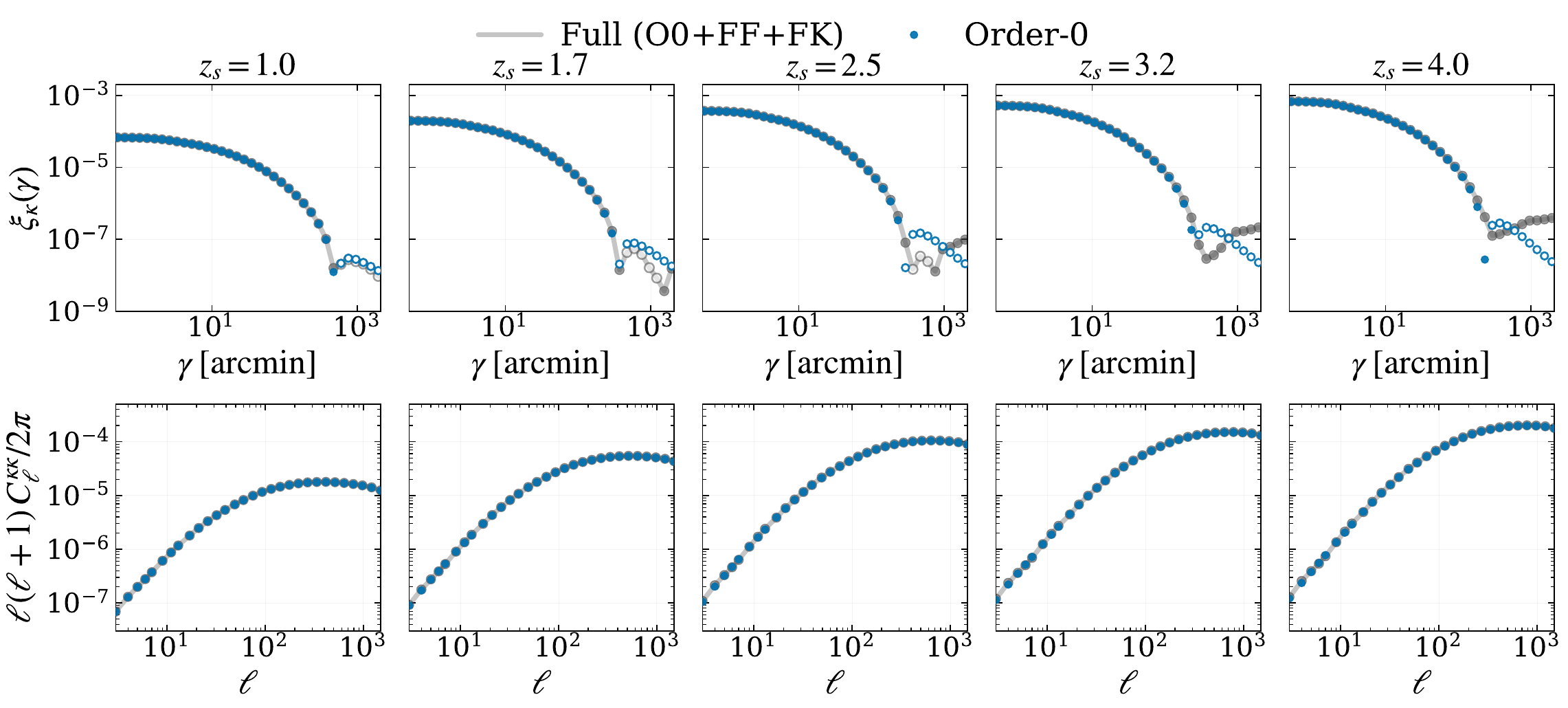}
    \caption{Source-redshift dependence of the convergence two-point
    statistics. The top row shows $\xi_\kappa(\gamma)$ and the bottom
    row shows its curved-sky transform $C_\ell^{\kappa\kappa}$ for
    $z_s=1,\,1.7,\,2.5,\,3.2,\,4.0$. Each panel compares
    Order-0$+$FF$+$FK (grey) with Order-0 (blue), at
    $\ell_{\max}=15360$ for FK. The fractional FK contribution increases
    with source redshift over the sampled separations within $2'$--$12'$.
    FF supplies a large-separation plateau in the convergence moment.
    Before the harmonic transform, this plateau is estimated from the
    largest sampled separation and subtracted. Markers are filled for
    positive values and hollow for negative values.}
    \label{fig: multiz kappa}
\end{figure*}

Third, this mixing gives small-scale structure a route into large-angle lensing
statistics. \edited{The three-point cumulant enters the two-point function in a
near-degenerate configuration, two short-wavelength legs sharing a line of
sight while the third follows the separation, so the contribution is sensitive
to small-scale modes of the matter bispectrum at every separation.} \edited{At $z_s=5$ and $\ell_{\max}=15360$, FK raises the convergence
correlation by $1.5$--$1.6\%$ and $\xi_+$ by $0.5$--$0.8\%$ over the
sampled separations within $2'$--$12'$, the band used in the discussion
of current cosmic-shear cuts~\citep{asgari2021kids,amon2022dark}.
The other two correlations respond differently, including a positive
convergence--tangential-shear correction. For convergence, the fractional
FK correction increases with source redshift and exceeds FF over these
sampled angles (Fig.~\ref{fig: multiz kappa}). Reduced shear is a separate
correction to the shape observable, rather than to propagation, and can
be comparable in magnitude~\citep{dodelson2006reduced,krause2010weak}.} The
nonlinear-propagation and non-Gaussianity corrections are physically distinct
and show up in different combinations of the data, a separation that later
analyses can exploit.

\begin{figure*}[t]
    \centering
    \includegraphics[width=\textwidth]{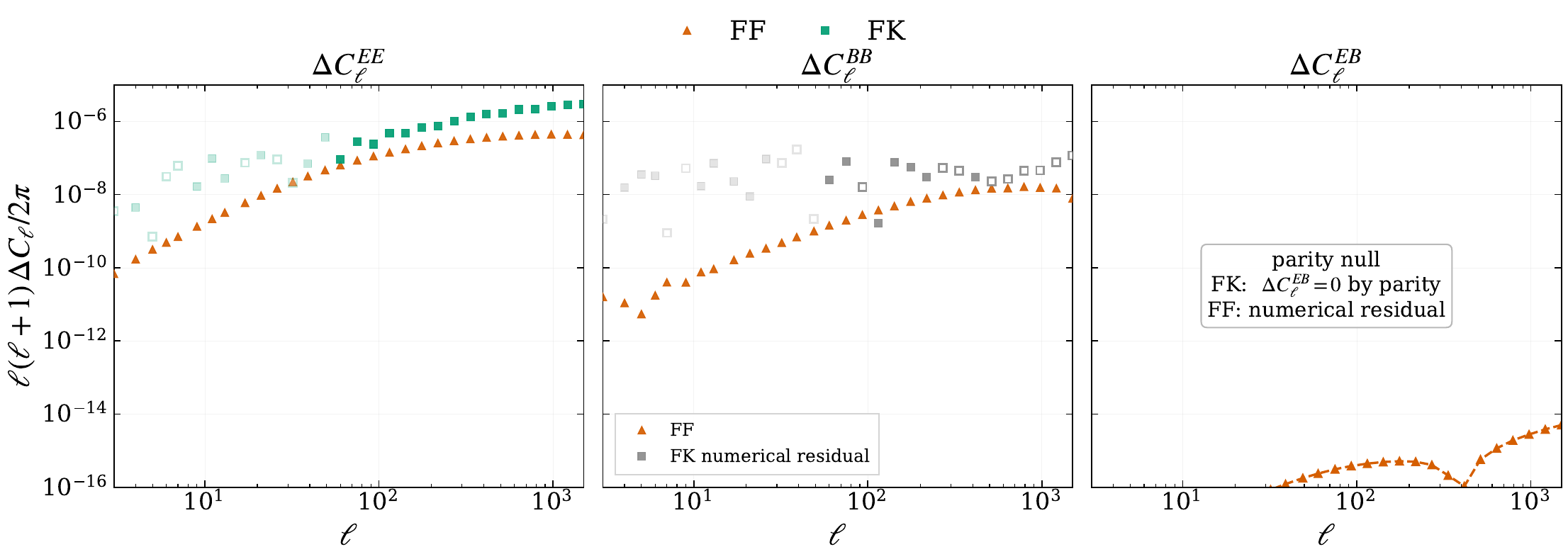}
    \caption{Polarization decomposition of the Order-2 corrections at
    $z_s=5$, shown as band powers $\ell(\ell+1)\Delta C_\ell/2\pi$.
    For this scalar example, FK contributes to EE while its physical
    BB and EB contributions vanish. The grey FK BB curve is a numerical
    residual. Sparse sampling over a finite angular range can
    mix E and B components, as confirmed by a pure-E control on the same
    angular grid. FF can contribute to both EE and BB, while its EB
    component vanishes by parity apart from a numerical residual.
    Fainter points below $\ell=50$ are not used for quantitative statements.
    Markers are filled for positive values and hollow for negative values.}
    \label{fig: cl EB}
\end{figure*}

In this scalar example, the polarization decomposition distinguishes
the two Order-2 channels.
For scalar perturbations and the linearized observable map used here,
the linear shear has no B component. FK pairs this linear shear with a
shear containing one nonlinear Sachs insertion, so its contribution to
$C_\ell^{BB}$ vanishes. Parity also removes $C_\ell^{EB}$. FF can
contribute BB at the same diagrammatic order because both external
shear fields can contain a nonlinear insertion. This contribution arises from
the nonlinear Sachs couplings with the linearized observable map used here.
The quadratic observable reconstruction, including scalar lens-lens image
rotation and the corresponding symmetric convergence/shear corrections,
is not included in this calculation. Our example therefore isolates
nonlinear Sachs evolution with a linearized observable map and does not
provide a complete second-order weak-lensing calculation. Figure~\ref{fig: cl EB} shows the calculated
polarization components and the numerical FK BB residual.

A further strength of the construction is its indifference to the detailed
statistics of the driving fields. Strongly correlated focusing and shearing
fields and discrete, white-noise-like sources are accommodated within the same
diagrammatic language, the only difference being the form of the connected
cumulants fed into the vertices. For white-noise sources the Dirac-$\delta$
correlations collapse the corresponding affine and angular integrals, reducing
the diagrams to closed forms with little additional effort. In this sense the
formalism naturally encompasses the line-of-sight stochastic-lensing
description of~\citet{fleury2015theory, fleury2017weak}, while generalising it
in two respects: it does not assume Gaussian driving fields, and it is
formulated over the full observational domain rather than along a single line
of sight.

\edited{Direct stochastic Sachs integrations check the two Order-2
channels at $z_s=5$. FK uses the corrected tabulated three-point cumulant
and agrees with the diagrammatic prediction within $1\%$ at four tested
angles. FF agrees within $0.7\%$ with a continuum reference using the
same local covariance as the simulation, a test distinct from the
nonlocal cosmological FF prediction (Fig.~\ref{fig: val mc}).
Independent quadrature of the FK cumulant also agrees within $1\%$.
The sampling and numerical diagnostics are reported in
Appendix~\ref{append: validation}. A realistic input cumulant would
call for N-body simulations with baryonic corrections, or a
separate-universe treatment of the squeezed response.}
The simulation route is demanding precisely because of the kinematics that
generate the effect: \edited{the FK leakage
originates in the collapsed configuration of the cumulant, where the leg at
the observed separation couples to two short-wavelength modes.} Capturing it
calls for a simulation box large enough to sample the
\edited{separation leg} together with a resolution fine enough to
resolve the small-scale modes that dominate the collapsed cumulant.

These results were obtained for a deliberately narrow demonstration scenario (scalar
perturbations, an unperturbed ray, a linear power spectrum, and a tree-level
bispectrum), but that slice is a bookkeeping choice rather than a limitation of
the formalism. \edited{In particular, the Limber reduction makes FK depend on a cumulant
with two coincident legs, whose value depends on the small-scale input.
A physical nonlinear prediction therefore depends on the assumed
small-scale matter statistics and any smoothing applied to the driving
fields. We quote no calibrated nonlinear amplitude here
(Section~\ref{subsec: cutoff}).} \edited{A further caveat is near-horizon: at the lowest multipoles a
manifestly gauge-invariant observable would also need the relativistic Poisson
source and the observed-redshift correction~\citep{clarkson2012mis}, fixed
here at sub-horizon, background level.} \edited{In the opposite, wide-angle limit the collapsed cumulant is genuinely
squeezed, one soft leg against two hard ones. The natural tool there is the
separate-universe response of the small-scale power to a long-wavelength
background \citep{barreira2017responses}, rather than the multipole sum used here, whose high-multipole
terms alternate in sign and largely cancel at those separations
(Section~\ref{subsec: cutoff}); that route would also carry over to the
nonlinear regime, though it would still need a named scale below which the
driving field is treated as unresolved.} These near-horizon refinements, and a
more accurate nonlinear three-point cumulant, are left to future work; the
toy-model driving field already suffices to illustrate the effect. Perturbed-path remapping, nonlinear matter statistics, and
primordial non-Gaussianity all enter through the driving-field cumulants inside the
same path integral, the perturbed path through the cumulants evaluated along the
corrected light cone. The selection rule also points to a natural next step.
Because an $n$-point observable in this map receives the $n$-point cumulant as the
leading-order term, higher-point lensing statistics are cleaner windows on
driving-field non-Gaussianity than the two-point function; the lensing
three-point function and the one-point convergence distribution are the obvious
targets. Finally, because the construction starts from the Sachs system rather
than a lensing-specific projection, the same machinery extends to other
light-cone observables, including the lensing of the stochastic gravitational-wave
background in the geometric-optics limit~\citep{DiegoEtAl2021}.

\begin{acknowledgments}
We are grateful to Ruth Durrer, Roy Maartens, Yan-Chuan Cai, Julien Larena \edited{and Robert Reischke} for valuable advice and discussions.
Throughout the course of this work, ZZ has been deeply grateful for and fondly remembers the late Nick Kaiser, who brilliantly guided his PhD journey, particularly in shaping his passion for GR and field theory.
ZZ would also like to express sincere gratitude for all the encouragement and feedback received during the `GR Effects in $\Lambda$SS 2026' workshop. 
This result is part of a project that has received funding from the 
European Research Council (ERC) under the European Union's Horizon 2020 
research and innovation programme (Grant agreement No. 948764; ZZ, PB). 
ZZ also acknowledges support from the RadioForegroundsPlus project 
HORIZON-CL4-2023-SPACE-01, GA 101135036.

\end{acknowledgments}

\section*{Code Availability}
\label{sec: code availability}
The full Python toolbox and analysis pipeline are publicly available as repositories at
\texttt{https://github.com/StatFieldTheory}.
The \textsc{Mathematica} scripts used for the symbolic derivations presented 
in this paper, based on the \textsc{xAct}\citep{martin2020xact} 
and \textsc{xPerm}\citep{martin2008xperm} tensor-algebra packages, 
are also included there.

\appendix

\section{The Newman--Penrose Formalism}
\label{append: NP formalism}

In General Relativity, the dynamics of a geodesic congruence is conventionally
described either through the geodesic-deviation equation or through a tensorial
decomposition of the covariant derivative of the tangent vector field. For null
congruences specifically, several equivalent formalisms have been developed:
one class works with the projected screen-space metric, for example the
``hatting'' operation in Wald's textbook, while another more prevalent in the
weak-lensing literature is rendered as a second-order equation for the Jacobi
map.

For the path-integral formulation pursued in this work, we require instead a
system of first-order evolution equations along each null geodesic.
It is also natural to express the Sachs-scalar evolution directly in the
parallel-transported screen-space basis, rather than projecting from a global
coordinate chart. These two requirements motivate a coordinate-free starting
point: the Newman--Penrose formalism, which replaces the use of the metric
with a complex null tetrad,
\begin{equation}
    \{k^\mu, e^\mu, \screenvec^{\mu}, \Bar{\screenvec}^{\mu}\}.
\end{equation}
Here $k^\mu$ is a real null vector, identified with the tangent to the photon
congruence; $e^\mu$ is an auxiliary real null vector pointing in the direction
``opposite'' to $k^\mu$; and $\screenvec^{\mu}$, $\Bar{\screenvec}^{\mu}$ are
complex-conjugate vectors spanning the screen plane perpendicular to the light
ray. Given two real, orthogonal, unit spacelike vectors $\mathbf{x}$ and
$\mathbf{y}$\footnote{Typically one takes
$\mathbf{x} = \hat{\mathbf{e}}_\theta$ and
$\mathbf{y} = \hat{\mathbf{e}}_\phi$.} in the screen plane, the complex
vectors are defined by
\bea
\screenvec^{\mu} & = \frac{1}{\sqrt{2}}(\mathbf{x}^\mu + \ii \, \mathbf{y}^\mu), \\
\Bar{\screenvec}^\mu & = \frac{1}{\sqrt{2}}(\mathbf{x}^\mu - \ii \, \mathbf{y}^\mu) .
\eea
By construction, the tetrad vectors satisfy the normalisation and orthogonality
conditions
\begin{equation}
    k^\mu e_\mu = -1,
    \quad
    \screenvec^{\mu} \Bar{\screenvec}_\mu = 1,
\end{equation}
with all other inner products vanishing. Besides being coordinate-free, this
basis makes the spin weight of a field manifest: each factor of $\screenvec$
raises the spin by one unit, while each $\Bar{\screenvec}$ lowers it by one.

The dynamics of a null geodesic congruence are encoded in the covariant
derivative of its tangent field, $\nabla_\nu k^\mu$. Observationally, however,
only the projection of this tensor onto the screen plane is of interest. It is
therefore natural to work with its contractions against the screen basis, which
define the complex scalars
\begin{align}
    \rho & = -\nabla_\nu k_\mu \screenvec^{\mu} \Bar{\screenvec}^\nu, \\
    \sigma & = \nabla_\nu k_\mu \screenvec^{\mu} \screenvec^{\nu},
\end{align}
known as the Newman--Penrose spin coefficients, with $\rho$ and $\sigma$ of spin weight $0$
and $2$ respectively. Their evolution along the congruence is governed by the
remarkably concise system
\bea
\frac{\diff{\rho}}{\diff{\lambda}} & = \rho^2 + \sigma \Bar{\sigma} - \Phi_{00}, \\
\frac{\diff{\sigma}}{\diff{\lambda}} & = (\rho + \Bar{\rho}) \sigma  + \Psi_{0}.
\eea
Here $\lambda$ is the affine parameter along the geodesic. The right-hand sides
introduce two driving fields. The first is the Ricci focusing scalar
\begin{equation}
    \Phi_{00} = -\frac{1}{2} R_{\mu \nu } k^{\mu} k^{\nu},
    \label{eq: appendix Phi 00 from Ricci}
\end{equation}
a real-valued field built from the Ricci tensor $R_{\mu\nu}$, which drives the
convergence of the beam. The second is the complex Weyl scalar
\begin{equation}
    \Psi_{0} = - C_{\alpha \beta \gamma \delta}
    k^\alpha \screenvec^{\beta} k^\gamma \screenvec^{\delta},
\end{equation}
which sources the tidal shearing induced by spacetime curvature. Here
$C_{\alpha \beta \gamma \delta}$ is the Weyl tensor, defined as the trace-free
part of the Riemann tensor:
\begin{equation}
    C_{\alpha \beta \gamma \delta}
    =
    R_{\alpha \beta \gamma \delta}
    - (g_{\alpha[\gamma} R_{\delta]\beta} - g_{\beta[\gamma} R_{\delta]\alpha})
    + \frac{1}{3} R g_{\alpha[\gamma} g_{\delta]\beta},
    \label{eq: appendix Weyl tensor def}
\end{equation}
where square brackets denote anti-symmetrisation of the enclosed indices,
e.g.\ $A_{[\alpha\beta]} = \tfrac{1}{2}(A_{\alpha\beta} - A_{\beta\alpha})$.

Throughout this work we adopt natural units in which $c = 1$. Via the Einstein
equations, $\Phi_{00}$ can equivalently be expressed in terms of the
stress--energy tensor:
\begin{equation}
    R_{\mu \nu } - \frac{1}{2} R g_{\mu \nu} = 8 \pi G \, T_{\mu\nu}
    \quad\Longrightarrow\quad
    \Phi_{00} = -4 \pi G \, T_{\mu\nu}\, k^{\mu} k^{\nu},
    \label{eq: appendix Phi00}
\end{equation}
where the $R\, g_{\mu\nu}$ contraction vanishes because $k^\mu$ is null. In the
same spirit, contracting Eq.~(\ref{eq: appendix Weyl tensor def}) with
$k^\alpha \screenvec^\beta k^\gamma \screenvec^\delta$ causes the Ricci and
trace terms to drop out, so that $\Psi_0$ reduces to its Riemann-tensor form:
\begin{equation}
    \Psi_{0}
    =
    -C_{\alpha \beta \gamma \delta} k^\alpha \screenvec^{\beta}
    k^\gamma \screenvec^{\delta}
    =
    -R_{\alpha \beta \gamma \delta} k^\alpha \screenvec^{\beta}
    k^\gamma \screenvec^{\delta}.
    \label{eq: appendix Psi0}
\end{equation}
Together, Eqs.~(\ref{eq: appendix Psi0}) and~(\ref{eq: appendix Phi00}) (or
Eq.~(\ref{eq: appendix Phi 00 from Ricci}) when direct contraction of the Ricci
tensor is more convenient) provide general expressions for the driving fields
that determine the evolution of weak-lensing observables. The main text quotes
these definitions in Eqs.~(\ref{eq: Phi 00 from Ricci})--(\ref{eq: Psi0}) before
turning to their explicit cosmological realisation in
Section~\ref{sec: cosmological setup}.

\section{Closed-form Jacobi solution on flat FLRW}
\label{append: D equals a chi}

On a flat FLRW background the mean Weyl source vanishes, and the isotropic
observer conditions set the source-averaged shear to zero. The remaining
expansion obeys the Riccati equation
\begin{equation}
    \Dot{\theta}^{\scriptscriptstyle\mathrm{(sa)}}(\lambda)
    = -\left[\theta^{\scriptscriptstyle\mathrm{(sa)}}(\lambda)\right]^2
    + \bar{\Phi}_{00}(\lambda).
\end{equation}
At the observer vertex, $\theta^{\scriptscriptstyle\mathrm{(sa)}}(\lambda)
\to 1/\lambda$ as $\lambda\to0^+$. Writing
$\theta^{\scriptscriptstyle\mathrm{(sa)}}=\Dot{\bar D}/\bar D$ gives the
Jacobi amplitude equation with regular vertex conditions:
\begin{equation}
    \Ddot{\bar D}(\lambda) = \bar{\Phi}_{00}(\lambda)\, \bar D(\lambda),
    \qquad \bar D(0) = 0,\quad \Dot{\bar D}(0) = 1,
\label{eq: Jacobi flrw}
\end{equation}
This equation admits the closed-form solution [see \cite{hogg1999distance} for
the expression of $\bar D$ as the angular diameter distance]
\begin{equation}
    \bar D(\lambda) = a(\lambda)\, \chi(\lambda),
\label{eq: D equals a chi}
\end{equation}
where $a$ is the scale factor and $\chi$ the comoving radial coordinate 
along the past null geodesic. Here the dot denotes $d/d\lambda$ (consistent 
with \S\ref{sec: sachs dynamics}) and 
$\bar{\Phi}_{00} = -(\mathcal{H}^2 - \mathcal{H}')/a^4$ is the FLRW limit 
of the Sachs scalar (Eq.~\ref{eq: driving bg}); $\mathcal{H} \equiv a'/a$ 
with the prime denoting $\partial_\tau$ (conformal time).

We verify Eq.~\eqref{eq: D equals a chi} by direct substitution. With $E_0 = 1$ by construction, 
the affine parameter satisfies $d\tau/d\lambda = -1/a^2$ and $d\chi/d\lambda = +1/a^2$ 
along the radial null geodesic, so that $da/d\lambda = -\mathcal{H}/a$ 
and $d^2 a/d\lambda^2 = \edited{-}(\mathcal{H}^2 - \mathcal{H}')/a^3$ after 
using $a'' = a(\mathcal{H}^2 + \mathcal{H}')$. 
Differentiating $\bar D = a\chi$ twice in $\lambda$ and substituting these expressions yields
\begin{equation}
    \Ddot{\bar D}(\lambda) = -\frac{\mathcal{H}^2 - \mathcal{H}'}{a^4} \, (a\chi)
    = \bar{\Phi}_{00}(\lambda)\, \bar D(\lambda),
\end{equation}
identically, with no use of the Friedmann equations: the proof is purely 
geometric and holds for any flat-FLRW $a(\tau)$. The vertex conditions 
$\bar D(0) = a(0)\chi(0) = 0$ and $\Dot{\bar D}(0) = a(0)\,(d\chi/d\lambda)(0) = 1$ 
follow directly from $a(0) = 1$ and $d\chi/d\lambda(0) = 1/a^2(0) = 1$.

Two consequences enter the analysis pipeline. First, since
$\theta^{\scriptscriptstyle\mathrm{(sa)}}=d\ln\bar D/d\lambda$, any
line-of-sight integral of this expansion scalar reduces to a difference of
logs:
\begin{equation}
    \int_{\lambda'}^{\lambda} \theta^{\scriptscriptstyle\mathrm{(sa)}}(\tau')\, d\tau'
    = \ln \bar D(\lambda) - \ln \bar D(\lambda').
\label{eq: theta sa integral}
\end{equation}
Second, the response propagator (Eq.~\ref{eq: explicit resp op}) reduces to the closed form
\begin{equation}
    \RespOp(\lambda, \lambda') = \edited{\Theta(\lambda' - \lambda)\, \left( \frac{\bar D(\lambda)}{\bar D(\lambda')} \right)^{\!2}},
\label{eq: R from D}
\end{equation}
which is finite at the vertex by construction (since $\bar D(0) = 0$ exactly and 
$\bar D(\lambda) \sim \lambda$ as $\lambda \to 0$). Working with $\bar D$ directly 
avoids the $1/\lambda$ integrand singularity of $\theta^{\scriptscriptstyle\mathrm{(sa)}}$ 
at the observer, eliminating any quadrature error in $\RespOp$ when $\bar D$ is tabulated from 
the ODE solve.

\section{Detailed First-Order Form of the $3\times 3$ Jacobi Map}
\label{append: jacobian blocks}

This appendix records the first-order Poisson-gauge form of the past-light-cone
Jacobian
$J^{I}_{\phantom{I}\alpha} = \partial y^{I}/\partial q^{\alpha}$, with
$y^{I} = (\chi,\hat r^{B})$ and $q^{\alpha} = (\lambda,\direc^{A})$,
used schematically in Subsection~\ref{subsec: affine to redshift}. This form
keeps the three-dimensional map explicit while showing how its transverse block
reduces to the usual weak-lensing Jacobi map.

We start from the perturbed ray. Writing
$k^\mu = k^{(0)\mu} + \varepsilon\,\delta k^\mu$ and similarly
$x^\mu = x^{(0)\mu} + \varepsilon\,\delta x^\mu$, the null-geodesic equation
$k^\nu\nabla_\nu k^\mu = 0$ splits into the background geodesic and the
first-order equation
\begin{multline}
    \frac{\diff \delta k^{\mu}}{\diff\lambda}
    \;=\; S^{\mu}(\tau,\vec x) \;-\; 2\,\Gamma^{(0)\mu}_{\phantom{(0)\mu}\nu\rho}\,k^{(0)\nu}\,\delta k^{\rho} , \\
    S^{\mu} \;\equiv\; -\,\Gamma^{(1)\mu}_{\phantom{(1)\mu}\nu\rho}\,k^{(0)\nu}\,k^{(0)\rho} .
    \label{eq: appendix delta k ODE}
\end{multline}
The explicit Poisson-gauge form of $S^{\mu}$, including the scalar potentials
$\{\Psi,\Phi\}$, \edited{the transverse shift vector $B_i$}, and the transverse-traceless mode $h_{ij}$,
follows by substituting the linearised Christoffels
$\Gamma^{(1)\mu}_{\phantom{(1)\mu}\nu\rho}$ and the background null tangent
into Eq.~(\ref{eq: appendix delta k ODE}). Its formal solution along the
background ray,
\begin{multline}
    \delta k^{\mu}(\lambda) \;=\; \int_{0}^{\lambda}\!\diff\lambda'\,
        \mathcal{G}^{\mu}_{\phantom\mu\nu}(\lambda,\lambda')\,
        S^{\nu}\bigl(\tau^{(0)}(\lambda'), \vec x^{(0)}(\lambda')\bigr) , \\
    \delta x^{\mu}(\lambda) \;=\; \int_{0}^{\lambda}\!\diff\lambda'\;\delta k^{\mu}(\lambda'),
    \label{eq: appendix delta x}
\end{multline}
uses the retarded propagator $\mathcal G$ for the homogeneous part of
Eq.~(\ref{eq: appendix delta k ODE}). In flat FLRW, $\mathcal G$ contributes a
single factor of $1/a^{2}$ along the ray, and the two nested integrals in
$\lambda$ become the standard light-cone kernels displayed below.

The spatial perturbation then separates cleanly into radial and angular pieces.
Using the background projector
$P^{ij} \equiv \delta^{ij} - \direc^{i}\direc^{j}$, the spatial
perturbation decomposes into
\be
    \delta\chi \;=\; \direc_{i}\,\delta x^{i},
    \qquad
    \delta x_{\perp}^{i} \;=\; P^{i}_{\phantom i j}\,\delta x^{j} ,
    \qquad
    \delta\hat r^{i} \;=\; \frac{\delta x_{\perp}^{i}}{\chi^{(0)}} + \mathcal{O}(\varepsilon^{2}) ,
    \label{eq: appendix radial transverse split}
\ee
and the four Jacobian blocks follow by differentiating these decompositions
in $\lambda$ (along the ray) or in $\direc^{A}$ (across observer directions).
With the same convention as Eq.~(\ref{eq: coord jacobian}) (upper index for the
output coordinate, lower index for the input coordinate), they are
\begin{align}
    \delta J^{\chi}_{\phantom{\chi}\lambda} &= \partial_{\lambda}\delta\chi
        \;=\; \direc_{i}\,\delta k^{i} ,
    \label{eq: appendix block chi lambda} \\
    \delta J^{A}_{\phantom{A}\lambda}      &= \partial_{\lambda}\delta\hat r^{A}
        \;=\; \frac{1}{\chi^{(0)}}\,P^{A}_{\phantom A i}\,\delta k^{i} ,
    \label{eq: appendix block A lambda} \\
    \delta J^{\chi}_{\phantom{\chi}A}       &= \partial_{\direc^{A}}\delta\chi
        \;=\; \direc_{i}\,\partial_{\direc^{A}}\delta x^{i} ,
    \label{eq: appendix block chi A} \\
    \delta J^{B}_{\phantom{B}A} &= \partial_{\direc^{A}}\delta\hat r^{B}
        \;=\; \frac{1}{\chi^{(0)}}\bigl(P^{B}_{\phantom B i}\,\partial_{\direc^{A}}\delta x^{i}
                                       \;-\;\delta^{B}_{A}\,\delta\chi\bigr) .
    \label{eq: appendix block AB}
\end{align}
The $2\times 2$ transverse block~(\ref{eq: appendix block AB}) coincides, up
to the rotation from the observer-sky basis to the screen basis, with the
standard Sachs Jacobi matrix $\mathcal J^{A}_{\phantom A B}(\lambda)$. This is
the block that carries the usual weak-lensing convergence and shear.

For the scalar part, substituting the Poisson-gauge Christoffels and setting the
shift and tensor
modes (\edited{$B_i = 0$}, $h_{ij} = 0$), the source $S^{\mu}$ reduces to
\begin{align}
    S^{\tau}\big|_{\rm scalar}
    &= \frac{E^{2}}{a^{2}}\Bigl[\,2\mathcal{H}(\Phi + \Psi)
         + 2\,\direc^{i}\partial_{i}\Psi
         + \partial_{\tau}(\Phi - \Psi)\,\Bigr],
    \label{eq: appendix S tau scalar}\\
    \direc_{i}\,S^{i}\big|_{\rm scalar}
    &= \frac{E^{2}}{a^{2}}\Bigl[\,\direc^{i}\partial_{i}(\Phi - \Psi)
         \;-\; 2\,\partial_{\tau}\Phi\,\Bigr],
    \label{eq: appendix S rad scalar}\\
    P^{A}_{\phantom A i}\,S^{i}\big|_{\rm scalar}
    &= -\,\frac{E^{2}}{a^{2}}\,\partial^{A}_{\perp}(\Phi + \Psi) .
    \label{eq: appendix S perp scalar}
\end{align}
Equation~(\ref{eq: appendix S tau scalar}) is the Sachs--Wolfe plus
integrated-Sachs--Wolfe integrand that drives $\delta\tau$, the radial
term~(\ref{eq: appendix S rad scalar}) gives the gauge-consistent
Shapiro shift of $\delta\chi$, and the transverse
term~(\ref{eq: appendix S perp scalar}) is the familiar gradient of the
lensing potential $\Phi + \Psi$ that sources the weak-lensing deflection.

The weak-lensing limit follows from the transverse source. Integrating
Eqs.~(\ref{eq: appendix delta x}) and
(\ref{eq: appendix S perp scalar}) twice along the background ray, converting
from $\lambda$ to $\chi$ via Eq.~(\ref{eq: dchi dlambda bg}), and dividing by
$\chi_{s} = \chi^{(0)}(\lambda_{s})$, one recovers the standard transverse
deflection and its sky-Jacobian,
\be
    \delta\hat r^{A}(\direc, \chi_{s})
    \;=\; -\,\int_{0}^{\chi_{s}}\!\frac{\chi_{s} - \chi}{\chi_{s}}\,
          \bigl[\partial^{A}_{\perp}(\Phi + \Psi)\bigr]
          \bigl(\tau^{(0)}(\chi),\,\chi\direc\bigr)\,\diff\chi ,
    \label{eq: appendix deflection}
\ee
\begin{multline}
    \delta J^{B}_{\phantom{B}A}(\direc, \chi_{s})
    \;=\; -\,\int_{0}^{\chi_{s}}\!\frac{(\chi_{s} - \chi)\,\chi}{\chi_{s}}\, \\
          \times \bigl[\partial^{B}_{\perp}\partial^{\perp}_{A}(\Phi + \Psi)\bigr]
          \bigl(\tau^{(0)}(\chi),\,\chi\direc\bigr)\,\diff\chi .
    \label{eq: appendix amplification}
\end{multline}
From Eq.~(\ref{eq: appendix amplification}) the convergence and shear of the
beam bundle are
\be
    \kappa \;=\; -\tfrac{1}{2}\,\delta J^{A}_{\phantom A A} ,
    \qquad
    \gamma_{1} \pm \ii\,\gamma_{2}
    \;=\; -\tfrac{1}{2}\bigl(\delta J^{1}_{\phantom 1 1} - \delta J^{2}_{\phantom 2 2}\bigr)
          \;\mp\;\ii\,\delta J^{1}_{\phantom 1 2} ,
    \label{eq: appendix convergence shear}
\ee
matching the standard CMB/large-scale-structure weak-lensing literature and,
by construction, the standard Sachs Jacobi map.

The same construction extends to the non-scalar perturbations. The shift-\edited{$B_i$} and
tensor-$h_{ij}$ contributions enter through $S^\mu$, while
Eqs.~(\ref{eq: appendix block chi lambda})--(\ref{eq: appendix block AB}) remain
unchanged as block identities. Equivalently, only the source terms in
Eqs.~(\ref{eq: appendix S tau scalar})--(\ref{eq: appendix S perp scalar})
gain additional \edited{$B_i$} and $h_{ij}$ terms.

\section{The Driving-Field Three-Point Cumulant}
\label{append: driving-field spectra}

This appendix builds the driving-field three-point cumulant that sources the FK
term, in three steps. We first reduce the full object to the single quantity the
two-point expansion actually consumes, then construct that quantity from the
present-day matter power spectrum, and finally resolve the one spin-$2$ subtlety
the reduction exposes. We introduce the equal-shell approximation below.

Start from the most general object the formalism produces, the cross-shell
three-point cumulant
$\cumulant^{(3)}_{abc}(\direc_1,\direc_2,\direc_3;\lambda_1,\lambda_2,\lambda_3)$,
a function of three directions and three affine labels. Its real driving-field
components $a,b,c\in\{\Phi_{00},\,\Psi_+,\,\Psi_\times\}$ are
the spin-$0$ Ricci focusing $\Phi_{00}$ and the real and imaginary parts of the
spin-$2$ Weyl shear $\Psi_0$ (Sec.~\ref{subsec: driving fields}). What enters the
\emph{two-point} expansion is far simpler, and the FK diagram
(Sec.~\ref{subsec: mixed hierarchy}) shows why. There the non-local $K$-vertex is
folded with response propagators, and each propagator carries a directional
$\delta$-function that pins a vertex leg to an external line
[Eq.~(\ref{eq: explicit resp op})]. Two of the three vertex directions are thereby
tied to the two observed directions, which differ by a single angle $\gamma$. The
object we actually need is therefore the one-angle, equal-shell reduction
$\zeta_{abc}(\gamma,\lambda)$: two directions coincident, the third offset by $\gamma$,
all on a common source shell $\lambda$. Although the statistic we ultimately need is
the one carried in the affine-parameter ($\lambda$) measure that the path integral
folds over, the same quantity is most naturally computed first in comoving space, as
a function of $\chi$, and only then converted into the corresponding $\lambda$-measure
form by applying the correct Jacobian.

Collapsing the three affine labels onto one shell is the three-point counterpart
of the Limber approximation. As in two-point lensing
projections~\citep{LemosEtAl2017, KitchingNonLimber2017, FangEtAl2020},
it is motivated by lensing response kernels that are broad compared with
the radial coherence length of the matter correlators.
Explicitly, the collapse replaces the cross-shell cumulant by its diagonal,
\be
\cumulant^{(3)}_{abc}(\gamma;\lambda_1,\lambda_2,\lambda_3)
  \approx \zeta_{abc}(\gamma,\lambda_1)\,
          \delta(\lambda_1-\lambda_2)\,\delta(\lambda_1-\lambda_3),
\label{eq: appendix equal shell approx}
\ee
turning the three radial integrals of the cross-shell vertex into one and making
the three-point cumulant analysis far easier to implement with the \textsc{canoes}
code~\citep{canoes2026}.

\paragraph{From the matter spectrum to $\zeta$.}
The cumulant follows from a single time-independent input, the present-day matter
power spectrum $P_\delta(k)$, through the chain
$P_\delta\!\to\!B_\delta\!\to\!B_\Phi\!\to\!(\Phi_{00},\Psi_0)\!\to\!\zeta$. At
tree level the matter bispectrum is the standard symmetrised second-order SPT
coupling \citep[e.g.][]{Bernardeau2002Review},
\be
B_\delta(k_1,k_2,k_3)
   = 2\,F_2^{(s)}(\mathbf{k}_1,\mathbf{k}_2)\,P_\delta(k_1)\,P_\delta(k_2)
   + \text{2 cyclic},
\label{eq: appendix bdelta tree}
\ee
with $F_2^{(s)}$ the usual mode-coupling kernel; the tree-level redshift evolution
is carried entirely by the growth factor,
\be
B_\delta(z)=D(z)^4\,B_\delta(0).
\label{eq: appendix bdelta growth}
\ee
The Poisson relation turns each matter leg into a potential leg,
\be
\Phi(\mathbf k)=\frac{A(a)}{k^2}\,\delta_m(\mathbf k),\qquad
A(a)=-\tfrac32\,\Omega_m H_0^2/a,
\label{eq: appendix poisson}
\ee
so the potential bispectrum
\be
B_\Phi(k_1,k_2,k_3)
   = \Bigl[\,\prod_{i=1}^{3}\frac{A(a)}{k_i^2}\,\Bigr]\,
     B_\delta(k_1,k_2,k_3)
\label{eq: appendix bphi}
\ee
inherits the negative Poisson cube $A(a)^3<0$. In the transverse scalar
approximation used for this equal-shell cumulant, the driving
fields are obtained from the trace and trace-free parts of the screen-space Hessian of the
lightcone potential [Eqs.~(\ref{eq: Phi00 scalar}) and~(\ref{eq: Psi0 scalar})].
Transformed to a spherical-harmonic basis, that Hessian reduces per multipole to a
simple angular multiplier on each leg,
\be
\begin{aligned}
\Phi_{00}:&\ +\frac{L^2}{\chi^2}\ \text{on } Y_{\ell m},\\
\Psi_0:&\ -\frac{\sqrt{L^2(L^2-2)}}{\chi^2}\ \text{on } {}_2Y_{\ell m},\\
& L^2\equiv\ell(\ell+1).
\end{aligned}
\label{eq: appendix screen hessian}
\ee
the Laplacian $L^2/\chi^2$ being the spin-$0$ trace (Ricci focusing, the
transverse-Laplacian term of Eq.~\ref{eq: Phi00 scalar}) and
$-\sqrt{L^2(L^2-2)}/\chi^2$ the spin-$2$ $\eth^2$ shear (the trace-free screen
Hessian of Eq.~\ref{eq: Psi0 scalar}); the shared $1/\chi^2$ converts the two
transverse screen derivatives at distance $\chi$, and the two moduli agree
at leading order for large $\ell$. It is convenient to
compute $\zeta_{abc}$ in the field basis: labelling the spin-$0$ leg
$T=\Phi_{00}$ and each spin-$2$ leg $P=\Psi_0$, the four projections
$XYZ\in\{TTT,TTP,TPP,PPP\}$ are the natural computational \emph{channels}.
Contracting the projected legs and summing over the parity-even multipole
triangles ($\ell_1+\ell_2+\ell_3$ even, selected by the scalar Wigner-$3j$) gives
each channel,
\be
\zeta_{XYZ}(\gamma,\chi)
   = \sum_{\ell_1\ell_2\ell_3}\mathcal G_{\ell_1\ell_2\ell_3}\,
     P_{\ell_3}(\cos\gamma)\,B^{XYZ}_{\ell_1\ell_2\ell_3}(\chi),
\label{eq: appendix zeta squeezed}
\ee
with $\mathcal G$ the parity-even Gaunt weight and each reduced bispectrum
$B^{XYZ}$ a single radial integral over $P_\delta$
[Eqs.~(\ref{eq: appendix bdelta tree})--(\ref{eq: appendix screen hessian})],
evaluated in closed form by an FFTlog double-Bessel method, with an exact
Wigner-3j sum at low multipoles ($\ell\le60$) and the Limber approximation
above.\footnote{The
collapsed FK geometry weights the near-degenerate triangles $\ell_1\!\simeq\!
\ell_2$ most heavily, but the sum runs over all parity-allowed triples. For the
spin-$2$ channels the scalar $w_{3j}(\ell_1,\ell_2,\ell_3;0,0,0)$ enters as a
triangle/parity selector; the spin weight is carried separately by the
spin-weighted-harmonic contraction.}

The high-$\ell$ branch makes $B^{XYZ}$ explicit and is the most transparent. With
$\chi=\chi(\lambda)$ the comoving distance to the shell and $k_i=\ell_i/\chi$ the
\edited{comoving} wavenumber that an angular scale $\ell_i$ probes at that distance,
\be
B^{XYZ}_{\ell_1\ell_2\ell_3}(\chi)
   = \frac{D(z)^4}{\chi^4}\,
     \mathcal K^{XYZ}_{\ell_1\ell_2\ell_3}(z)\,
     B_\delta\!\Bigl(\tfrac{\ell_1}{\chi},\tfrac{\ell_2}{\chi},\tfrac{\ell_3}{\chi}\Bigr).
\label{eq: appendix limber reduced bispectrum}
\ee
Reading the factors one by one: $B_\delta$ is the matter bispectrum of
Eq.~(\ref{eq: appendix bdelta tree}), the three-point clustering of the density
field, evaluated at the three wavenumbers that the multipole triangle
$(\ell_1,\ell_2,\ell_3)$ picks out at distance $\chi$. The growth factor $D(z)$
measures how much structure has assembled by redshift $z$; it enters at the fourth
power because the tree-level bispectrum is built from two power spectra, each
growing as $D^2$. The kernel $\mathcal K^{XYZ}_{\ell_1\ell_2\ell_3}(z)$ converts
each matter leg into the field that does the lensing, combining the gravitational
map from density to potential [$A(a)=-\tfrac32\Omega_m H_0^2/a$] with the on-sky
derivatives of Eq.~(\ref{eq: appendix screen hessian}) that isolate the spin-$0$
focusing ($\Phi_{00}$) or spin-$2$ shear ($\Psi_0$) part of each leg. The
$1/\chi^4$ is purely geometric. Because the matter correlations fade over a radial
distance much shorter than the depth of the source shell, the three points are
effectively pinned to a common distance; collapsing the two now-redundant radial
integrals leaves a single integral along the line of sight, and the surviving
$1/\chi^4$ is the three-point analogue of the $1/\chi^2$ in the familiar two-point
Limber projection.

\paragraph{The spin-2 modulus channels.}
The path integral contracts the $K$-vertex as the local cumulant
$\zeta_{abc}=\langle\varphi_a\varphi_b\varphi_c\rangle$ with the three real driving
fields $\varphi\in\{\Phi_{00},\Psi_+,\Psi_\times\}$, so the full cumulant needs every
component of this triple. Parity removes most of them: reflection about the
separation axis makes $\Psi_\times$ parity-odd, so any component with an odd number
of $\Psi_\times$ legs vanishes.

The field-basis channels of Eq.~(\ref{eq: appendix zeta squeezed}), with
$T=\Phi_{00}$ and $P=\Psi_0$, pair the two spin-$2$ legs \emph{without conjugation}
($\Psi_0\Psi_0$), so they deliver $\zeta_{TTT}$, $\zeta_{TTP}$, and the parity-even
\emph{differences},
\be
\zeta_{TPP}=\langle\Phi_{00}(\Psi_+^2-\Psi_\times^2)\rangle,\qquad
\zeta_{PPP}=\langle\Psi_+^3-3\Psi_+\Psi_\times^2\rangle.
\label{eq: appendix unconjugated channels}
\ee
Resolving the individual $\langle\Phi_{00}\Psi_+^2\rangle$ and
$\langle\Phi_{00}\Psi_\times^2\rangle$ (and likewise the $\Psi_+^3$,
$\Psi_+\Psi_\times^2$ pieces) needs the complementary \emph{sums}, supplied by two
modulus channels in which one spin-$2$ leg is conjugated,
\be
\zeta_{B}=\langle\Phi_{00}\,|\Psi_0|^2\rangle,\qquad
\zeta_{D}=\langle\Psi_0\,|\Psi_0|^2\rangle,
\label{eq: appendix modulus channels}
\ee
with $|\Psi_0|^2=\Psi_+^2+\Psi_\times^2$. Difference and sum reconstruct each
component,
\begin{equation}
\langle\Phi_{00}\Psi_+^2\rangle=\tfrac12(\zeta_{TPP}+\zeta_{B}),\qquad
\langle\Phi_{00}\Psi_\times^2\rangle=\tfrac12(\zeta_{B}-\zeta_{TPP}),
\label{eq: appendix modulus reconstruction}
\end{equation}
and likewise $\langle\Psi_+^3\rangle,\langle\Psi_+\Psi_\times^2\rangle$ from
$\zeta_{PPP},\zeta_{D}$.

\begin{figure*}[t!]
    \centering
    \includegraphics[width=\textwidth]{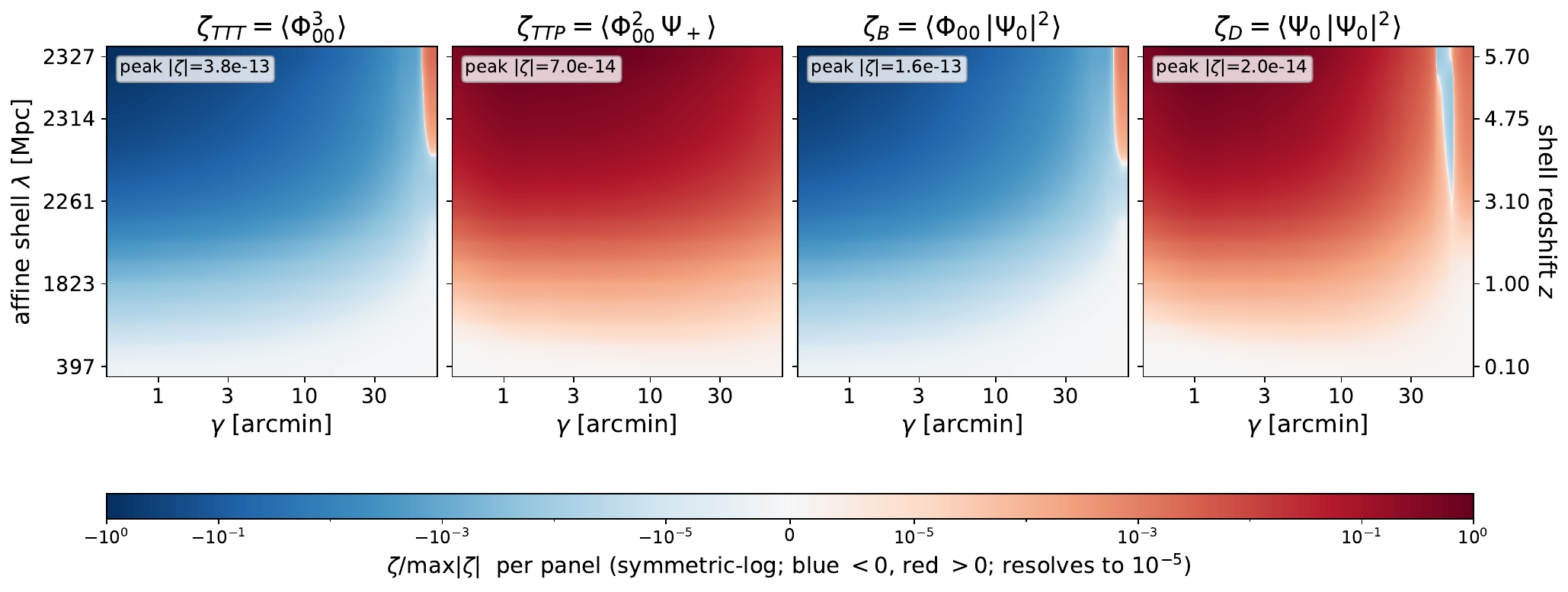}
    \caption{Equal-shell driving-field three-point cumulants in the
    collapsed configuration $(1,\cos\gamma,\cos\gamma)$. The panels
    show $\zeta_{TTT}$, $\zeta_{TTP}$, and the modulus channels
    $\zeta_B=\langle\Phi_{00}|\Psi_0|^2\rangle$ and
    $\zeta_D=\langle\Psi_0|\Psi_0|^2\rangle$ defined in
    Eq.~(\ref{eq: appendix modulus channels}). The channels
    $\zeta_{TPP}$ and $\zeta_{PPP}$ are not shown. Each panel is
    normalized to its own peak absolute amplitude and uses the same
    symmetric-logarithmic colour scale. Rows correspond to the $16$
    driving shells, labelled by affine distance and shell redshift.
    In the higher-redshift shells, TTT and B change sign between
    $56.27'$ and $71.26'$, while D changes sign between $44.43'$ and
    $56.27'$. TTP has no sign change over the displayed range.
    The input table uses $\ell_{\max}=15360$.}
    \label{fig: zeta slices}
\end{figure*}


\section{Workflow validation}
\label{append: validation}

\begin{figure}[H]
  \centering
  \includegraphics[width=\columnwidth]{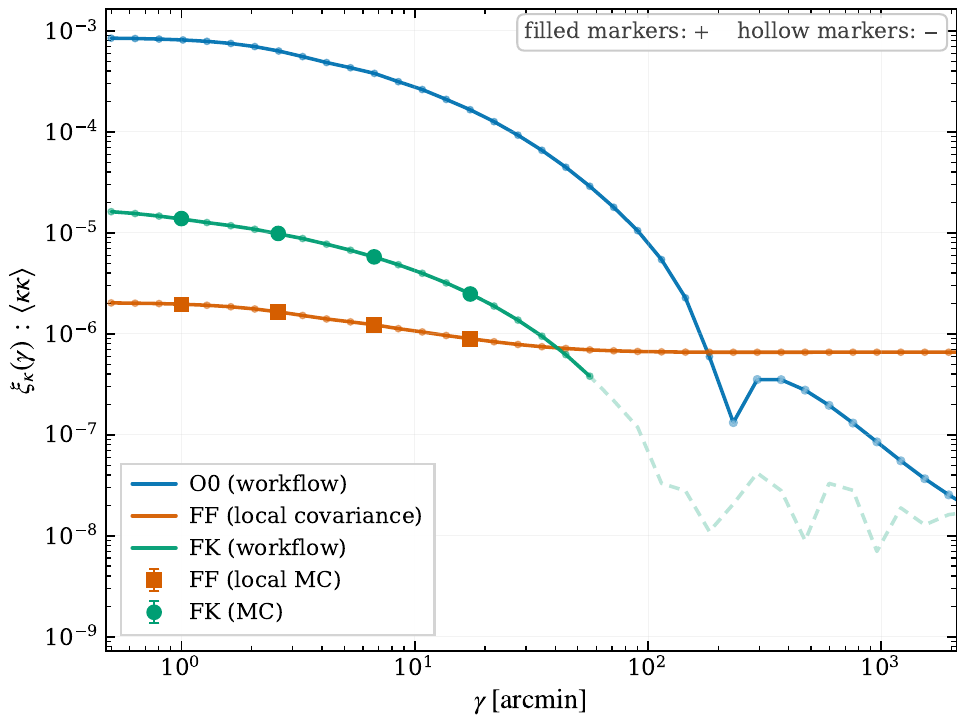}
  \caption{\emph{Workflow validation.} The convergence two-point function
  $\xi_\kappa(\gamma)$ split into its Order-$0$ (O0), nonlinear-propagation (FF)
  and three-point cumulant (FK) channels: the analytic workflow (lines) against a direct
  Monte-Carlo of the stochastic Sachs equation with matched input statistics
  (markers). The FF line uses the simulation's local covariance, distinct
  from the nonlocal cosmological FF prediction. The Order-0 curve is shown
  for reference. FK markers are extrapolated to vanishing
  coloured-noise correlation length, with error bars showing the standard error from
  independent realisation blocks. Marker angles are
  $1',\,2.6',\,6.7',\,17.3'$. Filled and
  hollow markers denote positive and negative values.}
  \label{fig: val mc}
\end{figure}

\edited{We validate that the workflow correctly propagates matched input
driving-field statistics into lensing observables. Both Order-2 channels are checked against direct
Monte-Carlo integration of the stochastic Sachs equation
(Fig.~\ref{fig: val mc}); FK agrees within $1\%$ and FF within $0.7\%$
at $\gamma=1',\,2.6',\,6.7',\,17.3'$. FK uses the corrected tabulated
cumulant at $\ell_{\max}=15360$. The FF test uses the local covariance
actually sampled by the simulation, including its positive-semidefinite
adjustment, and a matched continuum Order-$F^2$ reference. It is separate
from the nonlocal cosmological FF prediction of
Section~\ref{subsec: nonlinear propagation insight}. Sampling standard
errors are $0.36$--$2.10\%$ for FK and $0.27$--$0.30\%$ for FF,
relative to their respective references. As a second, independent
check of FK, we contract the tabulated cumulant $\zeta_{abc}$ with the FK
kernel by direct quadrature, independently of the workflow code. This
reproduces the workflow result within $1\%$ at the same four separations.}
The tests use the fiducial flat-$\Lambda$CDM
cosmology ($\Omega_m=0.316$, $h=0.671$, $n_s=0.97$) with a single source plane
at $z_s=5$.

\edited{The two channels require different estimators because their stochastic
noise enters differently. For FF, the nonlinear Sachs coupling
can be switched on and off for the same random draw, and differencing the
resulting matched runs cancels the common leading fluctuation. The FK signal,
by contrast, is a property of the input statistics: it is the response of
$\langle\kappa\kappa\rangle$ to the three-point cumulant of the driving field.
A naive estimate of a small third moment is affected by Gaussian
fluctuations, motivating a variance-controlled FK estimator.
For the coloured-noise regularization used here, the pointwise skewness
scales as $\sigma_\lambda^{-1/2}$ when the integrated cumulants are held
fixed. The FK estimate is therefore obtained at two finite correlation
lengths along the ray and linearly extrapolated to zero. Error bars come
from independent realisation blocks. The FF estimator also retains higher
propagation orders, which we check by varying the Sachs coupling.
Integration-step, cumulant-injection, finite-correlation and source-grid
checks assess numerical sensitivities separately from the sampling errors.}

\setlength{\bibsep}{-0.6pt}

\begin{thebibliography}{60}%
\makeatletter
\providecommand \@ifxundefined [1]{%
 \@ifx{#1\undefined}
}%
\providecommand \@ifnum [1]{%
 \ifnum #1\expandafter \@firstoftwo
 \else \expandafter \@secondoftwo
 \fi
}%
\providecommand \@ifx [1]{%
 \ifx #1\expandafter \@firstoftwo
 \else \expandafter \@secondoftwo
 \fi
}%
\providecommand \natexlab [1]{#1}%
\providecommand \enquote  [1]{``#1''}%
\providecommand \bibnamefont  [1]{#1}%
\providecommand \bibfnamefont [1]{#1}%
\providecommand \citenamefont [1]{#1}%
\providecommand \href@noop [0]{\@secondoftwo}%
\providecommand \href [0]{\begingroup \@sanitize@url \@href}%
\providecommand \@href[1]{\@@startlink{#1}\@@href}%
\providecommand \@@href[1]{\endgroup#1\@@endlink}%
\providecommand \@sanitize@url [0]{\catcode `\\12\catcode `\$12\catcode `\&12\catcode `\#12\catcode `\^12\catcode `\_12\catcode `\%12\relax}%
\providecommand \@@startlink[1]{}%
\providecommand \@@endlink[0]{}%
\providecommand \url  [0]{\begingroup\@sanitize@url \@url }%
\providecommand \@url [1]{\endgroup\@href {#1}{\urlprefix }}%
\providecommand \urlprefix  [0]{URL }%
\providecommand \Eprint [0]{\href }%
\providecommand \doibase [0]{https://doi.org/}%
\providecommand \selectlanguage [0]{\@gobble}%
\providecommand \bibinfo  [0]{\@secondoftwo}%
\providecommand \bibfield  [0]{\@secondoftwo}%
\providecommand \translation [1]{[#1]}%
\providecommand \BibitemOpen [0]{}%
\providecommand \bibitemStop [0]{}%
\providecommand \bibitemNoStop [0]{.\EOS\space}%
\providecommand \EOS [0]{\spacefactor3000\relax}%
\providecommand \BibitemShut  [1]{\csname bibitem#1\endcsname}%
\let\auto@bib@innerbib\@empty
\bibitem [{\citenamefont {Asgari}\ \emph {et~al.}(2021)\citenamefont {Asgari}, \citenamefont {Lin}, \citenamefont {Joachimi}, \citenamefont {Giblin}, \citenamefont {Heymans}, \citenamefont {Hildebrandt}, \citenamefont {Kannawadi}, \citenamefont {St{\"o}lzner}, \citenamefont {Tr{\"o}ster}, \citenamefont {van~den Busch} \emph {et~al.}}]{asgari2021kids}%
  \BibitemOpen
  \bibfield  {author} {\bibinfo {author} {\bibfnamefont {M.}~\bibnamefont {Asgari}}, \bibinfo {author} {\bibfnamefont {C.-A.}\ \bibnamefont {Lin}}, \bibinfo {author} {\bibfnamefont {B.}~\bibnamefont {Joachimi}}, \bibinfo {author} {\bibfnamefont {B.}~\bibnamefont {Giblin}}, \bibinfo {author} {\bibfnamefont {C.}~\bibnamefont {Heymans}}, \bibinfo {author} {\bibfnamefont {H.}~\bibnamefont {Hildebrandt}}, \bibinfo {author} {\bibfnamefont {A.}~\bibnamefont {Kannawadi}}, \bibinfo {author} {\bibfnamefont {B.}~\bibnamefont {St{\"o}lzner}}, \bibinfo {author} {\bibfnamefont {T.}~\bibnamefont {Tr{\"o}ster}}, \bibinfo {author} {\bibfnamefont {J.~L.}\ \bibnamefont {van~den Busch}}, \emph {et~al.},\ }\bibfield  {title} {\bibinfo {title} {Kids-1000 cosmology: Cosmic shear constraints and comparison between two point statistics},\ }\href@noop {} {\bibfield  {journal} {\bibinfo  {journal} {Astronomy \& Astrophysics}\ }\textbf {\bibinfo {volume} {645}},\ \bibinfo {pages} {A104} (\bibinfo {year} {2021})}\BibitemShut {NoStop}%
\bibitem [{\citenamefont {Amon}\ \emph {et~al.}(2022)\citenamefont {Amon}, \citenamefont {Gruen}, \citenamefont {Troxel}, \citenamefont {MacCrann}, \citenamefont {Dodelson}, \citenamefont {Choi}, \citenamefont {Doux}, \citenamefont {Secco}, \citenamefont {Samuroff}, \citenamefont {Krause} \emph {et~al.}}]{amon2022dark}%
  \BibitemOpen
  \bibfield  {author} {\bibinfo {author} {\bibfnamefont {A.}~\bibnamefont {Amon}}, \bibinfo {author} {\bibfnamefont {D.}~\bibnamefont {Gruen}}, \bibinfo {author} {\bibfnamefont {M.~A.}\ \bibnamefont {Troxel}}, \bibinfo {author} {\bibfnamefont {N.}~\bibnamefont {MacCrann}}, \bibinfo {author} {\bibfnamefont {S.}~\bibnamefont {Dodelson}}, \bibinfo {author} {\bibfnamefont {A.}~\bibnamefont {Choi}}, \bibinfo {author} {\bibfnamefont {C.}~\bibnamefont {Doux}}, \bibinfo {author} {\bibfnamefont {L.~F.}\ \bibnamefont {Secco}}, \bibinfo {author} {\bibfnamefont {S.}~\bibnamefont {Samuroff}}, \bibinfo {author} {\bibfnamefont {E.}~\bibnamefont {Krause}}, \emph {et~al.},\ }\bibfield  {title} {\bibinfo {title} {Dark energy survey year 3 results: Cosmology from cosmic shear and robustness to data calibration},\ }\href@noop {} {\bibfield  {journal} {\bibinfo  {journal} {Physical Review D}\ }\textbf {\bibinfo {volume} {105}},\ \bibinfo {pages} {023514} (\bibinfo {year} {2022})}\BibitemShut {NoStop}%
\bibitem [{\citenamefont {Ivezi{\'c}}\ \emph {et~al.}(2019)\citenamefont {Ivezi{\'c}}, \citenamefont {Kahn}, \citenamefont {Tyson}, \citenamefont {Abel}, \citenamefont {Acosta}, \citenamefont {Allsman}, \citenamefont {Alonso}, \citenamefont {AlSayyad}, \citenamefont {Anderson}, \citenamefont {Andrew} \emph {et~al.}}]{ivezic2019lsst}%
  \BibitemOpen
  \bibfield  {author} {\bibinfo {author} {\bibfnamefont {{\v{Z}}.}~\bibnamefont {Ivezi{\'c}}}, \bibinfo {author} {\bibfnamefont {S.~M.}\ \bibnamefont {Kahn}}, \bibinfo {author} {\bibfnamefont {J.~A.}\ \bibnamefont {Tyson}}, \bibinfo {author} {\bibfnamefont {B.}~\bibnamefont {Abel}}, \bibinfo {author} {\bibfnamefont {E.}~\bibnamefont {Acosta}}, \bibinfo {author} {\bibfnamefont {R.}~\bibnamefont {Allsman}}, \bibinfo {author} {\bibfnamefont {D.}~\bibnamefont {Alonso}}, \bibinfo {author} {\bibfnamefont {Y.}~\bibnamefont {AlSayyad}}, \bibinfo {author} {\bibfnamefont {S.~F.}\ \bibnamefont {Anderson}}, \bibinfo {author} {\bibfnamefont {J.}~\bibnamefont {Andrew}}, \emph {et~al.},\ }\bibfield  {title} {\bibinfo {title} {Lsst: from science drivers to reference design and anticipated data products},\ }\href@noop {} {\bibfield  {journal} {\bibinfo  {journal} {The Astrophysical Journal}\ }\textbf {\bibinfo {volume} {873}},\ \bibinfo {pages} {111} (\bibinfo {year} {2019})}\BibitemShut {NoStop}%
\bibitem [{\citenamefont {Laureijs}\ \emph {et~al.}(2011)\citenamefont {Laureijs}, \citenamefont {Amiaux}, \citenamefont {Arduini}, \citenamefont {Augueres}, \citenamefont {Brinchmann}, \citenamefont {Cole}, \citenamefont {Cropper}, \citenamefont {Dabin}, \citenamefont {Duvet}, \citenamefont {Ealet} \emph {et~al.}}]{laureijs2011euclid}%
  \BibitemOpen
  \bibfield  {author} {\bibinfo {author} {\bibfnamefont {R.}~\bibnamefont {Laureijs}}, \bibinfo {author} {\bibfnamefont {J.}~\bibnamefont {Amiaux}}, \bibinfo {author} {\bibfnamefont {S.}~\bibnamefont {Arduini}}, \bibinfo {author} {\bibfnamefont {J.-L.}\ \bibnamefont {Augueres}}, \bibinfo {author} {\bibfnamefont {J.}~\bibnamefont {Brinchmann}}, \bibinfo {author} {\bibfnamefont {R.}~\bibnamefont {Cole}}, \bibinfo {author} {\bibfnamefont {M.}~\bibnamefont {Cropper}}, \bibinfo {author} {\bibfnamefont {C.}~\bibnamefont {Dabin}}, \bibinfo {author} {\bibfnamefont {L.}~\bibnamefont {Duvet}}, \bibinfo {author} {\bibfnamefont {A.}~\bibnamefont {Ealet}}, \emph {et~al.},\ }\bibfield  {title} {\bibinfo {title} {Euclid definition study report},\ }\href@noop {} {\bibfield  {journal} {\bibinfo  {journal} {arXiv preprint arXiv:1110.3193}\ } (\bibinfo {year} {2011})}\BibitemShut {NoStop}%
\bibitem [{\citenamefont {Spergel}\ \emph {et~al.}(2015)\citenamefont {Spergel}, \citenamefont {Gehrels}, \citenamefont {Baltay}, \citenamefont {Bennett}, \citenamefont {Breckinridge}, \citenamefont {Donahue}, \citenamefont {Dressler}, \citenamefont {Gaudi}, \citenamefont {Greene}, \citenamefont {Guyon} \emph {et~al.}}]{spergel2015wide}%
  \BibitemOpen
  \bibfield  {author} {\bibinfo {author} {\bibfnamefont {D.}~\bibnamefont {Spergel}}, \bibinfo {author} {\bibfnamefont {N.}~\bibnamefont {Gehrels}}, \bibinfo {author} {\bibfnamefont {C.}~\bibnamefont {Baltay}}, \bibinfo {author} {\bibfnamefont {D.}~\bibnamefont {Bennett}}, \bibinfo {author} {\bibfnamefont {J.}~\bibnamefont {Breckinridge}}, \bibinfo {author} {\bibfnamefont {M.}~\bibnamefont {Donahue}}, \bibinfo {author} {\bibfnamefont {A.}~\bibnamefont {Dressler}}, \bibinfo {author} {\bibfnamefont {B.~S.}\ \bibnamefont {Gaudi}}, \bibinfo {author} {\bibfnamefont {T.}~\bibnamefont {Greene}}, \bibinfo {author} {\bibfnamefont {O.}~\bibnamefont {Guyon}}, \emph {et~al.},\ }\bibfield  {title} {\bibinfo {title} {Wide-field infrarred survey telescope-astrophysics focused telescope assets wfirst-afta 2015 report},\ }\href@noop {} {\bibfield  {journal} {\bibinfo  {journal} {ArXiv e-prints}\ ,\ \bibinfo {pages} {arXiv}} (\bibinfo {year} {2015})}\BibitemShut {NoStop}%
\bibitem [{\citenamefont {Bartelmann}\ and\ \citenamefont {Schneider}(2001)}]{bartelmann2001weak}%
  \BibitemOpen
  \bibfield  {author} {\bibinfo {author} {\bibfnamefont {M.}~\bibnamefont {Bartelmann}}\ and\ \bibinfo {author} {\bibfnamefont {P.}~\bibnamefont {Schneider}},\ }\bibfield  {title} {\bibinfo {title} {Weak gravitational lensing},\ }\href@noop {} {\bibfield  {journal} {\bibinfo  {journal} {Physics Reports}\ }\textbf {\bibinfo {volume} {340}},\ \bibinfo {pages} {291} (\bibinfo {year} {2001})}\BibitemShut {NoStop}%
\bibitem [{\citenamefont {Kilbinger}(2015)}]{kilbinger2015cosmology}%
  \BibitemOpen
  \bibfield  {author} {\bibinfo {author} {\bibfnamefont {M.}~\bibnamefont {Kilbinger}},\ }\bibfield  {title} {\bibinfo {title} {Cosmology with cosmic shear observations: a review},\ }\href@noop {} {\bibfield  {journal} {\bibinfo  {journal} {Reports on Progress in Physics}\ }\textbf {\bibinfo {volume} {78}},\ \bibinfo {pages} {086901} (\bibinfo {year} {2015})}\BibitemShut {NoStop}%
\bibitem [{\citenamefont {Mandelbaum}(2018)}]{mandelbaum2018weak}%
  \BibitemOpen
  \bibfield  {author} {\bibinfo {author} {\bibfnamefont {R.}~\bibnamefont {Mandelbaum}},\ }\bibfield  {title} {\bibinfo {title} {Weak lensing for precision cosmology},\ }\href@noop {} {\bibfield  {journal} {\bibinfo  {journal} {Annual Review of Astronomy and Astrophysics}\ }\textbf {\bibinfo {volume} {56}},\ \bibinfo {pages} {393} (\bibinfo {year} {2018})}\BibitemShut {NoStop}%
\bibitem [{\citenamefont {{Cooray}}\ and\ \citenamefont {{Hu}}(2002)}]{cooray2002second}%
  \BibitemOpen
  \bibfield  {author} {\bibinfo {author} {\bibfnamefont {A.}~\bibnamefont {{Cooray}}}\ and\ \bibinfo {author} {\bibfnamefont {W.}~\bibnamefont {{Hu}}},\ }\bibfield  {title} {\bibinfo {title} {{Second-Order Corrections to Weak Lensing by Large-Scale Structure}},\ }\href {https://doi.org/10.1086/340892} {\bibfield  {journal} {\bibinfo  {journal} {\apj}\ }\textbf {\bibinfo {volume} {574}},\ \bibinfo {pages} {19} (\bibinfo {year} {2002})},\ \Eprint {https://arxiv.org/abs/astro-ph/0202411} {arXiv:astro-ph/0202411 [astro-ph]} \BibitemShut {NoStop}%
\bibitem [{\citenamefont {Dodelson}\ \emph {et~al.}(2006)\citenamefont {Dodelson}, \citenamefont {Shapiro},\ and\ \citenamefont {White}}]{dodelson2006reduced}%
  \BibitemOpen
  \bibfield  {author} {\bibinfo {author} {\bibfnamefont {S.}~\bibnamefont {Dodelson}}, \bibinfo {author} {\bibfnamefont {C.}~\bibnamefont {Shapiro}},\ and\ \bibinfo {author} {\bibfnamefont {M.}~\bibnamefont {White}},\ }\bibfield  {title} {\bibinfo {title} {Reduced shear power spectrum},\ }\href@noop {} {\bibfield  {journal} {\bibinfo  {journal} {Physical Review D—Particles, Fields, Gravitation, and Cosmology}\ }\textbf {\bibinfo {volume} {73}},\ \bibinfo {pages} {023009} (\bibinfo {year} {2006})}\BibitemShut {NoStop}%
\bibitem [{\citenamefont {Bernardeau}\ \emph {et~al.}(2010)\citenamefont {Bernardeau}, \citenamefont {Bonvin},\ and\ \citenamefont {Vernizzi}}]{bernardeau2010full}%
  \BibitemOpen
  \bibfield  {author} {\bibinfo {author} {\bibfnamefont {F.}~\bibnamefont {Bernardeau}}, \bibinfo {author} {\bibfnamefont {C.}~\bibnamefont {Bonvin}},\ and\ \bibinfo {author} {\bibfnamefont {F.}~\bibnamefont {Vernizzi}},\ }\bibfield  {title} {\bibinfo {title} {Full-sky lensing shear at second order},\ }\href@noop {} {\bibfield  {journal} {\bibinfo  {journal} {Physical Review D—Particles, Fields, Gravitation, and Cosmology}\ }\textbf {\bibinfo {volume} {81}},\ \bibinfo {pages} {083002} (\bibinfo {year} {2010})}\BibitemShut {NoStop}%
\bibitem [{\citenamefont {Krause}\ and\ \citenamefont {Hirata}(2010)}]{krause2010weak}%
  \BibitemOpen
  \bibfield  {author} {\bibinfo {author} {\bibfnamefont {E.}~\bibnamefont {Krause}}\ and\ \bibinfo {author} {\bibfnamefont {C.~M.}\ \bibnamefont {Hirata}},\ }\bibfield  {title} {\bibinfo {title} {Weak lensing power spectra for precision cosmology-multiple-deflection, reduced shear, and lensing bias corrections},\ }\href@noop {} {\bibfield  {journal} {\bibinfo  {journal} {Astronomy \& Astrophysics}\ }\textbf {\bibinfo {volume} {523}},\ \bibinfo {pages} {A28} (\bibinfo {year} {2010})}\BibitemShut {NoStop}%
\bibitem [{\citenamefont {Pratten}\ and\ \citenamefont {Lewis}(2016)}]{PrattenLewis2016}%
  \BibitemOpen
  \bibfield  {author} {\bibinfo {author} {\bibfnamefont {G.}~\bibnamefont {Pratten}}\ and\ \bibinfo {author} {\bibfnamefont {A.}~\bibnamefont {Lewis}},\ }\bibfield  {title} {\bibinfo {title} {Impact of post-born lensing on the {CMB}},\ }\href {https://doi.org/10.1088/1475-7516/2016/08/047} {\bibfield  {journal} {\bibinfo  {journal} {Journal of Cosmology and Astroparticle Physics}\ }\textbf {\bibinfo {volume} {2016}}\bibinfo  {number} { (08)},\ \bibinfo {pages} {047}}\BibitemShut {NoStop}%
\bibitem [{\citenamefont {Fabbian}\ \emph {et~al.}(2018)\citenamefont {Fabbian}, \citenamefont {Calabrese},\ and\ \citenamefont {Carbone}}]{fabbian2018cmb}%
  \BibitemOpen
\bibfield  {number} {  }\bibfield  {author} {\bibinfo {author} {\bibfnamefont {G.}~\bibnamefont {Fabbian}}, \bibinfo {author} {\bibfnamefont {M.}~\bibnamefont {Calabrese}},\ and\ \bibinfo {author} {\bibfnamefont {C.}~\bibnamefont {Carbone}},\ }\bibfield  {title} {\bibinfo {title} {Cmb weak-lensing beyond the born approximation: a numerical approach},\ }\href@noop {} {\bibfield  {journal} {\bibinfo  {journal} {Journal of Cosmology and Astroparticle Physics}\ }\textbf {\bibinfo {volume} {2018}}\bibinfo  {number} { (02)},\ \bibinfo {pages} {050}}\BibitemShut {NoStop}%
\bibitem [{\citenamefont {Marozzi}\ \emph {et~al.}(2018)\citenamefont {Marozzi}, \citenamefont {Fanizza}, \citenamefont {Di~Dio},\ and\ \citenamefont {Durrer}}]{marozzi2018cmb}%
  \BibitemOpen
\bibfield  {number} {  }\bibfield  {author} {\bibinfo {author} {\bibfnamefont {G.}~\bibnamefont {Marozzi}}, \bibinfo {author} {\bibfnamefont {G.}~\bibnamefont {Fanizza}}, \bibinfo {author} {\bibfnamefont {E.}~\bibnamefont {Di~Dio}},\ and\ \bibinfo {author} {\bibfnamefont {R.}~\bibnamefont {Durrer}},\ }\bibfield  {title} {\bibinfo {title} {Cmb lensing beyond the leading order: Temperature and polarization anisotropies},\ }\href@noop {} {\bibfield  {journal} {\bibinfo  {journal} {Physical Review D}\ }\textbf {\bibinfo {volume} {98}},\ \bibinfo {pages} {023535} (\bibinfo {year} {2018})}\BibitemShut {NoStop}%
\bibitem [{\citenamefont {Fry}(1994)}]{fry1994gravity}%
  \BibitemOpen
  \bibfield  {author} {\bibinfo {author} {\bibfnamefont {J.}~\bibnamefont {Fry}},\ }\bibfield  {title} {\bibinfo {title} {Gravity, bias, and the galaxy three-point correlation function},\ }\href@noop {} {\bibfield  {journal} {\bibinfo  {journal} {Physical Review Letters}\ }\textbf {\bibinfo {volume} {73}},\ \bibinfo {pages} {215} (\bibinfo {year} {1994})}\BibitemShut {NoStop}%
\bibitem [{\citenamefont {Scoccimarro}\ \emph {et~al.}(2001)\citenamefont {Scoccimarro}, \citenamefont {Feldman}, \citenamefont {Fry},\ and\ \citenamefont {Frieman}}]{scoccimarro2001bispectrum}%
  \BibitemOpen
  \bibfield  {author} {\bibinfo {author} {\bibfnamefont {R.}~\bibnamefont {Scoccimarro}}, \bibinfo {author} {\bibfnamefont {H.~A.}\ \bibnamefont {Feldman}}, \bibinfo {author} {\bibfnamefont {J.~N.}\ \bibnamefont {Fry}},\ and\ \bibinfo {author} {\bibfnamefont {J.~A.}\ \bibnamefont {Frieman}},\ }\bibfield  {title} {\bibinfo {title} {The bispectrum of iras redshift catalogs},\ }\href@noop {} {\bibfield  {journal} {\bibinfo  {journal} {The Astrophysical Journal}\ }\textbf {\bibinfo {volume} {546}},\ \bibinfo {pages} {652} (\bibinfo {year} {2001})}\BibitemShut {NoStop}%
\bibitem [{\citenamefont {Jeong}\ and\ \citenamefont {Komatsu}(2009)}]{jeong2009primordial}%
  \BibitemOpen
  \bibfield  {author} {\bibinfo {author} {\bibfnamefont {D.}~\bibnamefont {Jeong}}\ and\ \bibinfo {author} {\bibfnamefont {E.}~\bibnamefont {Komatsu}},\ }\bibfield  {title} {\bibinfo {title} {Primordial non-gaussianity, scale-dependent bias, and the bispectrum of galaxies},\ }\href@noop {} {\bibfield  {journal} {\bibinfo  {journal} {The Astrophysical Journal}\ }\textbf {\bibinfo {volume} {703}},\ \bibinfo {pages} {1230} (\bibinfo {year} {2009})}\BibitemShut {NoStop}%
\bibitem [{\citenamefont {Halder}\ \emph {et~al.}(2021)\citenamefont {Halder}, \citenamefont {Friedrich}, \citenamefont {Seitz},\ and\ \citenamefont {Varga}}]{halder2021integrated}%
  \BibitemOpen
  \bibfield  {author} {\bibinfo {author} {\bibfnamefont {A.}~\bibnamefont {Halder}}, \bibinfo {author} {\bibfnamefont {O.}~\bibnamefont {Friedrich}}, \bibinfo {author} {\bibfnamefont {S.}~\bibnamefont {Seitz}},\ and\ \bibinfo {author} {\bibfnamefont {T.~N.}\ \bibnamefont {Varga}},\ }\bibfield  {title} {\bibinfo {title} {The integrated three-point correlation function of cosmic shear},\ }\href@noop {} {\bibfield  {journal} {\bibinfo  {journal} {Monthly Notices of the Royal Astronomical Society}\ }\textbf {\bibinfo {volume} {506}},\ \bibinfo {pages} {2780} (\bibinfo {year} {2021})}\BibitemShut {NoStop}%
\bibitem [{\citenamefont {{Lemos}}\ \emph {et~al.}(2017)\citenamefont {{Lemos}}, \citenamefont {{Challinor}},\ and\ \citenamefont {{Efstathiou}}}]{LemosEtAl2017}%
  \BibitemOpen
  \bibfield  {author} {\bibinfo {author} {\bibfnamefont {P.}~\bibnamefont {{Lemos}}}, \bibinfo {author} {\bibfnamefont {A.}~\bibnamefont {{Challinor}}},\ and\ \bibinfo {author} {\bibfnamefont {G.}~\bibnamefont {{Efstathiou}}},\ }\bibfield  {title} {\bibinfo {title} {{The effect of Limber and flat-sky approximations on galaxy weak lensing}},\ }\href {https://doi.org/10.1088/1475-7516/2017/05/014} {\bibfield  {journal} {\bibinfo  {journal} {\jcap}\ }\textbf {\bibinfo {volume} {2017}},\ \bibinfo {eid} {014} (\bibinfo {year} {2017})},\ \Eprint {https://arxiv.org/abs/1704.01054} {arXiv:1704.01054 [astro-ph.CO]} \BibitemShut {NoStop}%
\bibitem [{\citenamefont {{Fang}}\ \emph {et~al.}(2020)\citenamefont {{Fang}}, \citenamefont {{Krause}}, \citenamefont {{Eifler}},\ and\ \citenamefont {{MacCrann}}}]{FangEtAl2020}%
  \BibitemOpen
  \bibfield  {author} {\bibinfo {author} {\bibfnamefont {X.}~\bibnamefont {{Fang}}}, \bibinfo {author} {\bibfnamefont {E.}~\bibnamefont {{Krause}}}, \bibinfo {author} {\bibfnamefont {T.}~\bibnamefont {{Eifler}}},\ and\ \bibinfo {author} {\bibfnamefont {N.}~\bibnamefont {{MacCrann}}},\ }\bibfield  {title} {\bibinfo {title} {{Beyond Limber: efficient computation of angular power spectra for galaxy clustering and weak lensing}},\ }\href {https://doi.org/10.1088/1475-7516/2020/05/010} {\bibfield  {journal} {\bibinfo  {journal} {\jcap}\ }\textbf {\bibinfo {volume} {2020}},\ \bibinfo {eid} {010} (\bibinfo {year} {2020})},\ \Eprint {https://arxiv.org/abs/1911.11947} {arXiv:1911.11947 [astro-ph.CO]} \BibitemShut {NoStop}%
\bibitem [{\citenamefont {{Kitching}}\ \emph {et~al.}(2017)\citenamefont {{Kitching}}, \citenamefont {{Alsing}}, \citenamefont {{Heavens}}, \citenamefont {{Jimenez}}, \citenamefont {{McEwen}},\ and\ \citenamefont {{Verde}}}]{KitchingNonLimber2017}%
  \BibitemOpen
  \bibfield  {author} {\bibinfo {author} {\bibfnamefont {T.~D.}\ \bibnamefont {{Kitching}}}, \bibinfo {author} {\bibfnamefont {J.}~\bibnamefont {{Alsing}}}, \bibinfo {author} {\bibfnamefont {A.~F.}\ \bibnamefont {{Heavens}}}, \bibinfo {author} {\bibfnamefont {R.}~\bibnamefont {{Jimenez}}}, \bibinfo {author} {\bibfnamefont {J.~D.}\ \bibnamefont {{McEwen}}},\ and\ \bibinfo {author} {\bibfnamefont {L.}~\bibnamefont {{Verde}}},\ }\bibfield  {title} {\bibinfo {title} {{The limits of cosmic shear}},\ }\href {https://doi.org/10.1093/mnras/stx1039} {\bibfield  {journal} {\bibinfo  {journal} {\mnras}\ }\textbf {\bibinfo {volume} {469}},\ \bibinfo {pages} {2737} (\bibinfo {year} {2017})},\ \Eprint {https://arxiv.org/abs/1611.04954} {arXiv:1611.04954 [astro-ph.CO]} \BibitemShut {NoStop}%
\bibitem [{\citenamefont {{Cuesta-Lazaro}}\ \emph {et~al.}(2018)\citenamefont {{Cuesta-Lazaro}}, \citenamefont {{Quera-Bofarull}}, \citenamefont {{Reischke}},\ and\ \citenamefont {{Sch{\"a}fer}}}]{2018MNRAS.477..741C}%
  \BibitemOpen
  \bibfield  {author} {\bibinfo {author} {\bibfnamefont {C.}~\bibnamefont {{Cuesta-Lazaro}}}, \bibinfo {author} {\bibfnamefont {A.}~\bibnamefont {{Quera-Bofarull}}}, \bibinfo {author} {\bibfnamefont {R.}~\bibnamefont {{Reischke}}},\ and\ \bibinfo {author} {\bibfnamefont {B.~M.}\ \bibnamefont {{Sch{\"a}fer}}},\ }\bibfield  {title} {\bibinfo {title} {{Gravitational corrections to light propagation in a perturbed FLRW universe and corresponding weak-lensing spectra}},\ }\href {https://doi.org/10.1093/mnras/sty672} {\bibfield  {journal} {\bibinfo  {journal} {\mnras}\ }\textbf {\bibinfo {volume} {477}},\ \bibinfo {pages} {741} (\bibinfo {year} {2018})},\ \Eprint {https://arxiv.org/abs/1801.03325} {arXiv:1801.03325 [astro-ph.CO]} \BibitemShut {NoStop}%
\bibitem [{\citenamefont {{Hilbert, S.}}\ \emph {et~al.}(2009)\citenamefont {{Hilbert, S.}}, \citenamefont {{Hartlap, J.}}, \citenamefont {{White, S. D. M.}},\ and\ \citenamefont {{Schneider, P.}}}]{HilbertEtAl2009RayTracing}%
  \BibitemOpen
  \bibfield  {author} {\bibinfo {author} {\bibnamefont {{Hilbert, S.}}}, \bibinfo {author} {\bibnamefont {{Hartlap, J.}}}, \bibinfo {author} {\bibnamefont {{White, S. D. M.}}},\ and\ \bibinfo {author} {\bibnamefont {{Schneider, P.}}},\ }\bibfield  {title} {\bibinfo {title} {Ray-tracing through the millennium simulation: Born corrections and lens-lens coupling in cosmic shear and galaxy-galaxy lensing},\ }\href {https://doi.org/10.1051/0004-6361/200811054} {\bibfield  {journal} {\bibinfo  {journal} {A\&A}\ }\textbf {\bibinfo {volume} {499}},\ \bibinfo {pages} {31} (\bibinfo {year} {2009})}\BibitemShut {NoStop}%
\bibitem [{\citenamefont {{Petri}}\ \emph {et~al.}(2013)\citenamefont {{Petri}}, \citenamefont {{Haiman}}, \citenamefont {{Hui}}, \citenamefont {{May}},\ and\ \citenamefont {{Kratochvil}}}]{PetriEtAl2013}%
  \BibitemOpen
  \bibfield  {author} {\bibinfo {author} {\bibfnamefont {A.}~\bibnamefont {{Petri}}}, \bibinfo {author} {\bibfnamefont {Z.}~\bibnamefont {{Haiman}}}, \bibinfo {author} {\bibfnamefont {L.}~\bibnamefont {{Hui}}}, \bibinfo {author} {\bibfnamefont {M.}~\bibnamefont {{May}}},\ and\ \bibinfo {author} {\bibfnamefont {J.~M.}\ \bibnamefont {{Kratochvil}}},\ }\bibfield  {title} {\bibinfo {title} {{Cosmology with Minkowski functionals and moments of the weak lensing convergence field}},\ }\href {https://doi.org/10.1103/PhysRevD.88.123002} {\bibfield  {journal} {\bibinfo  {journal} {\prd}\ }\textbf {\bibinfo {volume} {88}},\ \bibinfo {eid} {123002} (\bibinfo {year} {2013})},\ \Eprint {https://arxiv.org/abs/1309.4460} {arXiv:1309.4460 [astro-ph.CO]} \BibitemShut {NoStop}%
\bibitem [{\citenamefont {{Takahashi}}\ \emph {et~al.}(2017)\citenamefont {{Takahashi}}, \citenamefont {{Hamana}}, \citenamefont {{Shirasaki}}, \citenamefont {{Namikawa}}, \citenamefont {{Nishimichi}}, \citenamefont {{Osato}},\ and\ \citenamefont {{Shiroyama}}}]{TakahashiEtAl2017Nbody}%
  \BibitemOpen
  \bibfield  {author} {\bibinfo {author} {\bibfnamefont {R.}~\bibnamefont {{Takahashi}}}, \bibinfo {author} {\bibfnamefont {T.}~\bibnamefont {{Hamana}}}, \bibinfo {author} {\bibfnamefont {M.}~\bibnamefont {{Shirasaki}}}, \bibinfo {author} {\bibfnamefont {T.}~\bibnamefont {{Namikawa}}}, \bibinfo {author} {\bibfnamefont {T.}~\bibnamefont {{Nishimichi}}}, \bibinfo {author} {\bibfnamefont {K.}~\bibnamefont {{Osato}}},\ and\ \bibinfo {author} {\bibfnamefont {K.}~\bibnamefont {{Shiroyama}}},\ }\bibfield  {title} {\bibinfo {title} {{Full-sky Gravitational Lensing Simulation for Large-area Galaxy Surveys and Cosmic Microwave Background Experiments}},\ }\href {https://doi.org/10.3847/1538-4357/aa943d} {\bibfield  {journal} {\bibinfo  {journal} {\apj}\ }\textbf {\bibinfo {volume} {850}},\ \bibinfo {eid} {24} (\bibinfo {year} {2017})},\ \Eprint {https://arxiv.org/abs/1706.01472} {arXiv:1706.01472 [astro-ph.CO]} \BibitemShut {NoStop}%
\bibitem [{\citenamefont {{Sachs}}(1961)}]{sachs1961gravitational}%
  \BibitemOpen
  \bibfield  {author} {\bibinfo {author} {\bibfnamefont {R.}~\bibnamefont {{Sachs}}},\ }\bibfield  {title} {\bibinfo {title} {{Gravitational Waves in General Relativity. VI. The Outgoing Radiation Condition}},\ }\href {https://doi.org/10.1098/rspa.1961.0202} {\bibfield  {journal} {\bibinfo  {journal} {Proceedings of the Royal Society of London Series A}\ }\textbf {\bibinfo {volume} {264}},\ \bibinfo {pages} {309} (\bibinfo {year} {1961})}\BibitemShut {NoStop}%
\bibitem [{\citenamefont {{Martin}}\ \emph {et~al.}(1973)\citenamefont {{Martin}}, \citenamefont {{Siggia}},\ and\ \citenamefont {{Rose}}}]{MartinSiggiaRose1973}%
  \BibitemOpen
  \bibfield  {author} {\bibinfo {author} {\bibfnamefont {P.~C.}\ \bibnamefont {{Martin}}}, \bibinfo {author} {\bibfnamefont {E.~D.}\ \bibnamefont {{Siggia}}},\ and\ \bibinfo {author} {\bibfnamefont {H.~A.}\ \bibnamefont {{Rose}}},\ }\bibfield  {title} {\bibinfo {title} {{Statistical Dynamics of Classical Systems}},\ }\href {https://doi.org/10.1103/PhysRevA.8.423} {\bibfield  {journal} {\bibinfo  {journal} {\pra}\ }\textbf {\bibinfo {volume} {8}},\ \bibinfo {pages} {423} (\bibinfo {year} {1973})}\BibitemShut {NoStop}%
\bibitem [{\citenamefont {{Janssen}}(1976)}]{Janssen1976}%
  \BibitemOpen
  \bibfield  {author} {\bibinfo {author} {\bibfnamefont {H.-K.}\ \bibnamefont {{Janssen}}},\ }\bibfield  {title} {\bibinfo {title} {{On a Lagrangean for classical field dynamics and renormalization group calculations of dynamical critical properties}},\ }\href {https://doi.org/10.1007/BF01316547} {\bibfield  {journal} {\bibinfo  {journal} {Zeitschrift fur Physik B Condensed Matter}\ }\textbf {\bibinfo {volume} {23}},\ \bibinfo {pages} {377} (\bibinfo {year} {1976})}\BibitemShut {NoStop}%
\bibitem [{\citenamefont {{Krommes}}(2002)}]{Krommes2002Review}%
  \BibitemOpen
  \bibfield  {author} {\bibinfo {author} {\bibfnamefont {J.~A.}\ \bibnamefont {{Krommes}}},\ }\bibfield  {title} {\bibinfo {title} {{Fundamental statistical descriptions of plasma turbulence in magnetic fields}},\ }\href {https://doi.org/10.1016/S0370-1573(01)00066-7} {\bibfield  {journal} {\bibinfo  {journal} {\physrep}\ }\textbf {\bibinfo {volume} {360}},\ \bibinfo {pages} {1} (\bibinfo {year} {2002})}\BibitemShut {NoStop}%
\bibitem [{\citenamefont {{Bernardeau}}\ \emph {et~al.}(2002)\citenamefont {{Bernardeau}}, \citenamefont {{Colombi}}, \citenamefont {{Gazta{\~n}aga}},\ and\ \citenamefont {{Scoccimarro}}}]{Bernardeau2002Review}%
  \BibitemOpen
  \bibfield  {author} {\bibinfo {author} {\bibfnamefont {F.}~\bibnamefont {{Bernardeau}}}, \bibinfo {author} {\bibfnamefont {S.}~\bibnamefont {{Colombi}}}, \bibinfo {author} {\bibfnamefont {E.}~\bibnamefont {{Gazta{\~n}aga}}},\ and\ \bibinfo {author} {\bibfnamefont {R.}~\bibnamefont {{Scoccimarro}}},\ }\bibfield  {title} {\bibinfo {title} {{Large-scale structure of the Universe and cosmological perturbation theory}},\ }\href {https://doi.org/10.1016/S0370-1573(02)00135-7} {\bibfield  {journal} {\bibinfo  {journal} {\physrep}\ }\textbf {\bibinfo {volume} {367}},\ \bibinfo {pages} {1} (\bibinfo {year} {2002})},\ \Eprint {https://arxiv.org/abs/astro-ph/0112551} {arXiv:astro-ph/0112551 [astro-ph]} \BibitemShut {NoStop}%
\bibitem [{\citenamefont {{Baumann}}\ \emph {et~al.}(2012)\citenamefont {{Baumann}}, \citenamefont {{Nicolis}}, \citenamefont {{Senatore}},\ and\ \citenamefont {{Zaldarriaga}}}]{2012JCAP...07..051B}%
  \BibitemOpen
  \bibfield  {author} {\bibinfo {author} {\bibfnamefont {D.}~\bibnamefont {{Baumann}}}, \bibinfo {author} {\bibfnamefont {A.}~\bibnamefont {{Nicolis}}}, \bibinfo {author} {\bibfnamefont {L.}~\bibnamefont {{Senatore}}},\ and\ \bibinfo {author} {\bibfnamefont {M.}~\bibnamefont {{Zaldarriaga}}},\ }\bibfield  {title} {\bibinfo {title} {{Cosmological non-linearities as an effective fluid}},\ }\href {https://doi.org/10.1088/1475-7516/2012/07/051} {\bibfield  {journal} {\bibinfo  {journal} {\jcap}\ }\textbf {\bibinfo {volume} {2012}},\ \bibinfo {eid} {051} (\bibinfo {year} {2012})},\ \Eprint {https://arxiv.org/abs/1004.2488} {arXiv:1004.2488 [astro-ph.CO]} \BibitemShut {NoStop}%
\bibitem [{\citenamefont {{Garny}}\ \emph {et~al.}(2023)\citenamefont {{Garny}}, \citenamefont {{Laxhuber}},\ and\ \citenamefont {{Scoccimarro}}}]{2023PhRvD.107f3539G}%
  \BibitemOpen
  \bibfield  {author} {\bibinfo {author} {\bibfnamefont {M.}~\bibnamefont {{Garny}}}, \bibinfo {author} {\bibfnamefont {D.}~\bibnamefont {{Laxhuber}}},\ and\ \bibinfo {author} {\bibfnamefont {R.}~\bibnamefont {{Scoccimarro}}},\ }\bibfield  {title} {\bibinfo {title} {{Perturbation theory with dispersion and higher cumulants: Framework and linear theory}},\ }\href {https://doi.org/10.1103/PhysRevD.107.063539} {\bibfield  {journal} {\bibinfo  {journal} {\prd}\ }\textbf {\bibinfo {volume} {107}},\ \bibinfo {eid} {063539} (\bibinfo {year} {2023})},\ \Eprint {https://arxiv.org/abs/2210.08088} {arXiv:2210.08088 [astro-ph.CO]} \BibitemShut {NoStop}%
\bibitem [{\citenamefont {{Crocce}}\ and\ \citenamefont {{Scoccimarro}}(2006)}]{2006PhRvD..73f3519C}%
  \BibitemOpen
  \bibfield  {author} {\bibinfo {author} {\bibfnamefont {M.}~\bibnamefont {{Crocce}}}\ and\ \bibinfo {author} {\bibfnamefont {R.}~\bibnamefont {{Scoccimarro}}},\ }\bibfield  {title} {\bibinfo {title} {{Renormalized cosmological perturbation theory}},\ }\href {https://doi.org/10.1103/PhysRevD.73.063519} {\bibfield  {journal} {\bibinfo  {journal} {\prd}\ }\textbf {\bibinfo {volume} {73}},\ \bibinfo {eid} {063519} (\bibinfo {year} {2006})},\ \Eprint {https://arxiv.org/abs/astro-ph/0509418} {arXiv:astro-ph/0509418 [astro-ph]} \BibitemShut {NoStop}%
\bibitem [{\citenamefont {{Valageas}}(2008)}]{2008A&A...484...79V}%
  \BibitemOpen
  \bibfield  {author} {\bibinfo {author} {\bibfnamefont {P.}~\bibnamefont {{Valageas}}},\ }\bibfield  {title} {\bibinfo {title} {{Expansion schemes for gravitational clustering: computing two-point and three-point functions}},\ }\href {https://doi.org/10.1051/0004-6361:20079071} {\bibfield  {journal} {\bibinfo  {journal} {\aap}\ }\textbf {\bibinfo {volume} {484}},\ \bibinfo {pages} {79} (\bibinfo {year} {2008})},\ \Eprint {https://arxiv.org/abs/0711.3407} {arXiv:0711.3407 [astro-ph]} \BibitemShut {NoStop}%
\bibitem [{\citenamefont {{Izumi}}\ and\ \citenamefont {{Soda}}(2007)}]{2007PhRvD..76h3517I}%
  \BibitemOpen
  \bibfield  {author} {\bibinfo {author} {\bibfnamefont {K.}~\bibnamefont {{Izumi}}}\ and\ \bibinfo {author} {\bibfnamefont {J.}~\bibnamefont {{Soda}}},\ }\bibfield  {title} {\bibinfo {title} {{Renormalized Newtonian cosmic evolution with primordial non-Gaussianity}},\ }\href {https://doi.org/10.1103/PhysRevD.76.083517} {\bibfield  {journal} {\bibinfo  {journal} {\prd}\ }\textbf {\bibinfo {volume} {76}},\ \bibinfo {eid} {083517} (\bibinfo {year} {2007})},\ \Eprint {https://arxiv.org/abs/0706.1604} {arXiv:0706.1604 [astro-ph]} \BibitemShut {NoStop}%
\bibitem [{\citenamefont {Zhang}(2026)}]{SFTwick2026}%
  \BibitemOpen
  \bibfield  {author} {\bibinfo {author} {\bibfnamefont {Z.}~\bibnamefont {Zhang}},\ }\href {https://arxiv.org/abs/2606.19480} {\bibinfo {title} {sft-wick: A formalism and package for feynman-diagram expansion and evaluation in stochastic field theories}} (\bibinfo {year} {2026}),\ \Eprint {https://arxiv.org/abs/2606.19480} {arXiv:2606.19480 [physics.comp-ph]} \BibitemShut {NoStop}%
\bibitem [{\citenamefont {{van Daalen}}\ \emph {et~al.}(2011)\citenamefont {{van Daalen}}, \citenamefont {{Schaye}}, \citenamefont {{Booth}},\ and\ \citenamefont {{Dalla Vecchia}}}]{vanDaalen2011}%
  \BibitemOpen
  \bibfield  {author} {\bibinfo {author} {\bibfnamefont {M.~P.}\ \bibnamefont {{van Daalen}}}, \bibinfo {author} {\bibfnamefont {J.}~\bibnamefont {{Schaye}}}, \bibinfo {author} {\bibfnamefont {C.~M.}\ \bibnamefont {{Booth}}},\ and\ \bibinfo {author} {\bibfnamefont {C.}~\bibnamefont {{Dalla Vecchia}}},\ }\bibfield  {title} {\bibinfo {title} {{The effects of galaxy formation on the matter power spectrum: a challenge for precision cosmology}},\ }\href {https://doi.org/10.1111/j.1365-2966.2011.18981.x} {\bibfield  {journal} {\bibinfo  {journal} {\mnras}\ }\textbf {\bibinfo {volume} {415}},\ \bibinfo {pages} {3649} (\bibinfo {year} {2011})},\ \Eprint {https://arxiv.org/abs/1104.1174} {arXiv:1104.1174 [astro-ph.CO]} \BibitemShut {NoStop}%
\bibitem [{\citenamefont {{Wagner}}\ \emph {et~al.}(2015)\citenamefont {{Wagner}}, \citenamefont {{Schmidt}}, \citenamefont {{Chiang}},\ and\ \citenamefont {{Komatsu}}}]{wagner2015squeezed}%
  \BibitemOpen
  \bibfield  {author} {\bibinfo {author} {\bibfnamefont {C.}~\bibnamefont {{Wagner}}}, \bibinfo {author} {\bibfnamefont {F.}~\bibnamefont {{Schmidt}}}, \bibinfo {author} {\bibfnamefont {C.-T.}\ \bibnamefont {{Chiang}}},\ and\ \bibinfo {author} {\bibfnamefont {E.}~\bibnamefont {{Komatsu}}},\ }\bibfield  {title} {\bibinfo {title} {{The angle-averaged squeezed limit of nonlinear matter N-point functions}},\ }\href {https://doi.org/10.1088/1475-7516/2015/08/042} {\bibfield  {journal} {\bibinfo  {journal} {\jcap}\ }\textbf {\bibinfo {volume} {2015}},\ \bibinfo {pages} {042} (\bibinfo {year} {2015})},\ \Eprint {https://arxiv.org/abs/1503.03487} {arXiv:1503.03487 [astro-ph.CO]} \BibitemShut {NoStop}%
\bibitem [{\citenamefont {{Li}}\ \emph {et~al.}(2014)\citenamefont {{Li}}, \citenamefont {{Hu}},\ and\ \citenamefont {{Takada}}}]{li2014supersample}%
  \BibitemOpen
  \bibfield  {author} {\bibinfo {author} {\bibfnamefont {Y.}~\bibnamefont {{Li}}}, \bibinfo {author} {\bibfnamefont {W.}~\bibnamefont {{Hu}}},\ and\ \bibinfo {author} {\bibfnamefont {M.}~\bibnamefont {{Takada}}},\ }\bibfield  {title} {\bibinfo {title} {{Super-sample covariance in simulations}},\ }\href {https://doi.org/10.1103/PhysRevD.89.083519} {\bibfield  {journal} {\bibinfo  {journal} {\prd}\ }\textbf {\bibinfo {volume} {89}},\ \bibinfo {eid} {083519} (\bibinfo {year} {2014})},\ \Eprint {https://arxiv.org/abs/1401.0385} {arXiv:1401.0385 [astro-ph.CO]} \BibitemShut {NoStop}%
\bibitem [{\citenamefont {{Barreira}}\ and\ \citenamefont {{Schmidt}}(2017)}]{barreira2017responses}%
  \BibitemOpen
  \bibfield  {author} {\bibinfo {author} {\bibfnamefont {A.}~\bibnamefont {{Barreira}}}\ and\ \bibinfo {author} {\bibfnamefont {F.}~\bibnamefont {{Schmidt}}},\ }\bibfield  {title} {\bibinfo {title} {{Responses in large-scale structure}},\ }\href {https://doi.org/10.1088/1475-7516/2017/06/053} {\bibfield  {journal} {\bibinfo  {journal} {\jcap}\ }\textbf {\bibinfo {volume} {2017}},\ \bibinfo {eid} {053} (\bibinfo {year} {2017})},\ \Eprint {https://arxiv.org/abs/1703.09212} {arXiv:1703.09212 [astro-ph.CO]} \BibitemShut {NoStop}%
\bibitem [{\citenamefont {{Diego}}\ \emph {et~al.}(2021)\citenamefont {{Diego}}, \citenamefont {{Broadhurst}},\ and\ \citenamefont {{Smoot}}}]{DiegoEtAl2021}%
  \BibitemOpen
  \bibfield  {author} {\bibinfo {author} {\bibfnamefont {J.~M.}\ \bibnamefont {{Diego}}}, \bibinfo {author} {\bibfnamefont {T.}~\bibnamefont {{Broadhurst}}},\ and\ \bibinfo {author} {\bibfnamefont {G.~F.}\ \bibnamefont {{Smoot}}},\ }\bibfield  {title} {\bibinfo {title} {{Evidence for lensing of gravitational waves from LIGO-Virgo data}},\ }\href {https://doi.org/10.1103/PhysRevD.104.103529} {\bibfield  {journal} {\bibinfo  {journal} {\prd}\ }\textbf {\bibinfo {volume} {104}},\ \bibinfo {eid} {103529} (\bibinfo {year} {2021})},\ \Eprint {https://arxiv.org/abs/2106.06545} {arXiv:2106.06545 [gr-qc]} \BibitemShut {NoStop}%
\bibitem [{\citenamefont {{Gunn}}(1967)}]{1967ApJ...150..737G}%
  \BibitemOpen
  \bibfield  {author} {\bibinfo {author} {\bibfnamefont {J.~E.}\ \bibnamefont {{Gunn}}},\ }\bibfield  {title} {\bibinfo {title} {{On the Propagation of Light in Inhomogeneous Cosmologies. I. Mean Effects}},\ }\href {https://doi.org/10.1086/149378} {\bibfield  {journal} {\bibinfo  {journal} {\apj}\ }\textbf {\bibinfo {volume} {150}},\ \bibinfo {pages} {737} (\bibinfo {year} {1967})}\BibitemShut {NoStop}%
\bibitem [{\citenamefont {Ellis}\ \emph {et~al.}(2012)\citenamefont {Ellis}, \citenamefont {Maartens},\ and\ \citenamefont {MacCallum}}]{ellis2012relativistic}%
  \BibitemOpen
  \bibfield  {author} {\bibinfo {author} {\bibfnamefont {G.~F.}\ \bibnamefont {Ellis}}, \bibinfo {author} {\bibfnamefont {R.}~\bibnamefont {Maartens}},\ and\ \bibinfo {author} {\bibfnamefont {M.~A.}\ \bibnamefont {MacCallum}},\ }\href@noop {} {\emph {\bibinfo {title} {Relativistic cosmology}}}\ (\bibinfo  {publisher} {Cambridge University Press},\ \bibinfo {year} {2012})\BibitemShut {NoStop}%
\bibitem [{\citenamefont {{Newman}}\ and\ \citenamefont {{Penrose}}(1962)}]{1962JMP.....3..566N}%
  \BibitemOpen
  \bibfield  {author} {\bibinfo {author} {\bibfnamefont {E.}~\bibnamefont {{Newman}}}\ and\ \bibinfo {author} {\bibfnamefont {R.}~\bibnamefont {{Penrose}}},\ }\bibfield  {title} {\bibinfo {title} {{An Approach to Gravitational Radiation by a Method of Spin Coefficients}},\ }\href {https://doi.org/10.1063/1.1724257} {\bibfield  {journal} {\bibinfo  {journal} {Journal of Mathematical Physics}\ }\textbf {\bibinfo {volume} {3}},\ \bibinfo {pages} {566} (\bibinfo {year} {1962})}\BibitemShut {NoStop}%
\bibitem [{\citenamefont {{Wald}}(1984)}]{1984ucp..book.....W}%
  \BibitemOpen
  \bibfield  {author} {\bibinfo {author} {\bibfnamefont {R.~M.}\ \bibnamefont {{Wald}}},\ }\href@noop {} {\emph {\bibinfo {title} {{General Relativity}}}}\ (\bibinfo {year} {1984})\BibitemShut {NoStop}%
\bibitem [{\citenamefont {{Fanizza}}\ \emph {et~al.}(2022)\citenamefont {{Fanizza}}, \citenamefont {{Di Dio}}, \citenamefont {{Durrer}},\ and\ \citenamefont {{Marozzi}}}]{2022JCAP...08..052F}%
  \BibitemOpen
  \bibfield  {author} {\bibinfo {author} {\bibfnamefont {G.}~\bibnamefont {{Fanizza}}}, \bibinfo {author} {\bibfnamefont {E.}~\bibnamefont {{Di Dio}}}, \bibinfo {author} {\bibfnamefont {R.}~\bibnamefont {{Durrer}}},\ and\ \bibinfo {author} {\bibfnamefont {G.}~\bibnamefont {{Marozzi}}},\ }\bibfield  {title} {\bibinfo {title} {{The gauge invariant cosmological Jacobi map from weak lensing at leading order}},\ }\href {https://doi.org/10.1088/1475-7516/2022/08/052} {\bibfield  {journal} {\bibinfo  {journal} {\jcap}\ }\textbf {\bibinfo {volume} {2022}},\ \bibinfo {eid} {052} (\bibinfo {year} {2022})},\ \Eprint {https://arxiv.org/abs/2201.11552} {arXiv:2201.11552 [astro-ph.CO]} \BibitemShut {NoStop}%
\bibitem [{\citenamefont {{Pueblas}}\ and\ \citenamefont {{Scoccimarro}}(2009)}]{2009PhRvD..80d3504P}%
  \BibitemOpen
  \bibfield  {author} {\bibinfo {author} {\bibfnamefont {S.}~\bibnamefont {{Pueblas}}}\ and\ \bibinfo {author} {\bibfnamefont {R.}~\bibnamefont {{Scoccimarro}}},\ }\bibfield  {title} {\bibinfo {title} {{Generation of vorticity and velocity dispersion by orbit crossing}},\ }\href {https://doi.org/10.1103/PhysRevD.80.043504} {\bibfield  {journal} {\bibinfo  {journal} {\prd}\ }\textbf {\bibinfo {volume} {80}},\ \bibinfo {eid} {043504} (\bibinfo {year} {2009})},\ \Eprint {https://arxiv.org/abs/0809.4606} {arXiv:0809.4606 [astro-ph]} \BibitemShut {NoStop}%
\bibitem [{\citenamefont {{Bertschinger}}(1996)}]{1996clss.conf..273B}%
  \BibitemOpen
  \bibfield  {author} {\bibinfo {author} {\bibfnamefont {E.}~\bibnamefont {{Bertschinger}}},\ }\bibfield  {title} {\bibinfo {title} {{Cosmological Dynamics}},\ }in\ \href@noop {} {\emph {\bibinfo {booktitle} {Cosmology and Large Scale Structure}}},\ \bibinfo {editor} {edited by\ \bibinfo {editor} {\bibfnamefont {R.}~\bibnamefont {{Schaeffer}}}, \bibinfo {editor} {\bibfnamefont {J.}~\bibnamefont {{Silk}}}, \bibinfo {editor} {\bibfnamefont {M.}~\bibnamefont {{Spiro}}},\ and\ \bibinfo {editor} {\bibfnamefont {J.}~\bibnamefont {{Zinn-Justin}}}}\ (\bibinfo {year} {1996})\ p.\ \bibinfo {pages} {273}\BibitemShut {NoStop}%
\bibitem [{\citenamefont {{Planck Collaboration}}\ \emph {et~al.}(2016)\citenamefont {{Planck Collaboration}}, \citenamefont {{Ade}}, \citenamefont {{Aghanim}}, \citenamefont {{Arnaud}}, \citenamefont {{Ashdown}}, \citenamefont {{Aumont}}, \citenamefont {{Baccigalupi}}, \citenamefont {{Banday}}, \citenamefont {{Barreiro}}, \citenamefont {{Bartlett}}, \citenamefont {{Bartolo}}, \citenamefont {{Battaner}}, \citenamefont {{Battye}}, \citenamefont {{Benabed}}, \citenamefont {{Beno{\^\i}t}}, \citenamefont {{Benoit-L{\'e}vy}}, \citenamefont {{Bernard}}, \citenamefont {{Bersanelli}}, \citenamefont {{Bielewicz}}, \citenamefont {{Bock}}, \citenamefont {{Bonaldi}}, \citenamefont {{Bonavera}}, \citenamefont {{Bond}}, \citenamefont {{Borrill}}, \citenamefont {{Bouchet}}, \citenamefont {{Boulanger}}, \citenamefont {{Bucher}}, \citenamefont {{Burigana}}, \citenamefont {{Butler}}, \citenamefont {{Calabrese}}, \citenamefont {{Cardoso}}, \citenamefont {{Catalano}}, \citenamefont {{Challinor}}, \citenamefont {{Chamballu}},
  \citenamefont {{Chary}}, \citenamefont {{Chiang}}, \citenamefont {{Chluba}}, \citenamefont {{Christensen}}, \citenamefont {{Church}}, \citenamefont {{Clements}}, \citenamefont {{Colombi}}, \citenamefont {{Colombo}}, \citenamefont {{Combet}}, \citenamefont {{Coulais}}, \citenamefont {{Crill}}, \citenamefont {{Curto}}, \citenamefont {{Cuttaia}}, \citenamefont {{Danese}}, \citenamefont {{Davies}}, \citenamefont {{Davis}}, \citenamefont {{de Bernardis}}, \citenamefont {{de Rosa}}, \citenamefont {{de Zotti}}, \citenamefont {{Delabrouille}}, \citenamefont {{D{\'e}sert}}, \citenamefont {{Di Valentino}}, \citenamefont {{Dickinson}}, \citenamefont {{Diego}}, \citenamefont {{Dolag}}, \citenamefont {{Dole}}, \citenamefont {{Donzelli}}, \citenamefont {{Dor{\'e}}}, \citenamefont {{Douspis}}, \citenamefont {{Ducout}}, \citenamefont {{Dunkley}}, \citenamefont {{Dupac}}, \citenamefont {{Efstathiou}}, \citenamefont {{Elsner}}, \citenamefont {{En{\ss}lin}}, \citenamefont {{Eriksen}}, \citenamefont {{Farhang}}, \citenamefont
  {{Fergusson}}, \citenamefont {{Finelli}}, \citenamefont {{Forni}}, \citenamefont {{Frailis}}, \citenamefont {{Fraisse}}, \citenamefont {{Franceschi}}, \citenamefont {{Frejsel}}, \citenamefont {{Galeotta}}, \citenamefont {{Galli}}, \citenamefont {{Ganga}}, \citenamefont {{Gauthier}}, \citenamefont {{Gerbino}}, \citenamefont {{Ghosh}}, \citenamefont {{Giard}}, \citenamefont {{Giraud-H{\'e}raud}}, \citenamefont {{Giusarma}}, \citenamefont {{Gjerl{\o}w}}, \citenamefont {{Gonz{\'a}lez-Nuevo}}, \citenamefont {{G{\'o}rski}}, \citenamefont {{Gratton}}, \citenamefont {{Gregorio}}, \citenamefont {{Gruppuso}}, \citenamefont {{Gudmundsson}}, \citenamefont {{Hamann}}, \citenamefont {{Hansen}}, \citenamefont {{Hanson}}, \citenamefont {{Harrison}}, \citenamefont {{Helou}}, \citenamefont {{Henrot-Versill{\'e}}}, \citenamefont {{Hern{\'a}ndez-Monteagudo}}, \citenamefont {{Herranz}}, \citenamefont {{Hildebrandt}}, \citenamefont {{Hivon}}, \citenamefont {{Hobson}}, \citenamefont {{Holmes}}, \citenamefont {{Hornstrup}},
  \citenamefont {{Hovest}}, \citenamefont {{Huang}}, \citenamefont {{Huffenberger}}, \citenamefont {{Hurier}}, \citenamefont {{Jaffe}}, \citenamefont {{Jaffe}}, \citenamefont {{Jones}}, \citenamefont {{Juvela}}, \citenamefont {{Keih{\"a}nen}}, \citenamefont {{Keskitalo}}, \citenamefont {{Kisner}}, \citenamefont {{Kneissl}}, \citenamefont {{Knoche}}, \citenamefont {{Knox}}, \citenamefont {{Kunz}}, \citenamefont {{Kurki-Suonio}}, \citenamefont {{Lagache}}, \citenamefont {{L{\"a}hteenm{\"a}ki}}, \citenamefont {{Lamarre}}, \citenamefont {{Lasenby}}, \citenamefont {{Lattanzi}}, \citenamefont {{Lawrence}}, \citenamefont {{Leahy}}, \citenamefont {{Leonardi}}, \citenamefont {{Lesgourgues}}, \citenamefont {{Levrier}}, \citenamefont {{Lewis}}, \citenamefont {{Liguori}}, \citenamefont {{Lilje}}, \citenamefont {{Linden-V{\o}rnle}}, \citenamefont {{L{\'o}pez-Caniego}}, \citenamefont {{Lubin}}, \citenamefont {{Mac{\'\i}as-P{\'e}rez}}, \citenamefont {{Maggio}}, \citenamefont {{Maino}}, \citenamefont {{Mandolesi}},
  \citenamefont {{Mangilli}}, \citenamefont {{Marchini}}, \citenamefont {{Maris}}, \citenamefont {{Martin}}, \citenamefont {{Martinelli}}, \citenamefont {{Mart{\'\i}nez-Gonz{\'a}lez}}, \citenamefont {{Masi}}, \citenamefont {{Matarrese}}, \citenamefont {{McGehee}}, \citenamefont {{Meinhold}}, \citenamefont {{Melchiorri}}, \citenamefont {{Melin}}, \citenamefont {{Mendes}}, \citenamefont {{Mennella}}, \citenamefont {{Migliaccio}}, \citenamefont {{Millea}}, \citenamefont {{Mitra}}, \citenamefont {{Miville-Desch{\^e}nes}}, \citenamefont {{Moneti}}, \citenamefont {{Montier}}, \citenamefont {{Morgante}}, \citenamefont {{Mortlock}}, \citenamefont {{Moss}}, \citenamefont {{Munshi}}, \citenamefont {{Murphy}}, \citenamefont {{Naselsky}}, \citenamefont {{Nati}}, \citenamefont {{Natoli}}, \citenamefont {{Netterfield}}, \citenamefont {{N{\o}rgaard-Nielsen}}, \citenamefont {{Noviello}}, \citenamefont {{Novikov}}, \citenamefont {{Novikov}}, \citenamefont {{Oxborrow}}, \citenamefont {{Paci}}, \citenamefont {{Pagano}},
  \citenamefont {{Pajot}}, \citenamefont {{Paladini}}, \citenamefont {{Paoletti}}, \citenamefont {{Partridge}}, \citenamefont {{Pasian}}, \citenamefont {{Patanchon}}, \citenamefont {{Pearson}}, \citenamefont {{Perdereau}}, \citenamefont {{Perotto}}, \citenamefont {{Perrotta}}, \citenamefont {{Pettorino}}, \citenamefont {{Piacentini}}, \citenamefont {{Piat}}, \citenamefont {{Pierpaoli}}, \citenamefont {{Pietrobon}}, \citenamefont {{Plaszczynski}}, \citenamefont {{Pointecouteau}}, \citenamefont {{Polenta}}, \citenamefont {{Popa}}, \citenamefont {{Pratt}},\ and\ \citenamefont {{Pr{\'e}zeau}}}]{planck2015parameters}%
  \BibitemOpen
  \bibfield  {author} {\bibinfo {author} {\bibnamefont {{Planck Collaboration}}}, \bibinfo {author} {\bibfnamefont {P.~A.~R.}\ \bibnamefont {{Ade}}}, \bibinfo {author} {\bibfnamefont {N.}~\bibnamefont {{Aghanim}}}, \bibinfo {author} {\bibfnamefont {M.}~\bibnamefont {{Arnaud}}}, \bibinfo {author} {\bibfnamefont {M.}~\bibnamefont {{Ashdown}}}, \bibinfo {author} {\bibfnamefont {J.}~\bibnamefont {{Aumont}}}, \bibinfo {author} {\bibfnamefont {C.}~\bibnamefont {{Baccigalupi}}}, \bibinfo {author} {\bibfnamefont {A.~J.}\ \bibnamefont {{Banday}}}, \bibinfo {author} {\bibfnamefont {R.~B.}\ \bibnamefont {{Barreiro}}}, \bibinfo {author} {\bibfnamefont {J.~G.}\ \bibnamefont {{Bartlett}}}, \bibinfo {author} {\bibfnamefont {N.}~\bibnamefont {{Bartolo}}}, \bibinfo {author} {\bibfnamefont {E.}~\bibnamefont {{Battaner}}}, \bibinfo {author} {\bibfnamefont {R.}~\bibnamefont {{Battye}}}, \bibinfo {author} {\bibfnamefont {K.}~\bibnamefont {{Benabed}}}, \bibinfo {author} {\bibfnamefont {A.}~\bibnamefont {{Beno{\^\i}t}}}, \bibinfo
  {author} {\bibfnamefont {A.}~\bibnamefont {{Benoit-L{\'e}vy}}}, \bibinfo {author} {\bibfnamefont {J.-P.}\ \bibnamefont {{Bernard}}}, \bibinfo {author} {\bibfnamefont {M.}~\bibnamefont {{Bersanelli}}}, \bibinfo {author} {\bibfnamefont {P.}~\bibnamefont {{Bielewicz}}}, \bibinfo {author} {\bibfnamefont {J.~J.}\ \bibnamefont {{Bock}}}, \bibinfo {author} {\bibfnamefont {A.}~\bibnamefont {{Bonaldi}}}, \bibinfo {author} {\bibfnamefont {L.}~\bibnamefont {{Bonavera}}}, \bibinfo {author} {\bibfnamefont {J.~R.}\ \bibnamefont {{Bond}}}, \bibinfo {author} {\bibfnamefont {J.}~\bibnamefont {{Borrill}}}, \bibinfo {author} {\bibfnamefont {F.~R.}\ \bibnamefont {{Bouchet}}}, \bibinfo {author} {\bibfnamefont {F.}~\bibnamefont {{Boulanger}}}, \bibinfo {author} {\bibfnamefont {M.}~\bibnamefont {{Bucher}}}, \bibinfo {author} {\bibfnamefont {C.}~\bibnamefont {{Burigana}}}, \bibinfo {author} {\bibfnamefont {R.~C.}\ \bibnamefont {{Butler}}}, \bibinfo {author} {\bibfnamefont {E.}~\bibnamefont {{Calabrese}}}, \bibinfo {author}
  {\bibfnamefont {J.-F.}\ \bibnamefont {{Cardoso}}}, \bibinfo {author} {\bibfnamefont {A.}~\bibnamefont {{Catalano}}}, \bibinfo {author} {\bibfnamefont {A.}~\bibnamefont {{Challinor}}}, \bibinfo {author} {\bibfnamefont {A.}~\bibnamefont {{Chamballu}}}, \bibinfo {author} {\bibfnamefont {R.-R.}\ \bibnamefont {{Chary}}}, \bibinfo {author} {\bibfnamefont {H.~C.}\ \bibnamefont {{Chiang}}}, \bibinfo {author} {\bibfnamefont {J.}~\bibnamefont {{Chluba}}}, \bibinfo {author} {\bibfnamefont {P.~R.}\ \bibnamefont {{Christensen}}}, \bibinfo {author} {\bibfnamefont {S.}~\bibnamefont {{Church}}}, \bibinfo {author} {\bibfnamefont {D.~L.}\ \bibnamefont {{Clements}}}, \bibinfo {author} {\bibfnamefont {S.}~\bibnamefont {{Colombi}}}, \bibinfo {author} {\bibfnamefont {L.~P.~L.}\ \bibnamefont {{Colombo}}}, \bibinfo {author} {\bibfnamefont {C.}~\bibnamefont {{Combet}}}, \bibinfo {author} {\bibfnamefont {A.}~\bibnamefont {{Coulais}}}, \bibinfo {author} {\bibfnamefont {B.~P.}\ \bibnamefont {{Crill}}}, \bibinfo {author} {\bibfnamefont
  {A.}~\bibnamefont {{Curto}}}, \bibinfo {author} {\bibfnamefont {F.}~\bibnamefont {{Cuttaia}}}, \bibinfo {author} {\bibfnamefont {L.}~\bibnamefont {{Danese}}}, \bibinfo {author} {\bibfnamefont {R.~D.}\ \bibnamefont {{Davies}}}, \bibinfo {author} {\bibfnamefont {R.~J.}\ \bibnamefont {{Davis}}}, \bibinfo {author} {\bibfnamefont {P.}~\bibnamefont {{de Bernardis}}}, \bibinfo {author} {\bibfnamefont {A.}~\bibnamefont {{de Rosa}}}, \bibinfo {author} {\bibfnamefont {G.}~\bibnamefont {{de Zotti}}}, \bibinfo {author} {\bibfnamefont {J.}~\bibnamefont {{Delabrouille}}}, \bibinfo {author} {\bibfnamefont {F.-X.}\ \bibnamefont {{D{\'e}sert}}}, \bibinfo {author} {\bibfnamefont {E.}~\bibnamefont {{Di Valentino}}}, \bibinfo {author} {\bibfnamefont {C.}~\bibnamefont {{Dickinson}}}, \bibinfo {author} {\bibfnamefont {J.~M.}\ \bibnamefont {{Diego}}}, \bibinfo {author} {\bibfnamefont {K.}~\bibnamefont {{Dolag}}}, \bibinfo {author} {\bibfnamefont {H.}~\bibnamefont {{Dole}}}, \bibinfo {author} {\bibfnamefont {S.}~\bibnamefont
  {{Donzelli}}}, \bibinfo {author} {\bibfnamefont {O.}~\bibnamefont {{Dor{\'e}}}}, \bibinfo {author} {\bibfnamefont {M.}~\bibnamefont {{Douspis}}}, \bibinfo {author} {\bibfnamefont {A.}~\bibnamefont {{Ducout}}}, \bibinfo {author} {\bibfnamefont {J.}~\bibnamefont {{Dunkley}}}, \bibinfo {author} {\bibfnamefont {X.}~\bibnamefont {{Dupac}}}, \bibinfo {author} {\bibfnamefont {G.}~\bibnamefont {{Efstathiou}}}, \bibinfo {author} {\bibfnamefont {F.}~\bibnamefont {{Elsner}}}, \bibinfo {author} {\bibfnamefont {T.~A.}\ \bibnamefont {{En{\ss}lin}}}, \bibinfo {author} {\bibfnamefont {H.~K.}\ \bibnamefont {{Eriksen}}}, \bibinfo {author} {\bibfnamefont {M.}~\bibnamefont {{Farhang}}}, \bibinfo {author} {\bibfnamefont {J.}~\bibnamefont {{Fergusson}}}, \bibinfo {author} {\bibfnamefont {F.}~\bibnamefont {{Finelli}}}, \bibinfo {author} {\bibfnamefont {O.}~\bibnamefont {{Forni}}}, \bibinfo {author} {\bibfnamefont {M.}~\bibnamefont {{Frailis}}}, \bibinfo {author} {\bibfnamefont {A.~A.}\ \bibnamefont {{Fraisse}}}, \bibinfo {author}
  {\bibfnamefont {E.}~\bibnamefont {{Franceschi}}}, \bibinfo {author} {\bibfnamefont {A.}~\bibnamefont {{Frejsel}}}, \bibinfo {author} {\bibfnamefont {S.}~\bibnamefont {{Galeotta}}}, \bibinfo {author} {\bibfnamefont {S.}~\bibnamefont {{Galli}}}, \bibinfo {author} {\bibfnamefont {K.}~\bibnamefont {{Ganga}}}, \bibinfo {author} {\bibfnamefont {C.}~\bibnamefont {{Gauthier}}}, \bibinfo {author} {\bibfnamefont {M.}~\bibnamefont {{Gerbino}}}, \bibinfo {author} {\bibfnamefont {T.}~\bibnamefont {{Ghosh}}}, \bibinfo {author} {\bibfnamefont {M.}~\bibnamefont {{Giard}}}, \bibinfo {author} {\bibfnamefont {Y.}~\bibnamefont {{Giraud-H{\'e}raud}}}, \bibinfo {author} {\bibfnamefont {E.}~\bibnamefont {{Giusarma}}}, \bibinfo {author} {\bibfnamefont {E.}~\bibnamefont {{Gjerl{\o}w}}}, \bibinfo {author} {\bibfnamefont {J.}~\bibnamefont {{Gonz{\'a}lez-Nuevo}}}, \bibinfo {author} {\bibfnamefont {K.~M.}\ \bibnamefont {{G{\'o}rski}}}, \bibinfo {author} {\bibfnamefont {S.}~\bibnamefont {{Gratton}}}, \bibinfo {author} {\bibfnamefont
  {A.}~\bibnamefont {{Gregorio}}}, \bibinfo {author} {\bibfnamefont {A.}~\bibnamefont {{Gruppuso}}}, \bibinfo {author} {\bibfnamefont {J.~E.}\ \bibnamefont {{Gudmundsson}}}, \bibinfo {author} {\bibfnamefont {J.}~\bibnamefont {{Hamann}}}, \bibinfo {author} {\bibfnamefont {F.~K.}\ \bibnamefont {{Hansen}}}, \bibinfo {author} {\bibfnamefont {D.}~\bibnamefont {{Hanson}}}, \bibinfo {author} {\bibfnamefont {D.~L.}\ \bibnamefont {{Harrison}}}, \bibinfo {author} {\bibfnamefont {G.}~\bibnamefont {{Helou}}}, \bibinfo {author} {\bibfnamefont {S.}~\bibnamefont {{Henrot-Versill{\'e}}}}, \bibinfo {author} {\bibfnamefont {C.}~\bibnamefont {{Hern{\'a}ndez-Monteagudo}}}, \bibinfo {author} {\bibfnamefont {D.}~\bibnamefont {{Herranz}}}, \bibinfo {author} {\bibfnamefont {S.~R.}\ \bibnamefont {{Hildebrandt}}}, \bibinfo {author} {\bibfnamefont {E.}~\bibnamefont {{Hivon}}}, \bibinfo {author} {\bibfnamefont {M.}~\bibnamefont {{Hobson}}}, \bibinfo {author} {\bibfnamefont {W.~A.}\ \bibnamefont {{Holmes}}}, \bibinfo {author}
  {\bibfnamefont {A.}~\bibnamefont {{Hornstrup}}}, \bibinfo {author} {\bibfnamefont {W.}~\bibnamefont {{Hovest}}}, \bibinfo {author} {\bibfnamefont {Z.}~\bibnamefont {{Huang}}}, \bibinfo {author} {\bibfnamefont {K.~M.}\ \bibnamefont {{Huffenberger}}}, \bibinfo {author} {\bibfnamefont {G.}~\bibnamefont {{Hurier}}}, \bibinfo {author} {\bibfnamefont {A.~H.}\ \bibnamefont {{Jaffe}}}, \bibinfo {author} {\bibfnamefont {T.~R.}\ \bibnamefont {{Jaffe}}}, \bibinfo {author} {\bibfnamefont {W.~C.}\ \bibnamefont {{Jones}}}, \bibinfo {author} {\bibfnamefont {M.}~\bibnamefont {{Juvela}}}, \bibinfo {author} {\bibfnamefont {E.}~\bibnamefont {{Keih{\"a}nen}}}, \bibinfo {author} {\bibfnamefont {R.}~\bibnamefont {{Keskitalo}}}, \bibinfo {author} {\bibfnamefont {T.~S.}\ \bibnamefont {{Kisner}}}, \bibinfo {author} {\bibfnamefont {R.}~\bibnamefont {{Kneissl}}}, \bibinfo {author} {\bibfnamefont {J.}~\bibnamefont {{Knoche}}}, \bibinfo {author} {\bibfnamefont {L.}~\bibnamefont {{Knox}}}, \bibinfo {author} {\bibfnamefont
  {M.}~\bibnamefont {{Kunz}}}, \bibinfo {author} {\bibfnamefont {H.}~\bibnamefont {{Kurki-Suonio}}}, \bibinfo {author} {\bibfnamefont {G.}~\bibnamefont {{Lagache}}}, \bibinfo {author} {\bibfnamefont {A.}~\bibnamefont {{L{\"a}hteenm{\"a}ki}}}, \bibinfo {author} {\bibfnamefont {J.-M.}\ \bibnamefont {{Lamarre}}}, \bibinfo {author} {\bibfnamefont {A.}~\bibnamefont {{Lasenby}}}, \bibinfo {author} {\bibfnamefont {M.}~\bibnamefont {{Lattanzi}}}, \bibinfo {author} {\bibfnamefont {C.~R.}\ \bibnamefont {{Lawrence}}}, \bibinfo {author} {\bibfnamefont {J.~P.}\ \bibnamefont {{Leahy}}}, \bibinfo {author} {\bibfnamefont {R.}~\bibnamefont {{Leonardi}}}, \bibinfo {author} {\bibfnamefont {J.}~\bibnamefont {{Lesgourgues}}}, \bibinfo {author} {\bibfnamefont {F.}~\bibnamefont {{Levrier}}}, \bibinfo {author} {\bibfnamefont {A.}~\bibnamefont {{Lewis}}}, \bibinfo {author} {\bibfnamefont {M.}~\bibnamefont {{Liguori}}}, \bibinfo {author} {\bibfnamefont {P.~B.}\ \bibnamefont {{Lilje}}}, \bibinfo {author} {\bibfnamefont
  {M.}~\bibnamefont {{Linden-V{\o}rnle}}}, \bibinfo {author} {\bibfnamefont {M.}~\bibnamefont {{L{\'o}pez-Caniego}}}, \bibinfo {author} {\bibfnamefont {P.~M.}\ \bibnamefont {{Lubin}}}, \bibinfo {author} {\bibfnamefont {J.~F.}\ \bibnamefont {{Mac{\'\i}as-P{\'e}rez}}}, \bibinfo {author} {\bibfnamefont {G.}~\bibnamefont {{Maggio}}}, \bibinfo {author} {\bibfnamefont {D.}~\bibnamefont {{Maino}}}, \bibinfo {author} {\bibfnamefont {N.}~\bibnamefont {{Mandolesi}}}, \bibinfo {author} {\bibfnamefont {A.}~\bibnamefont {{Mangilli}}}, \bibinfo {author} {\bibfnamefont {A.}~\bibnamefont {{Marchini}}}, \bibinfo {author} {\bibfnamefont {M.}~\bibnamefont {{Maris}}}, \bibinfo {author} {\bibfnamefont {P.~G.}\ \bibnamefont {{Martin}}}, \bibinfo {author} {\bibfnamefont {M.}~\bibnamefont {{Martinelli}}}, \bibinfo {author} {\bibfnamefont {E.}~\bibnamefont {{Mart{\'\i}nez-Gonz{\'a}lez}}}, \bibinfo {author} {\bibfnamefont {S.}~\bibnamefont {{Masi}}}, \bibinfo {author} {\bibfnamefont {S.}~\bibnamefont {{Matarrese}}}, \bibinfo {author}
  {\bibfnamefont {P.}~\bibnamefont {{McGehee}}}, \bibinfo {author} {\bibfnamefont {P.~R.}\ \bibnamefont {{Meinhold}}}, \bibinfo {author} {\bibfnamefont {A.}~\bibnamefont {{Melchiorri}}}, \bibinfo {author} {\bibfnamefont {J.-B.}\ \bibnamefont {{Melin}}}, \bibinfo {author} {\bibfnamefont {L.}~\bibnamefont {{Mendes}}}, \bibinfo {author} {\bibfnamefont {A.}~\bibnamefont {{Mennella}}}, \bibinfo {author} {\bibfnamefont {M.}~\bibnamefont {{Migliaccio}}}, \bibinfo {author} {\bibfnamefont {M.}~\bibnamefont {{Millea}}}, \bibinfo {author} {\bibfnamefont {S.}~\bibnamefont {{Mitra}}}, \bibinfo {author} {\bibfnamefont {M.-A.}\ \bibnamefont {{Miville-Desch{\^e}nes}}}, \bibinfo {author} {\bibfnamefont {A.}~\bibnamefont {{Moneti}}}, \bibinfo {author} {\bibfnamefont {L.}~\bibnamefont {{Montier}}}, \bibinfo {author} {\bibfnamefont {G.}~\bibnamefont {{Morgante}}}, \bibinfo {author} {\bibfnamefont {D.}~\bibnamefont {{Mortlock}}}, \bibinfo {author} {\bibfnamefont {A.}~\bibnamefont {{Moss}}}, \bibinfo {author} {\bibfnamefont
  {D.}~\bibnamefont {{Munshi}}}, \bibinfo {author} {\bibfnamefont {J.~A.}\ \bibnamefont {{Murphy}}}, \bibinfo {author} {\bibfnamefont {P.}~\bibnamefont {{Naselsky}}}, \bibinfo {author} {\bibfnamefont {F.}~\bibnamefont {{Nati}}}, \bibinfo {author} {\bibfnamefont {P.}~\bibnamefont {{Natoli}}}, \bibinfo {author} {\bibfnamefont {C.~B.}\ \bibnamefont {{Netterfield}}}, \bibinfo {author} {\bibfnamefont {H.~U.}\ \bibnamefont {{N{\o}rgaard-Nielsen}}}, \bibinfo {author} {\bibfnamefont {F.}~\bibnamefont {{Noviello}}}, \bibinfo {author} {\bibfnamefont {D.}~\bibnamefont {{Novikov}}}, \bibinfo {author} {\bibfnamefont {I.}~\bibnamefont {{Novikov}}}, \bibinfo {author} {\bibfnamefont {C.~A.}\ \bibnamefont {{Oxborrow}}}, \bibinfo {author} {\bibfnamefont {F.}~\bibnamefont {{Paci}}}, \bibinfo {author} {\bibfnamefont {L.}~\bibnamefont {{Pagano}}}, \bibinfo {author} {\bibfnamefont {F.}~\bibnamefont {{Pajot}}}, \bibinfo {author} {\bibfnamefont {R.}~\bibnamefont {{Paladini}}}, \bibinfo {author} {\bibfnamefont {D.}~\bibnamefont
  {{Paoletti}}}, \bibinfo {author} {\bibfnamefont {B.}~\bibnamefont {{Partridge}}}, \bibinfo {author} {\bibfnamefont {F.}~\bibnamefont {{Pasian}}}, \bibinfo {author} {\bibfnamefont {G.}~\bibnamefont {{Patanchon}}}, \bibinfo {author} {\bibfnamefont {T.~J.}\ \bibnamefont {{Pearson}}}, \bibinfo {author} {\bibfnamefont {O.}~\bibnamefont {{Perdereau}}}, \bibinfo {author} {\bibfnamefont {L.}~\bibnamefont {{Perotto}}}, \bibinfo {author} {\bibfnamefont {F.}~\bibnamefont {{Perrotta}}}, \bibinfo {author} {\bibfnamefont {V.}~\bibnamefont {{Pettorino}}}, \bibinfo {author} {\bibfnamefont {F.}~\bibnamefont {{Piacentini}}}, \bibinfo {author} {\bibfnamefont {M.}~\bibnamefont {{Piat}}}, \bibinfo {author} {\bibfnamefont {E.}~\bibnamefont {{Pierpaoli}}}, \bibinfo {author} {\bibfnamefont {D.}~\bibnamefont {{Pietrobon}}}, \bibinfo {author} {\bibfnamefont {S.}~\bibnamefont {{Plaszczynski}}}, \bibinfo {author} {\bibfnamefont {E.}~\bibnamefont {{Pointecouteau}}}, \bibinfo {author} {\bibfnamefont {G.}~\bibnamefont {{Polenta}}},
  \bibinfo {author} {\bibfnamefont {L.}~\bibnamefont {{Popa}}}, \bibinfo {author} {\bibfnamefont {G.~W.}\ \bibnamefont {{Pratt}}},\ and\ \bibinfo {author} {\bibfnamefont {G.}~\bibnamefont {{Pr{\'e}zeau}}},\ }\bibfield  {title} {\bibinfo {title} {{Planck 2015 results. XIII. Cosmological parameters}},\ }\href {https://doi.org/10.1051/0004-6361/201525830} {\bibfield  {journal} {\bibinfo  {journal} {\aap}\ }\textbf {\bibinfo {volume} {594}},\ \bibinfo {eid} {A13} (\bibinfo {year} {2016})},\ \Eprint {https://arxiv.org/abs/1502.01589} {arXiv:1502.01589 [astro-ph.CO]} \BibitemShut {NoStop}%
\bibitem [{\citenamefont {Zhang}\ \emph {et~al.}(2026)\citenamefont {Zhang}, \citenamefont {Bull},\ and\ \citenamefont {Nicola}}]{canoes2026}%
  \BibitemOpen
  \bibfield  {author} {\bibinfo {author} {\bibfnamefont {Z.}~\bibnamefont {Zhang}}, \bibinfo {author} {\bibfnamefont {P.}~\bibnamefont {Bull}},\ and\ \bibinfo {author} {\bibfnamefont {A.}~\bibnamefont {Nicola}},\ }\bibfield  {title} {\bibinfo {title} {{{\textsc{CANOES}}: Cosmological ANgular Observables and Estimator Suite}},\ }\href@noop {} {\bibfield  {journal} {\bibinfo  {journal} {(companion paper, in preparation)}\ } (\bibinfo {year} {2026})}\BibitemShut {NoStop}%
\bibitem [{\citenamefont {Clarkson}\ \emph {et~al.}(2012)\citenamefont {Clarkson}, \citenamefont {Ellis}, \citenamefont {Faltenbacher}, \citenamefont {Maartens}, \citenamefont {Umeh},\ and\ \citenamefont {Uzan}}]{clarkson2012mis}%
  \BibitemOpen
  \bibfield  {author} {\bibinfo {author} {\bibfnamefont {C.}~\bibnamefont {Clarkson}}, \bibinfo {author} {\bibfnamefont {G.~F.}\ \bibnamefont {Ellis}}, \bibinfo {author} {\bibfnamefont {A.}~\bibnamefont {Faltenbacher}}, \bibinfo {author} {\bibfnamefont {R.}~\bibnamefont {Maartens}}, \bibinfo {author} {\bibfnamefont {O.}~\bibnamefont {Umeh}},\ and\ \bibinfo {author} {\bibfnamefont {J.-P.}\ \bibnamefont {Uzan}},\ }\bibfield  {title} {\bibinfo {title} {(mis) interpreting supernovae observations in a lumpy universe},\ }\href@noop {} {\bibfield  {journal} {\bibinfo  {journal} {Monthly Notices of the Royal Astronomical Society}\ }\textbf {\bibinfo {volume} {426}},\ \bibinfo {pages} {1121} (\bibinfo {year} {2012})}\BibitemShut {NoStop}%
\bibitem [{\citenamefont {{Adamek}}\ \emph {et~al.}(2016)\citenamefont {{Adamek}}, \citenamefont {{Daverio}}, \citenamefont {{Durrer}},\ and\ \citenamefont {{Kunz}}}]{adamek2016gevolution}%
  \BibitemOpen
  \bibfield  {author} {\bibinfo {author} {\bibfnamefont {J.}~\bibnamefont {{Adamek}}}, \bibinfo {author} {\bibfnamefont {D.}~\bibnamefont {{Daverio}}}, \bibinfo {author} {\bibfnamefont {R.}~\bibnamefont {{Durrer}}},\ and\ \bibinfo {author} {\bibfnamefont {M.}~\bibnamefont {{Kunz}}},\ }\bibfield  {title} {\bibinfo {title} {{General relativity and cosmic structure formation}},\ }\href {https://doi.org/10.1038/nphys3673} {\bibfield  {journal} {\bibinfo  {journal} {Nature Physics}\ }\textbf {\bibinfo {volume} {12}},\ \bibinfo {pages} {346} (\bibinfo {year} {2016})},\ \Eprint {https://arxiv.org/abs/1509.01699} {arXiv:1509.01699 [astro-ph.CO]} \BibitemShut {NoStop}%
\bibitem [{\citenamefont {Fleury}\ \emph {et~al.}(2015)\citenamefont {Fleury}, \citenamefont {Larena},\ and\ \citenamefont {Uzan}}]{fleury2015theory}%
  \BibitemOpen
  \bibfield  {author} {\bibinfo {author} {\bibfnamefont {P.}~\bibnamefont {Fleury}}, \bibinfo {author} {\bibfnamefont {J.}~\bibnamefont {Larena}},\ and\ \bibinfo {author} {\bibfnamefont {J.-P.}\ \bibnamefont {Uzan}},\ }\bibfield  {title} {\bibinfo {title} {The theory of stochastic cosmological lensing},\ }\href@noop {} {\bibfield  {journal} {\bibinfo  {journal} {Journal of Cosmology and Astroparticle Physics}\ }\textbf {\bibinfo {volume} {2015}}\bibinfo  {number} { (11)},\ \bibinfo {pages} {022}}\BibitemShut {NoStop}%
\bibitem [{\citenamefont {Chisari}\ \emph {et~al.}(2019)\citenamefont {Chisari}, \citenamefont {Alonso}, \citenamefont {Krause}, \citenamefont {Leonard}, \citenamefont {Bull}, \citenamefont {Neveu}, \citenamefont {Villarreal}, \citenamefont {Singh}, \citenamefont {McClintock}, \citenamefont {Ellison} \emph {et~al.}}]{chisari2019core}%
  \BibitemOpen
\bibfield  {number} {  }\bibfield  {author} {\bibinfo {author} {\bibfnamefont {N.~E.}\ \bibnamefont {Chisari}}, \bibinfo {author} {\bibfnamefont {D.}~\bibnamefont {Alonso}}, \bibinfo {author} {\bibfnamefont {E.}~\bibnamefont {Krause}}, \bibinfo {author} {\bibfnamefont {C.~D.}\ \bibnamefont {Leonard}}, \bibinfo {author} {\bibfnamefont {P.}~\bibnamefont {Bull}}, \bibinfo {author} {\bibfnamefont {J.}~\bibnamefont {Neveu}}, \bibinfo {author} {\bibfnamefont {A.}~\bibnamefont {Villarreal}}, \bibinfo {author} {\bibfnamefont {S.}~\bibnamefont {Singh}}, \bibinfo {author} {\bibfnamefont {T.}~\bibnamefont {McClintock}}, \bibinfo {author} {\bibfnamefont {J.}~\bibnamefont {Ellison}}, \emph {et~al.},\ }\bibfield  {title} {\bibinfo {title} {Core cosmology library: Precision cosmological predictions for lsst},\ }\href@noop {} {\bibfield  {journal} {\bibinfo  {journal} {The Astrophysical Journal Supplement Series}\ }\textbf {\bibinfo {volume} {242}},\ \bibinfo {pages} {2} (\bibinfo {year} {2019})}\BibitemShut {NoStop}%
\bibitem [{\citenamefont {{Takahashi}}\ \emph {et~al.}(2020)\citenamefont {{Takahashi}}, \citenamefont {{Nishimichi}}, \citenamefont {{Namikawa}}, \citenamefont {{Taruya}}, \citenamefont {{Kayo}}, \citenamefont {{Osato}}, \citenamefont {{Kobayashi}},\ and\ \citenamefont {{Shirasaki}}}]{takahashi2020bihalofit}%
  \BibitemOpen
  \bibfield  {author} {\bibinfo {author} {\bibfnamefont {R.}~\bibnamefont {{Takahashi}}}, \bibinfo {author} {\bibfnamefont {T.}~\bibnamefont {{Nishimichi}}}, \bibinfo {author} {\bibfnamefont {T.}~\bibnamefont {{Namikawa}}}, \bibinfo {author} {\bibfnamefont {A.}~\bibnamefont {{Taruya}}}, \bibinfo {author} {\bibfnamefont {I.}~\bibnamefont {{Kayo}}}, \bibinfo {author} {\bibfnamefont {K.}~\bibnamefont {{Osato}}}, \bibinfo {author} {\bibfnamefont {Y.}~\bibnamefont {{Kobayashi}}},\ and\ \bibinfo {author} {\bibfnamefont {M.}~\bibnamefont {{Shirasaki}}},\ }\bibfield  {title} {\bibinfo {title} {{Fitting the Nonlinear Matter Bispectrum by the Halofit Approach}},\ }\href {https://doi.org/10.3847/1538-4357/ab908d} {\bibfield  {journal} {\bibinfo  {journal} {\apj}\ }\textbf {\bibinfo {volume} {895}},\ \bibinfo {eid} {113} (\bibinfo {year} {2020})},\ \Eprint {https://arxiv.org/abs/1911.07886} {arXiv:1911.07886 [astro-ph.CO]} \BibitemShut {NoStop}%
\bibitem [{\citenamefont {Fleury}\ \emph {et~al.}(2017)\citenamefont {Fleury}, \citenamefont {Larena},\ and\ \citenamefont {Uzan}}]{fleury2017weak}%
  \BibitemOpen
  \bibfield  {author} {\bibinfo {author} {\bibfnamefont {P.}~\bibnamefont {Fleury}}, \bibinfo {author} {\bibfnamefont {J.}~\bibnamefont {Larena}},\ and\ \bibinfo {author} {\bibfnamefont {J.-P.}\ \bibnamefont {Uzan}},\ }\bibfield  {title} {\bibinfo {title} {Weak gravitational lensing of finite beams},\ }\href@noop {} {\bibfield  {journal} {\bibinfo  {journal} {Physical Review Letters}\ }\textbf {\bibinfo {volume} {119}},\ \bibinfo {pages} {191101} (\bibinfo {year} {2017})}\BibitemShut {NoStop}%
\bibitem [{\citenamefont {Martin-Garcia}\ \emph {et~al.}(2020)\citenamefont {Martin-Garcia}, \citenamefont {Garc{\'\i}a-Parrado}, \citenamefont {Stecchina}, \citenamefont {Wardell}, \citenamefont {Pitrou}, \citenamefont {Brizuela} \emph {et~al.}}]{martin2020xact}%
  \BibitemOpen
  \bibfield  {author} {\bibinfo {author} {\bibfnamefont {J.~M.}\ \bibnamefont {Martin-Garcia}}, \bibinfo {author} {\bibfnamefont {A.}~\bibnamefont {Garc{\'\i}a-Parrado}}, \bibinfo {author} {\bibfnamefont {A.}~\bibnamefont {Stecchina}}, \bibinfo {author} {\bibfnamefont {B.}~\bibnamefont {Wardell}}, \bibinfo {author} {\bibfnamefont {C.}~\bibnamefont {Pitrou}}, \bibinfo {author} {\bibfnamefont {D.}~\bibnamefont {Brizuela}}, \emph {et~al.},\ }\href@noop {} {\bibinfo {title} {xact: Efficient tensor computer algebra for the wolfram language}} (\bibinfo {year} {2020})\BibitemShut {NoStop}%
\bibitem [{\citenamefont {Mart{\'\i}n-Garc{\'\i}a}(2008)}]{martin2008xperm}%
  \BibitemOpen
  \bibfield  {author} {\bibinfo {author} {\bibfnamefont {J.~M.}\ \bibnamefont {Mart{\'\i}n-Garc{\'\i}a}},\ }\bibfield  {title} {\bibinfo {title} {xperm: fast index canonicalization for tensor computer algebra},\ }\href@noop {} {\bibfield  {journal} {\bibinfo  {journal} {Computer physics communications}\ }\textbf {\bibinfo {volume} {179}},\ \bibinfo {pages} {597} (\bibinfo {year} {2008})}\BibitemShut {NoStop}%
\bibitem [{\citenamefont {Hogg}(1999)}]{hogg1999distance}%
  \BibitemOpen
  \bibfield  {author} {\bibinfo {author} {\bibfnamefont {D.~W.}\ \bibnamefont {Hogg}},\ }\bibfield  {title} {\bibinfo {title} {Distance measures in cosmology},\ }\href@noop {} {\bibfield  {journal} {\bibinfo  {journal} {arXiv preprint astro-ph/9905116}\ } (\bibinfo {year} {1999})}\BibitemShut {NoStop}%
\end{thebibliography}
%

\end{document}